\documentstyle[psfig]{mn}
\def\la{Ly$\alpha$}
\def\etal{et~al.}
\def\ang{\AA}               
\def\spose#1{\hbox to 0pt{#1\hss}}
\def\lta{\mathrel{\spose{\lower 3pt\hbox{$\mathchar"218$}}
     \raise 2.0pt\hbox{$\mathchar"13C$}}}
\def\gta{\mathrel{\spose{\lower 3pt\hbox{$\mathchar"218$}}
     \raise 2.0pt\hbox{$\mathchar"13E$}}}
\def\Ha{H$\alpha$}
\def\Hb{H$\beta$}

\def\clean{{\sc clean}}
\def\mx{{\sc mx}}
\def\uv{{\it uv}}
\def\vtess{{\sc vtess}}
\def\gcomb{{\sc gcombine}}
\def\aips{{\sc aips}}

\title[Radio galaxies at $z \sim 1$]{HST, radio and infrared observations
of 28 3CR radio galaxies at redshift $\mathbf {z \sim 1}$ --- I. The
observations}

\author[P.~N.~Best \etal]{P.~N.~Best,$^{1,2}$ M.~S.~Longair$^1$ and
H.~J.~A.~R\"ottgering$^2$ \\
$^1$ Cavendish Laboratory, Madingley Road,
Cambridge, CB3 0HE, England \\
$^2$ Sterrewacht, Huygens Laboratory, Postbus 9513, 2300 RA Leiden, The
Netherlands: E-mail pbest@strw.leidenuniv.nl} 

\pagerange{\pageref{firstpage}--\pageref{lastpage}}
\pubyear{1995}
\begin{document}
 
\label{firstpage}
 
\maketitle
 
\begin{abstract}

\noindent Hubble Space Telescope images are presented of a sample of 28
3CR radio galaxies with redshifts in the range $0.6 < z < 1.8$, together
with maps at comparable angular resolution of their radio structure, taken
using the Very Large Array. Infrared images of the fields, taken with the
United Kingdom InfraRed Telescope, are also presented. The optical images
display a spectacular range of structures. Many of the galaxies show
highly elongated optical emission aligned along the directions of the
radio axes, but this is not a universal effect; a small number of sources
are either symmetrical or misaligned.  Amongst those sources which do show
an alignment effect, the morphology of the optical emission varies
greatly, from a single bright elongated emission region to strings of
optical knots stretching from one radio hotspot to the other. The infrared
images display much less complexity. Although their significantly lower
angular resolution would wash out some of the smaller structures seen in
the HST images, it is clear that these galaxies are less aligned at
infrared wavelengths than in the optical. In this paper, we discuss the
galaxies individually, but defer a statistical analysis of the
multi-waveband properties of the complete sample of sources to later
papers in this series.
\end{abstract}


\section{Introduction}

High redshift radio galaxies provide well--defined samples of objects for
studies of the distant Universe, for investigating both the formation and
evolution of their stellar populations and, as a result of the strong
interactions of the radio components with their environment, the structure
of the interstellar and intergalactic medium at cosmological
redshifts. The 3CR sample of Laing, Riley and Longair \shortcite{lai83}
contains the brightest radio galaxies in the northern sky, selected at low
radio frequency, and provides a well--defined sample of these objects out
to redshifts $z \sim 2$. This sample provides the basis of the research
presented in this paper.

In 1987, McCarthy \etal\ and Chambers \etal \nocite{cha87,mcc87}
discovered that the optical emission of high redshift radio galaxies has a
strong tendency to be elongated and aligned along the radio axis. Many
different models have been proposed to explain the alignment, but none are
entirely satisfactory (see McCarthy 1993 \nocite{mcc93} for a general
review). The most popular models for this alignment are: (i) massive star
formation induced by shocks associated with the passage of the radio jets
\cite{mcc87,cha87,ree89,beg89,dey89,dal90}; (ii) scattering of light from
an obscured active galactic nucleus by electrons or dust
\cite{dis89,fab89a,cim93a,dis94a,dis96,cim96,dey96}; (iii) nebular
continuum emission from thermal gas \cite{dic95}. It seems likely that
some combination of all three proposals will be required to explain all of
the properties of these active galaxies.

The high--redshift 3CR radio sources are of particular importance for many
reasons. It has been known since the 1960's that the sources in the 3CR
sample exhibit strong cosmological evolution, of exactly the same form as
found for the more populous radio quiet quasars and the faint X--ray
sources (eg. Dunlop 1994).\nocite{dun94} Therefore the objects in the 3CR
sample must contain clues to the origin of these evolutionary effects. In
addition, the 3CR radio galaxies in the complete sample display a very
well defined infrared apparent magnitude {\it vs} redshift relation which
shows evidence for passive evolution of the stellar populations (Lilly and
Longair 1984; Best, Longair and R\"ottgering 1997b, submitted --- hereafter
Paper II).  \nocite{lil84a,bes97d}

Guaranteed--time observations with the Hubble Space Telescope (HST) were
awarded to a programme of imaging the large redshift 3CR radio galaxies,
and the sample was selected in the following way. From the revised 3CR
sample of Laing \etal\ \shortcite{lai83}, a complete subsample of 113
sources was selected which satisfied the selection criteria $z > 0.03$,
$|b| \ge 10^{\circ}$, and $10^{\circ} < \delta < 55^{\circ}$. Of these
sources, 24 are quasars and 89 are radio galaxies. The radio galaxies can
be further subdivided into FR\,I and FR\,II classes on the basis of their
radio structures \cite{fan74}, the FR\,II class being the classical double
radio sources with hotspots towards the leading edges of the diffuse radio
lobes.

\begin{figure}
\centerline{
\psfig{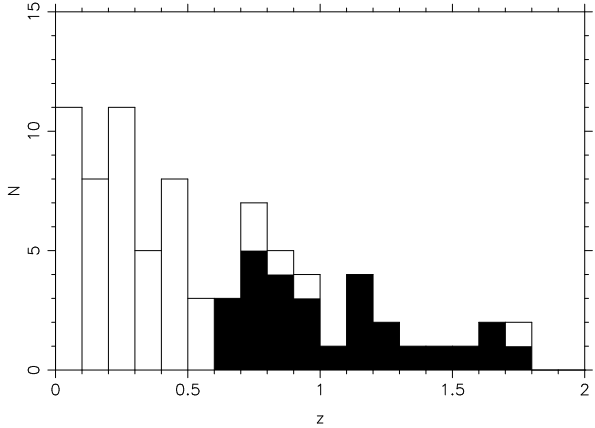}
}
\caption{\label{zhist} A histogram showing the redshift distribution of
the complete subsample of 3CR radio galaxies. Those galaxies which
constitute the HST programme are indicated by filled boxes.}
\end{figure}

For our programme, the FR\,II radio galaxies in the redshift range $0.6 <
z < 1.8$ were initially selected for observation. In fact, 5 radio
galaxies of the 33 in this redshift range were not observed. Two sources
(3C55 and 3C263.1) had originally been wrongly identified with galaxies at
lower redshifts; the redshifts of the revised identifications indicate
that they should have been included in the sample. In the other three
cases (3C175.1, 3C294 and 3C318) the galaxies were omitted at random from
the programme because of the constraints of observing time. As a result,
the sample selected is not quite complete, but the omission of these 5
galaxies should not introduce any serious selection effects.  The redshift
distribution for the complete subsample of 79 3CR FR\,II radio galaxies is
shown in Figure~\ref{zhist}, the 28 radio galaxies which constitute this
programme being indicated by filled boxes.

In the present paper, observations of the 28 radio galaxies are presented
at various wavelengths. In addition to the HST observations, all the radio
galaxies have been imaged at radio frequencies using the Very Large Array
interferometer (VLA) at 8.4\,GHz, and in the near infrared using the
United Kingdom InfraRed Telescope (UKIRT). The first results of this
programme were presented by Longair \etal\ \shortcite{lon95}, and a
detailed analysis of the optical morphologies of the 8 galaxies in the
redshift range $1 \lta z \lta 1.3$ was presented by Best \etal\
\shortcite{bes96a}. The interesting case of a jet--cloud interaction in
the source 3C34 has also been presented \cite{bes97b}.

In Section~2 we provide details of the observing techniques and the data
reduction. In Section~3 the main observational results of the programme
are presented in the form of images and radio maps of the galaxies in the
sample. The most important features of each source are noted, together
with a short discussion of the implications of these observations for our
understanding of that source. In Section~4 we briefly discuss the
significance of these results for understanding the astrophysics of
powerful radio galaxies, but defer most of the discussion to Papers II and
III of this series. In Paper II \cite{bes97d} we investigate the old
stellar populations of these galaxies: we compare their radial intensity
profiles with de Vaucouleurs' law, investigate the evolution of the
environments, luminosities and characteristic sizes of the galaxies with
cosmic epoch, and discuss the significance of our observations for the use
of these galaxies as cosmological probes. In Paper III (Best \etal, in
preparation) we investigate the multi--wavelength properties of the
sources, and discuss the alignment effect.

Throughout this paper, all positions are given in equinox J2000
coordinates, and we assume $H_0 = 50$\,km\,s$^{-1}$\,Mpc$^{-1}$ and
$\Omega_0 = 1$.

\section{Observations and data reduction}

\subsection{Hubble Space Telescope observations}
\label{hstobs}

\begin{table*}
\begin{center}
\begin{tabular}{lcccccclcclc}

Source & z     & Obs.   & Filter & Cent. & Exp. & Ap.  & WFPC  & Error & Emission              &\% flux  &Refs$^3$\\
       &       & Date   &        & Wave. & Time & Diam & Mag.$^1$&     & Lines $^2$            &in lines &        \\
       &       &        &        & [\ang]& [s]  &[$''$]&       &       &                            &         &        \\
\hspace{3mm}(1)&(2)&(3) &  (4)   &  (5)  &  (6) &  (7) &\hspace{2.5mm}(8)& (9)  &  (10)      & \hspace{2.0mm}(11)&(12) \\
3C13   & 1.351 &03/06/94& f785LP & 8620  & 1700 &   5  & 20.02 & 0.05  & [NeV], [OII], [NeIII]      &$\approx 35\%$& A  \\
       &       &20/09/95& f814W  & 7880  & 2800 &   5  & 20.42 & 0.04  & [NeV], [OII], [NeIII]      &$\approx 19\%$& A  \\
3C22   & 0.938 &07/09/94& f622W	 & 6160  & 1700 &   9  & 20.19 & 0.15  & MgII, [NeV]                &$\approx \hspace{1.9mm}5\%$& B \\
       &       &03/09/95& f814W  & 7880  & 1400 &   9  & 18.95 & 0.03  &[OII], [NeIII], \Hb , [OIII]&$\sim  16\%^4$& B  \\
3C34   & 0.690 &10/06/94& f555W  & 5400  & 1700 &   5  & 21.92 & 0.09  & MgII, [NeV], [OII], [NeIII]&$\approx 12\%$& C  \\
       &       &10/06/94& f785LP & 8620  & 1700 &   5  & 19.10 & 0.07  & \Hb , [OIII]               &$\approx 20\%$& C  \\
3C41   & 0.795 &29/07/94& f555W  & 5400  & 1700 &   9  & 21.68 & 0.09  & MgII, [NeV], [OII]         &$\approx \hspace{1.9mm}6\%$& D \\
       &       &29/07/94& f785LP & 8620  & 1700 &   9  & 19.00 & 0.04  & \Hb , [OIII]               &$\sim  11\%^4$& D  \\
3C49   & 0.621 &31/10/95& f555W  & 5400  & 1400 &   9  & 20.92 & 0.06  & MgII, [NeV], [OII]         &$\sim  10\%^5$& E  \\
       &       &31/10/95& f814W  & 7880  & 1400 &   9  & 19.07 & 0.04  & \Hb , [OIII]               &$\sim \hspace{1.9mm}8\%^4$& E  \\
3C65   & 1.176 &02/02/95& f675W  & 6700  & 1700 &   9  & 21.88 & 0.17  & MgII, [NeV]                &$\approx \hspace{1.9mm}2\%$&F,G\\
       &       &02/02/95& f814W  & 7880  & 1760 &   9  & 20.60 & 0.11  & [NeV], [OII], [NeIII]      &$\approx \hspace{1.9mm}5\%$&F,G\\
3C68.2 & 1.575 &15/07/94& f785LP & 8620  & 3400 &   9  & 21.05 & 0.11  & [NeV], [OII], [NeIII]      &$\sim \hspace{1.9mm}7\%^6$&    \\
3C217  & 0.897 &26/02/95& f622W  & 6160  & 1700 &   9  & 21.45 & 0.08  & [NeV]                      &$\sim \hspace{1.9mm}4\%^4$& E  \\
       &       &26/02/95& f814W  & 7880  & 1700 &   9  & 20.01 & 0.06  &[OII], [NeIII], \Hb , [OIII]&$\sim  11\%^5$& E  \\
3C226  & 0.820 &04/05/94& f555W  & 5400  & 1700 &   9  & 21.44 & 0.10  & MgII, [NeV],[OII]          &$\approx \hspace{1.9mm}7\%$&H,I\\
       &       &04/05/94& f785LP & 8620  & 1700 &   9  & 19.00 & 0.06  & \Hb , [OIII]               &$\sim  14\%^4$& H  \\
3C239  & 1.781 &21/05/94& f785LP & 8620  & 1700 &   5  & 20.67 & 0.08  & MgII, [NeV], [OII]         &$\sim \hspace{1.9mm}8\%^7$& J  \\
       &       &14/02/96& f814W  & 7880  & 2200 &   5  & 21.00 & 0.05  & MgII, [NeV]                &$\sim \hspace{1.9mm}6\%^7$& J  \\
3C241  & 1.617 &02/06/94& f785LP & 8620  & 1700 &   9  & 21.25 & 0.24  & [NeV], [OII], Ne[III]      &$\sim \hspace{1.9mm}5\%^8$& A  \\
       &       &29/04/96& f814W  & 7880  & 2400 &   9  & 21.38 & 0.09  & MgII, [NeV]                &$\approx \hspace{1.9mm}8\%$& A \\
3C247  & 0.749 &29/03/96& f555W  & 5400  & 2400 &   5  & 21.48 & 0.03  & MgII, [NeV], [OII], [NeIII]&$\sim  21\%^5$& H  \\
       &       &29/03/96& f814W  & 7880  & 2400 &   5  & 19.27 & 0.03  & \Hb , [OIII]               &$\sim  15\%^4$& H  \\
3C252  & 1.105 &27/11/94& f622W  & 6160  & 1700 &   9  & 21.14 & 0.07  & MgII                       &$\sim \hspace{1.9mm}6\%^4$& H  \\
       &       &27/11/94& f814W  & 7880  & 1700 &   9  & 20.13 & 0.07  & [NeV], [OII], [NeIII]      &$\sim  17\%^5$& H  \\
3C265  & 0.811 &29/05/94& f555W  & 5400  & 1700 &   9  & 19.92 & 0.05  & MgII, [NeV], [OII]         &$\approx \hspace{1.9mm}5\%$&K,L\\
       &       &29/05/94& f785LP & 8620  & 1700 &   9  & 17.86 & 0.03  & \Hb , [OIII]               &$\sim  21\%^4$& K,L\\
3C266  & 1.272 &15/03/95& f555W  & 5400  & 1800 &   9  & 21.53 & 0.06  & CII], [NeIV], MgII         &$\sim  12\%^4$& H  \\
       &       &15/03/95& f702W  & 6860  & 1700 &   9  & 20.81 & 0.07  & MgII, [NeV], [OII]         &$\sim  10\%^4$& H  \\
       &       &15/03/95& f814W  & 7880  & 1800 &   9  & 20.26 & 0.03  & [NeV], [OII], [NeIII]      &$\sim  16\%^5$& H  \\
3C267  & 1.144 &11/06/94& f702W  & 6860  & 1700 &   9  & 21.06 & 0.09  & MgII, [NeV], [OII], [NeIII]&$\sim  \hspace{1.9mm}8\%^5$& H  \\
       &       &11/06/94& f791W  & 7790  & 1800 &   9  & 20.59 & 0.06  & [NeV], [OII], [NeIII]      &$\sim  14\%^5$& H  \\
3C277.2& 0.766 &20/06/96& f555W  & 5400  & 2400 &   9  & 21.28 & 0.04  & MgII, [NeV], [OII]         &$\approx 11\%$& I  \\
       &       &20/06/96& f814W  & 7880  & 2400 &   9  & 19.51 & 0.03  & \Hb , [OIII]               &$\sim  \hspace{1.9mm}9\%^4$& I  \\
3C280  & 0.996 &23/08/94& f622W  & 6160  & 1700 &   9  & 20.80 & 0.05  & MgII, [NeV]                &$\sim  15\%^4$& M  \\
       &       &25/03/96& f814W  & 7880  & 2200 &   9  & 19.78 & 0.04  & [OII], [NeIII]             &$\sim  23\%^5$& M  \\
3C289  & 0.967 &19/03/95& f622W  & 6160  & 1700 &   9  & 21.45 & 0.10  & MgII, [NeV]                &$\sim  \hspace{1.9mm}4\%^4$& E  \\
       &       &19/03/95& f814W  & 7880  & 1800 &   9  & 20.01 & 0.13  & [OII], [NeIII]             &$\sim  \hspace{1.9mm}7\%^5$& E  \\
3C324  & 1.206 &27/04/94& f702W  & 6860  & 1700 &   9  & 21.24 & 0.07  & MgII, [NeV], [OII]         &$\approx 11\%$& I,N\\
       &       &27/04/94& f791W  & 7790  & 1800 &   9  & 20.39 & 0.08  & [NeV], [OII], [NeIII]      &$\approx 26\%$& I  \\
3C337  & 0.635 &24/08/95& f555W  & 5400  & 1400 &   9  & 22.05 & 0.13  & MgII, [NeV], [OII], [NeIII]&$\sim  \hspace{1.9mm}8\%^5$& E  \\
       &       &24/08/95& f814W  & 7880  & 1400 &   9  & 19.57 & 0.04  & \Hb , [OIII]               &$\sim  \hspace{1.9mm}5\%^4$& E  \\
3C340  & 0.775 &25/04/94& f555W  & 5400  & 1700 &   9  & 21.71 & 0.07  & MgII, [NeV], [OII]         &$\sim  \hspace{1.9mm}8\%^5$& H  \\
       &       &25/04/94& f785LP & 8620  & 1700 &   9  & 19.37 & 0.09  & \Hb , [OIII]               &$\sim  21\%^4$& H  \\
3C352  & 0.806 &22/02/95& f555W  & 5400  & 1700 &   5  & 21.79 & 0.10  & MgII, [NeV], [OII]         &$\approx \hspace{1.9mm}5\%$& L,O\\
       &       &22/02/95& f814W  & 7880  & 1800 &   5  & 19.95 & 0.04  & \Hb , [OIII]               &$\sim  13\%^4$& O  \\
3C356  & 1.079 &06/03/95& f622W  & 6160  & 1700 &   5  &21.54$^9$& 0.08& MgII                       &$\approx \hspace{1.9mm}3\%$& P  \\
       &       &06/03/95& f814W  & 7880  & 1700 &   5  &20.43$^9$& 0.07& [NeV], [OII], [NeIII]      &$\approx 18\%$& P  \\
3C368  & 1.132 &18/06/94& f702W  & 6860  & 1700 &   9&19.96$^{10}$&0.10& MgII, [NeV], [OII], [NeIII]&$\approx 24\%$&Q,R \\
       &       &18/06/94& f791W  & 7790  & 1800 &   9&19.34$^{10}$&0.10& [NeV], [OII], [NeIII]      &$\approx 38\%$&Q,R \\
3C437  & 1.480 &25/04/94& f785LP & 8620  & 3400 &   9  & 21.26 & 0.28  & [NeV], [OII], [NeIII]      &$\sim  12\%^5$& H  \\
3C441  & 0.708 &30/05/94& f555W  & 5400  & 1700 &   5  & 21.53 & 0.07  & MgII, [NeV], [OII], [NeIII]&$\approx \hspace{1.9mm}6\%$& D  \\
       &       &30/05/94& f785LP & 8620  & 1700 &   5  & 19.00 & 0.04  & \Hb , [OIII]               &$\sim  10\%^4$& D  \\
3C470  & 1.653 &19/10/94& f785LP & 8620  & 3500 &   5  & 21.69 & 0.38  & [NeV], [OII]               &$\sim  15\%^7$& H  \\
\end{tabular}
\end{center}
\caption{\label{hsttab} The parameters of the HST observations. The source
name is given in column (1), and its redshift in column (2). Column (3)
gives the date of the observation using the filter in column (4), the
central wavelength of which is given in column (5). Column (6) gives the
exposure time for this observation, measured in seconds. The aperture
diameter used for the photometry (see Section~\ref{hstobs}) is given in
column (7), and the WFPC2 magnitude within this aperture, and its error,
are given in columns (8) and (9). Column (10) lists the emission lines
expected to be contained within this filter and column (11) gives an
estimate of the percentage of flux through the filter that is associated
with line emission, based upon line luminosities taken from the references
in column (12).\ \ \ $^1$ All magnitudes have been corrected for galactic
extinction using the extinction maps of Burstein and Heiles (1982). \ \ \
$^2$ The emission lines in column (10) refer to the following prominent
lines: CII]~2326, [NeIV]~2426, MgII~2798, [NeV]~3346,3426, [OII]~3727,
[NeIII]~3869, \Hb ~4861, [OIII]~4959,5007.\ \ \ $^3$ References (see
bibliography for abbreviations): A -- SD84, B -- RLSE95, C -- BLR97a, D --
PLLD84 (continued on next page)}
\end{table*}
\nocite{spi82,spi84b,raw95,per84,sto95,lac95,dis94a,mcc88,dey96,bes97b}
\nocite{smi79,hip92,lac94,djo87,ham91,spi85b,bur82a}

\addtocounter{table}{-1}
\begin{table*}
\caption{cont.  E -- Spinrad (private communication), F --
SKR95, G -- LRED95, H -- M88, I -- dSACF94, J -- SFWSWL85, K -- DS96, L --
SJSGV79, M -- S82, N -- DDS96, O -- HM92, P -- LR94, Q -- DSPRS87, R --
HLFP91.\ \ \ $^4$ No spectral data have been published at the
longest\,/\,shortest wavelengths, and so the flux is estimated from the
[OII]~3727 line flux, assuming that the flux ratios of the other lines
relative to [OII] 3727 are: [OIII]~4959,5007 $\sim 3$, \Hb~4861 $\sim
0.3$, [NeV]~3426 $\sim 0.2$, MgII~2798 $\sim 0.25$, [NeIV]~2426 $\sim 0.1$
and CII]~2326 $\sim 0.25$ (McCarthy 1988).\ \ \ $^5$ Estimated from
[OII]~3727 line flux only.\ \ \ $^6$ No spectral data have been published
for 3C68.2, and so the line emission is estimated using mean line
fluxes. $^7$ No spectral data have been published at these wavelengths for
the highest redshift sources, and so the line emission is estimated using
the \la\ fluxes, and assuming a line ratio of \la\ / [OII]~3727 $\sim 5$
(McCarthy 1988).\ \ \ $^8$ Estimated from [NeV]~3426 flux only ---
[OII]~3727 falls at very low filter transmission, and so will not
dominate.\ \ \ $^9$ Note that the magnitudes quoted for 3C356 are for the
more northerly galaxy. The corresponding values for the southern object
are $22.32 \pm 0.11$ and $20.78 \pm 0.10$ respectively.\ \ \ $^{10}$ Note
that the magnitudes quoted for 3C368 are after best subtraction of the
M--star. Including the star, the magnitudes are $19.74\pm 0.05$ and
$19.05\pm 0.06$ respectively.}
\end{table*}

During cycle 4, 24 of the galaxies were imaged with the Wide--Field
Planetary Camera II (WFPC2) of the HST for one orbit, approximately 30
minutes, generally in each of two wavebands. During cycle 5, the remaining
four galaxies were observed, and longer observations were made of three of
the highest redshift ($z > 1.3$) galaxies, to achieve comparable
signal--to--noise ratios as for those at lower redshift. Details of the
filters and exposures times are given in Table~\ref{hsttab}.

Each orbit's observation was split into two exposures, and the first
calibration steps were carried out separately on each, according to the
standard Space Telescope Science Institute (STScI) calibration pipeline
\cite{lau89}. The two calibrated images were combined using the STSDAS
task \gcomb , removing well over 95\% of the cosmic rays. The remainder of
the cosmic ray events were distinguished from stars and galaxies by their
radial profiles and by comparison of the images in the two filters. These
were then removed individually, replacing each affected pixel with the
mean value of those surrounding it.

The galaxy fluxes were, in general, obtained through a circular aperture
of 9 arcsec diameter, although the presence of a companion galaxy within
this aperture in a minority of sources led to a 5 arcsec diameter being
adopted instead. The 9 arcsec diameter is sufficiently large as to enclose
virtually all of the galaxy light. Subtraction of the background flux was
performed using the average flux contained within four or more apertures
of the same size placed on blank areas of the sky, rather than through an
annulus surrounding the object, since the latter was often contaminated by
companion objects.  These sky apertures were placed as close as possible
to the source, at different position angles relative to it, to avoid
introducing any errors from residual gradients in the background
flux. Conversion of the counts to WFPC2 magnitudes was carried out
according to the prescription of Holtzman \etal\ \shortcite{hol95}. These
were then corrected for galactic extinction using the extinction maps of
Burstein and Heiles \shortcite{bur82a}, and the results are presented in
Table~\ref{hsttab}.

A number of effects contribute to the errors in the photometric
magnitudes: (i) the Poisson noise of the detected counts; (ii) accurate
determination of the mean sky background by measurement through different
blank apertures; (iii) sky noise within the source aperture; (iv)
uncertainties in the charge transfer efficiency of the WFPC2 CCD's ($\lta
2\%$, Holtzman \etal\ 1995); (v) uncertainties in the gain ratios of the
WFPC2 chips ($\sim 1\%$, Holtzman \etal\ 1995); (vi) errors in the
photometry due to aperture extrapolation from the 0.5 arcsec diameter
aperture used by Holtzman \etal\ to the apertures used in this paper. Of
these, it is found that the sky noise and the sky subtraction are usually
the dominant errors.

Throughout this paper where reference is made to colour differences
between components within a radio galaxy, these are quoted as the
difference between the two WFPC2 magnitudes. Conversion to ground--based
magnitudes would introduce uncertainties which are greater than the
differential colours under discussion. Readers who wish to make accurate
conversions to ground--based magnitudes are referred to the tables in the
paper by Holtzman \etal\ \shortcite{hol95}. As a rough guide to these
conversions, for a typical 3CR galaxy at redshift one the colour
differential f555W$-$f814W is approximately equivalent to $V-I-0.1$,
f555W$-$f785LP to $V-I+0.3$, f622W$-$f814W to $R-I+0.5$, f555W$-$K to
$V-K+0.1$, and f702W$-$K to $R-K-0.15$, although in each case the exact
conversion depends upon the spectral energy distribution of the component
being studied.

\subsection{Radio observations}
\label{vlaobs}

To complement the HST observations, the structures of the radio sources
were mapped using the A--array configuration of the VLA at a frequency of
8.4\,GHz, providing an angular resolution of $\sim$ 0.18 arcsec, comparable
to that of the HST observations. The sources smaller than 40 arcsec were
observed for 22--minutes using a bandwidth of 50\,MHz, whilst the larger
sources were observed for 44--minutes using a narrower bandwidth of 25\,MHz,
to avoid chromatic aberration problems. These data were taken on 27th
February, 1994, shortly after the conversion from D to A array. The data
from four of the antenna along the northern arm of the array were unusable
due to pointing\,/\,baseline errors. The rms noise was typically of order $60
\mu$Jy.

Radio sources larger than 10 arcsec in extent were also observed using the
B--array of the VLA, so that their extended radio structures could be
mapped; these exposures were for between 25 and 30 minutes, and were taken
on the 11th and 14th July, 1994. Similarly, sources over 25 arcsec in size
were observed in C--array configuration on 5th and 7th December,
1994. Sources were not observed using a particular array configuration if
another observer had already mapped them at 8.4~GHz in that
configuration, with an exposure time in excess of twenty minutes. Details
are provided in Table~\ref{vlatab}.

The observations were made using standard VLA procedures. The bright
sources 3C286 and 3C48 were used for primary flux calibration, whilst
accurate phase calibration was achieved by frequent short observations of
secondary calibrator sources within a few degrees of the target. The data
were reduced using the \aips\ software \cite{per89} provided by the
National Radio Astronomy Observatory (NRAO).  The data from each different
array were individually \clean ed using the \aips\ task \mx , and then
phase self--calibration was used to improve further the map quality. Care
was taken that only real features were included in the self--calibration
model. This self--calibrated dataset was once again \clean ed, producing
an improved map, and in a number of cases self--calibration was then
repeated. One side--effect of self--calibration is that the position of
the peak flux may be displaced by up to a pixel; this effect was minimised
by oversampling the resolution by a factor of four, and by comparing the
position of the self--calibrated peak with that of the original
data. Discrepancies were corrected by shifting the self--calibrated data
by hand, although none were greater than 0.2 arcsec.

\begin{table}
\begin{center}
\begin{tabular}{lrcccr}
Source & Radio               & VLA      & Exp.    &Flux dens.& Core\hspace{0.8mm}  \\ 
       & Size\hspace{1.3mm}  & Configs. & Time    & 8.4 GHz & Flux\hspace{0.8mm}  \\
       & [kpc]\hspace{0.5mm} &          & [min]   & [mJy]   & [mJy] \\
\hspace{2mm}(1)& (2) \hspace{1.7mm} &  (3)   & (4)     & (5)     & (6)\hspace{1.6mm}   \\
\\     				     	  
3C13   & 246 \hspace{0.9mm}  & A,B      & 22,29   & 212     & 0.18  \\
3C22   & 208 \hspace{0.9mm}  & A        & 22      & 339     & 7.02  \\
3C34   & 372 \hspace{0.9mm}  & A,C      & 44,30   & 208     & 1.20  \\
3C41   & 197 \hspace{0.9mm}  & A,B      & 22,25   & 975     & 0.57  \\
3C49   & 7 \hspace{0.9mm}    & A        & 22      & 493    &$<1.30$ \\
3C65   & 155 \hspace{0.9mm}  & A,B      & 22,25   & 427     & 0.52  \\
3C68.2 & 203 \hspace{0.9mm}  & A,B      & 22,29   & 84      & 0.13  \\
3C217  & 110 \hspace{0.9mm}  & A,B      & 22,27   & 247    &$<0.13$ \\
3C226  & 263 \hspace{0.9mm}  & A,C      & 22,19   & 288     & 2.83  \\
3C239  & 102 \hspace{0.9mm}  & A,B      & 22,27   & 160     & 0.44  \\
3C241  & 7 \hspace{0.9mm}    & A        & 22      & 164    &$<0.16$ \\
3C247  & 113 \hspace{0.9mm}  & A,B      & 22,27   & 492     & 4.21  \\
3C252  & 488 \hspace{0.9mm}  & A        & 44      & 76      & 1.79  \\
3C265  & 646 \hspace{0.9mm}  & A        & 44      & 358     & 1.68  \\
3C266  & 39 \hspace{0.9mm}   & A        & 22      & 150    &$<0.18$ \\
3C267  & 329 \hspace{0.9mm}  & A,B,C    & 22,27,19& 435     & 1.41  \\ 
3C277.2& 432 \hspace{0.9mm}  & A        & 44      & 148     & 0.36  \\
3C280  & 117 \hspace{0.9mm}  & B        & 27      & 968    &$<1.60$ \\
3C289  & 89 \hspace{0.9mm}   & A,B      & 22,27   & 356     & 1.37  \\
3C324  & 96 \hspace{0.9mm}   & A        & 22      & 228    &$<0.14$ \\
3C337  & 342 \hspace{0.9mm}  & A,B,C    & 44,27,30& 767    &$<0.11$ \\
3C340  & 371 \hspace{0.9mm}  & A,B,C    & 44,27,30& 811     & 0.83  \\
3C352  & 102 \hspace{0.9mm}  & B        & 25      & 295     & 3.18  \\
3C356  & 624 \hspace{0.9mm}  & B,C      & 25,30   & 204     & 0.22  \\
3C368  & 73 \hspace{0.9mm}   & A        & 22      & 80     &$<0.14$ \\
3C437  & 318 \hspace{0.9mm}  & A,B,C    & 22,25,30& 543    &$<0.11$ \\
3C441  & 266 \hspace{0.9mm}  & A,C      & 22,30   & 380    &$<0.21$ \\
3C470  & 211 \hspace{0.9mm}  & A,B      & 22,29   & 276     & 1.20  \\
\end{tabular}
\end{center}
\caption{\label{vlatab} The parameters of the VLA observations. Column (1)
gives the name of the source, and its size, measured from the edge of the
bright emission of one radio lobe to that of the other on our radio maps,
is given in column (2). The VLA configurations used to observe each source
are given in column (3) with the corresponding integration times, measured
in minutes, given in column (4). Column (5) gives the measured flux
density of the source at 8.4~GHz. This is measured from the lowest
resolution array map available; for sources of which we did not make an
observation at sufficiently low resolution to detect the extended flux,
the values are taken from the lower resolution maps of other
observers. The flux density of the radio source core is given in column
(6). Where a core is not detected, the $3\sigma$ upper limit is given.\ \
\ $^1$In the case of 3C356, this is the flux density of the northern radio
core candidate. The southern core candidate of that source has a flux
density of 0.95\,mJy.}
\end{table}

The \uv\ \,data from the different array configurations were then merged
by self--calibrating the lower resolution \uv\ \,data with that of the
highest resolution array. A combined map was produced and the data were
once again self--calibrated. The final map was then made using a hybrid
combination of the \aips\ tasks \mx\ and the maximum entropy technique,
\vtess\ \cite{lea91}. Whilst the \clean\ method is very good for
deconvolving point sources, or sources which can reasonably be described
as a combination of point sources, it is not particularly effective at
\clean ing diffuse extended structure. On the other hand, \vtess\ uses a
maximum entropy technique and tends to produce much smoother maps than \mx
. It struggles to deal with bright point--like sources of emission such as
the cores and the hot--spots. The technique used was therefore to \clean\
the map to remove the brighter point--like emission, then to use \vtess\
for the remaining diffuse emission, and finally to add back the \clean\
components to produce a final map of the source.

\subsection{Infrared observations}
\label{ukobs}

Infrared imaging of the galaxies was carried out using IRCAM3 of UKIRT on
the 6th to 8th August 1994 and the 2nd to 4th February 1995. IRCAM3 is an
infrared camera for the 1 to 5\,$\mu$m waveband, and incorporates a $256
\times 256$ SBRC InSb array, with 0.286 arcsec pixels. All the galaxies
were observed in the K--band ($2.2\,\mu$m), generally for 54 minutes, and
the majority were also observed at $1.2\,\mu$m through the J waveband (see
Table~\ref{uktab}). For the August 1994 run, the first two nights and half
of the final night were photometric, and the seeing was typically about 
1 arcsec. Short service observations were later taken of those sources
observed during the non--photometric conditions of the third night, to
provide photometric magnitudes. Conditions were photometric throughout the
February 1995 run, but the observations were hampered by high wind
speeds. The seeing therefore varied from about 0.8 to 1.3 arcsec,
depending upon the orientation of the telescope, and was frequently
elliptical with its long axis aligned roughly east--west. A proportion of
this ellipticity may be attributable to RA judder of the telescope. Those
images which were affected by elliptical seeing are indicated in
Table~\ref{uktab}.

The observations were made using a 9 point jittering technique, with
offsets of 15 arcseconds between each 1 minute exposure. After subtraction
of a dark frame, and removal of known bad pixels, the images were median
filtered to construct an accurate sky flat--field. The nine flat--fielded
images were then accurately registered using the peak positions of two or
more bright unresolved objects visible on all images, and were
summed. This provided a mosaiced image of approximately 100\,$\times$\,100
arcsec$^2$, although the highest signal--to--noise ratio is only available
in the central 45\,$\times$\,45 arcsec$^2$. In general six such mosaiced
images were made, enabling accurate removal of cosmic ray events before
these were registered and summed. These images were aligned with the HST
data by using several unresolved sources visible on both images. In most
cases this was possible to an accuracy significantly better than one pixel
size, 0.286 arcsec.

Flux calibration was achieved by frequent observations of the UKIRT Faint
Standards. Photometry was performed through either a 9 arcsec or 5 arcsec
diameter aperture, matching the apertures adopted for the HST data in
Table~\ref{hsttab}. As with the HST observations, the background flux was
measured using apertures placed in empty regions of sky, within the
central 45 arcsec square. Accurate subtraction of the background flux is
the main source of the quoted errors in the photometric magnitudes,
presented in Table~\ref{uktab}. The extinction maps of Burstein and Heiles
(1982) were used to correct for galactic extinction.

\begin{table*}
\begin{center}
\begin{tabular}{lccclcccclcc}
Source & \multicolumn{6}{c}{K--Band}                     &     \multicolumn{5}{c}{J--Band}        \\ 
       & Observ.  & Exp. & Ap. &Mag.$^1$& Error & Seeing & Observ. & Exp.&Mag.$^1$& Error & Seeing\\
       & Date     & Time & Diam &       &       &        & Date    & Time &       &       &       \\
       &          & [min]&[$''$]&       &       &        &         &[min] &       &       &       \\ 
\hspace{2mm}(1)&(2)& (3) & (4)  &\hspace{1.7mm}(5)&(6)&(7)&(8)     & (9)  &\hspace{1.7mm}(10)& (11) & (12)\\
\\ 
3C13   & 06/08/94 & 54   &  5   & 17.52 & 0.06  &   o    &07/08/94 &  54  & 18.53 & 0.06  &   o   \\
3C22   & 08/08/94 & 27   &  9   & 15.40 & 0.15  &   e    &08/08/94 &  27  & 16.01 & 0.20  &   e   \\
3C34   & 06/08/94 & 54   &  5   & 16.43 & 0.05  &   o    &07/08/94 &  54  & 18.15 & 0.06  &   o   \\
3C41   & 07/08/94 & 36   &  9   & 15.68 & 0.04  &   o    &08/08/94 &  27  & 18.47 & 0.25  &   o   \\
3C49   & 08/08/94 & 27   &  9   & 16.15 & 0.15  &   o    &  ---    &  --- &  ---  & ---   &  ---  \\
3C65   & Service  & 54   &  9   & 16.59 & 0.07  &   o    &04/02/95 &  54  & 18.41 & 0.06  &   o   \\
3C68.2 & 07/08/94 & 54   &  9   & 17.49 & 0.12  &  se    &04/02/95 &  54  & 19.15 & 0.18  &   o   \\
3C217  & 02/02/95 & 54   &  9   & 17.52 & 0.08  &   o    &04/02/95 &  45  & 18.65 & 0.06  &  se   \\
3C226  & 03/02/95 & 54   &  9   & 16.52 & 0.05  &   o    &04/02/95 &  45  & 18.05 & 0.05  &   o   \\
3C239  & 02/02/95 & 54   &  5   & 17.83 & 0.06  &   o    &04/02/95 &  27  & 18.96 & 0.08  &   o   \\
3C241  & 03/02/95 & 54   &  9   & 17.45 & 0.08  &   o    &04/02/95 &  27  & 18.86 & 0.10  &   o   \\ 
3C247  & 04/02/95 & 54   &  5   & 15.96 & 0.02  &  se    &  ---    &  --- & ---   & ---   &  ---  \\
3C252  & 02/02/95 & 54   &  9   & 17.32 & 0.07  &   o    &  ---    &  --- & ---   & ---   &  ---  \\
3C265  & 02/02/95 & 54   &  9   & 16.03 & 0.04  &  se    &03/02/95 &  45  & 17.20 & 0.05  &  se   \\
3C266  & 04/02/95 & 54   &  9   & 17.65 & 0.09  &   o    &  ---    &  --- & ---   & ---   &  ---  \\
3C267  & 03/02/95 & 54   &  9   & 17.21 & 0.05  &   o    &04/02/95 &  45  & 18.74 & 0.06  &   o   \\
3C277.2& 03/02/95 & 54   &  9   & 16.96 & 0.05  &   e    &04/02/95 &  18  & 18.32 & 0.07  &   e   \\
3C280  & 02/02/95 & 54   &  9   & 16.70 & 0.04  &   o    &04/02/95 &  45  & 18.07 & 0.05  &   o   \\
3C289  & 02/02/95 & 54   &  9   & 16.66 & 0.07  &   o    &04/02/95 &  18  & 18.19 & 0.06  &   o   \\
3C324  & 06/08/94 & 54   &  9   & 16.99 & 0.06  &   o    &07/08/94 &  63  & 18.58 & 0.11  &   o   \\ 
3C337  & 03/02/95 & 45   &  9   & 16.57 & 0.05  &  se    &03/02/95 &  27  & 18.08 & 0.05  &   o   \\
3C340  & 06/08/94 & 54   &  9   & 16.91 & 0.08  &   e    &07/08/94 &  45  & 18.30 & 0.09  &  se   \\ 
3C352  & 08/08/94 & 45   &  5   & 16.92 & 0.05  &  se    &  ---    &  --- & ---   & ---   &  ---  \\
3C356  & 08/08/94 & 54   &  5   &17.50$^2$& 0.06&   o    &  ---    &  --- & ---   & ---   &  ---  \\ 
3C368  & 06/08/94 & 54   &  9   &17.03$^3$&0.15 &  se    &07/08/94 &  54  &18.43$^3$&0.15 &   e   \\
3C437  & 08/08/94 & 54   &  9   & 17.74 & 0.20  &  se    &  ---    &  --- & ---   & ---   &  ---  \\
3C441  & 06/08/94 & 54   &  5   & 16.42 & 0.04  &   e    &07/08/94 &  54  & 18.00 & 0.06  &  se   \\
3C470  & 08/08/94 & 45   &  5   & 18.02 & 0.15  &   o    &  ---    & ---  & ---   & ---   &  ---  \\
\end{tabular}  
\end{center}
\caption{\label{uktab} The parameters of the UKIRT observations. The
source name is given in column (1). Columns (2) and (3) give the
observation date and exposure times for the K--band ($2.2\,\mu$m)
observations. The photometric magnitude, measured through the aperture
given in column (4), is given in column (5), with its error in column
(6). Column (7) indicates the extent to which the K--band image was
affected by elliptical seeing and RA judder of the telescope: `o' means
that the effective seeing was essentially round, `se' that it was slightly
elliptical, and `e' that it was clearly elliptical. These were determined
by investigating the ellipticities of stars within the fields. Columns (8)
and (9) give the observation date and exposure time of the J--band
($1.2\,\mu$m) observation, whilst the photometry of this, and its
associated error are given in columns (10) and (11). Column 12 indicates
the ellipticity of the seeing for the J--band observation, with symbols as
for column 7.\ \ \ $^1$ All magnitudes have been corrected for galactic
extinction using the extinction maps of Burstein and Heiles (1982).\ \ \
$^2$~The value quoted for 3C356 is for the more northerly galaxy. The
corresponding value for the southern object is $17.21 \pm 0.05$. $^3$~Note
that in the case of 3C368, the magnitudes quoted are after the best
possible subtraction of the M--star. Before subtraction of this, the
magnitudes are $16.67\pm 0.06$ in the K--band and $17.83\pm 0.06$ in the
J--band.}
\end{table*}

\section{Images of the radio galaxies}

In this section we present images of the galaxies at optical, infrared and
radio wavebands. Unless extra information is provided by displaying the
galaxy as observed through both HST filters separately, we have combined
the two images. Overlaid upon these HST images are contours of the radio
emission, as observed in our VLA observations.  Adjacent to the HST
images, and displayed at the same scale, are the infrared images: in cases
where the infrared J and K--band morphologies are similar we display only
the K--band image. 

Registering the radio and optical images could not be carried out with
high accuracy because of uncertainties in the alignment of the radio and
optical reference frames. For sources in which there was a $\gta 5 \sigma$
detection of a radio core, the frames were aligned by assuming that the
centre of the infrared images lay directly over the radio core. Unlike the
optical images, which could not be used for this procedure because of the
possibility of dust extinction affecting the central peak, the infrared
images are sharply peaked towards the centre of the galaxy. For the
sources without a radio core, the absolute positioning of the HST and
UKIRT frames was found using one or more unsaturated stars which were
present on the frames, and also present in the APM database \cite{mad90};
in general at least four such objects were available. Then, the optical
and radio images were overlaid assuming that the two reference frames were
accurately registered. The astrometric errors introduced in this procedure
are estimated to be less than about 1 arcsecond. The forthcoming release
of the Hipparcos data should enable this problem to be somewhat alleviated
in the future.

The sources are presented in numerical (right ascension) order, and
important details about each source are provided. As a guide to
interpreting the HST images, it is a useful rule of thumb that a standard
$L^*$ elliptical galaxy at a redshift of one possesses a very red colour,
and will have a low-surface brightness in optical (rest--frame
ultraviolet) images. Only regions of high surface brightness will stand
out prominently in the images \cite{gia96a,dic96}. As a result, if the 3CR
radio galaxies are standard elliptical galaxies, little structure would be
observed in the HST images; a good example of this is the case of 3C34
(see below) at redshift $z = 0.69$, which shows very little optical
activity, is diffuse, and is barely visible in the bluer of the HST
images. By contrast, the majority of the HST images of the 3CR galaxies
show a wealth of bright structures.

\subsection*{3C13}

\begin{figure}
\centerline{
\psfig{figure=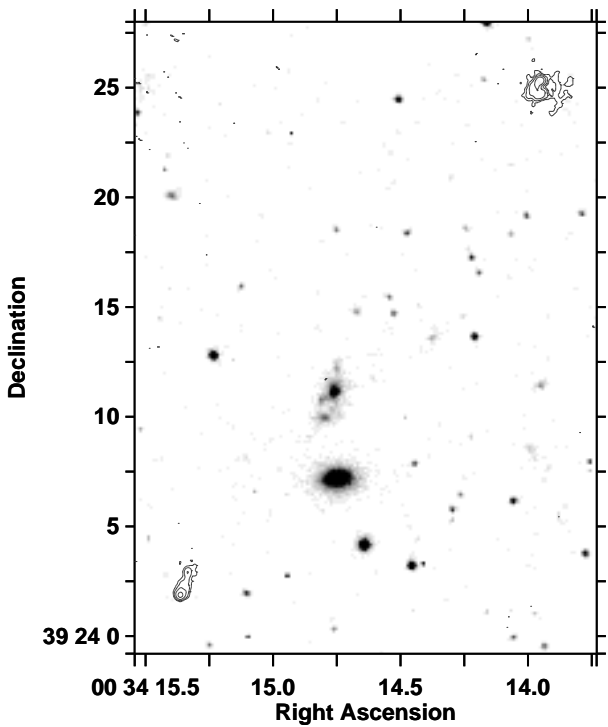,angle=90,clip=,width=8cm}
} 
\caption{\label{fig3c13} (a -- above) The sum of the HST images of 3C13
taken through the f785LP and f814W filters. Overlaid are contours of the
radio emission from the A and B array observations. Contour levels are
160\,$\mu$Jy beam$^{-1}$\,$\times (1,4,16,64)$.  (b,c,d -- upper, centre,
lower right) Enlarged HST, J-band and K-band images, respectively, of the
central regions of 3C13, displayed on the same scale.}
\end{figure}

\begin{figure}
\centerline{
\psfig{figure=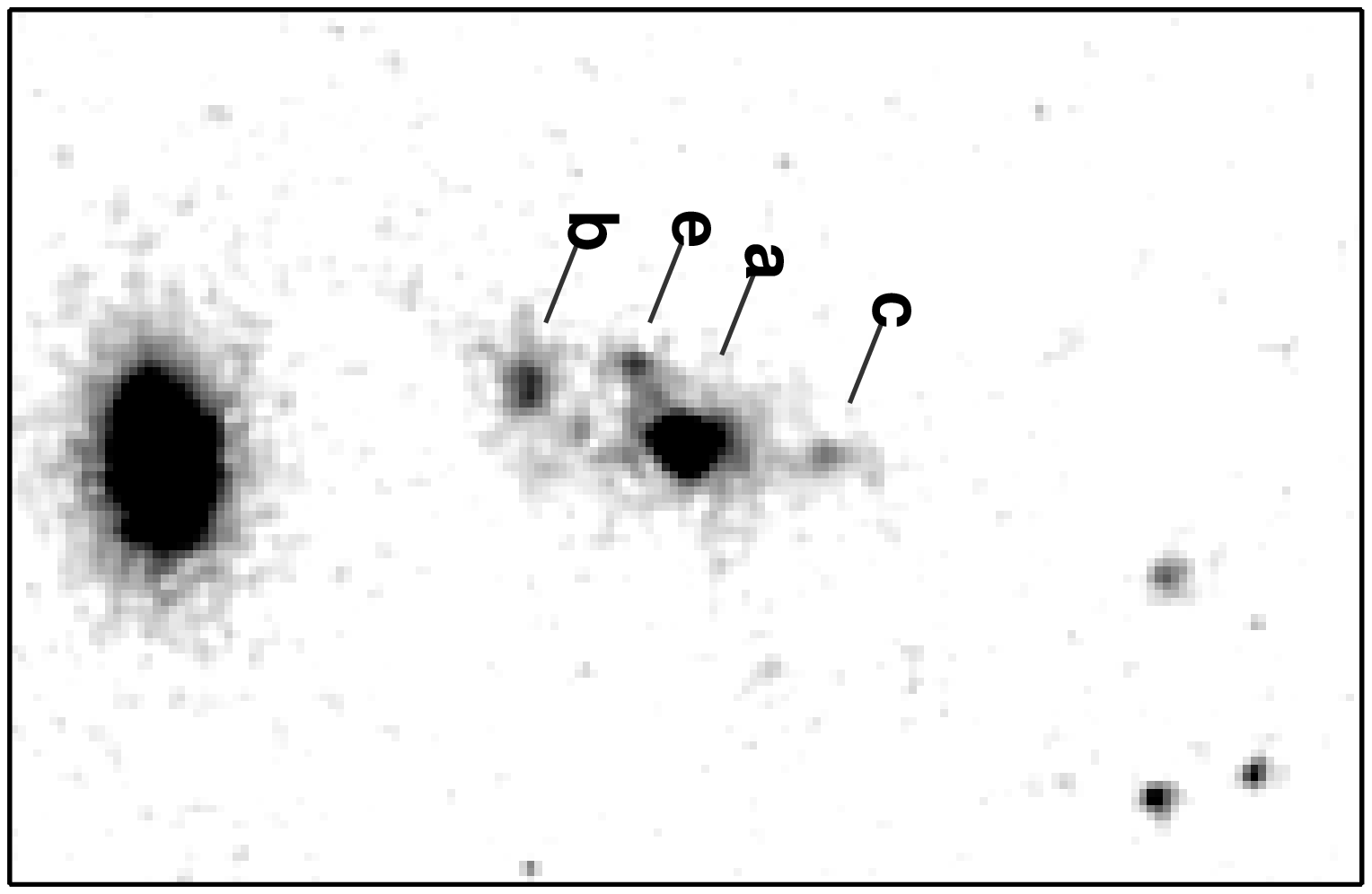,angle=90,clip=,width=5cm} 
}
\smallskip
\centerline{
\psfig{figure=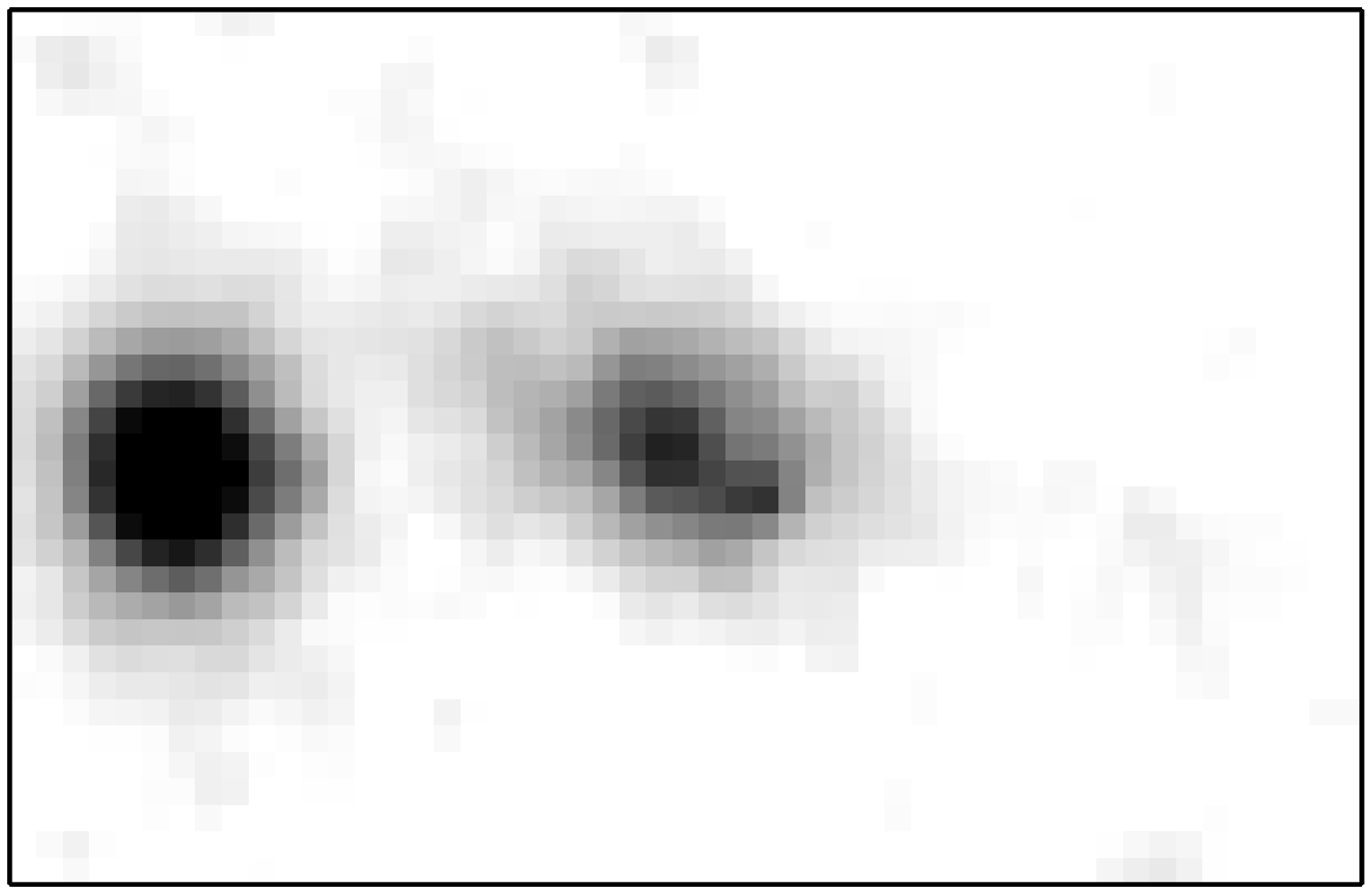,angle=90,clip=,width=5cm}
}
\smallskip
\centerline{
\psfig{figure=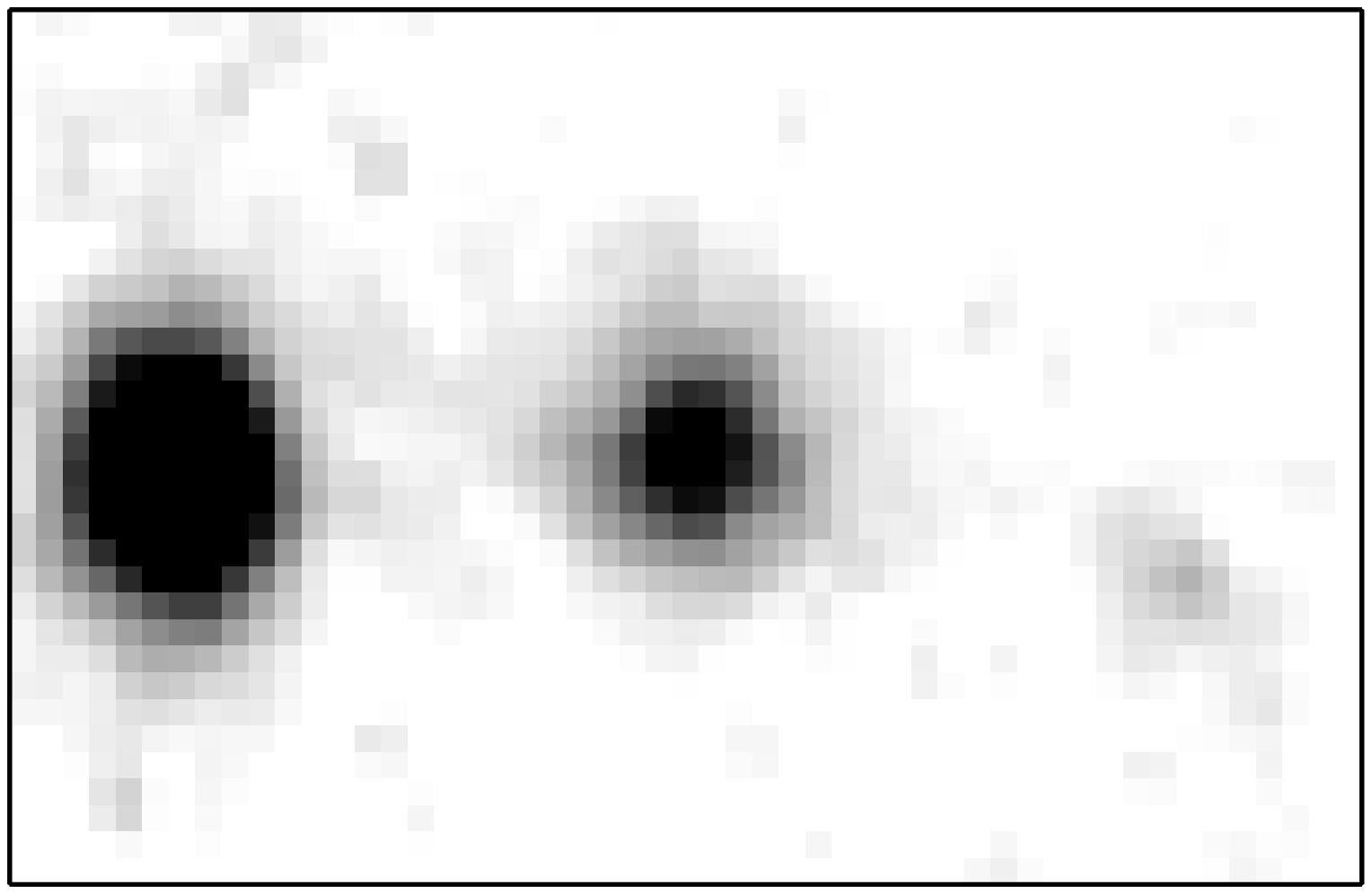,angle=90,clip=,width=5cm}
}
\end{figure}

The HST image of the host galaxy associated with 3C13
(Figure~\ref{fig3c13}a,b), at redshift $z = 1.351$, consists of a
bright central component (`a') overlying diffuse emission elongated along
the radio axis. A second, fainter, component (`b') lies just over an
arcsec to the south, with a fainter component (`e') just to the north of
it. A fourth component (`c') lies 0.5 arcsec to the north of `a'. [NB. For
consistency we have labelled the components following Le F\`evre \etal\
\shortcite{fev88c}. We do not detect anything corresponding to their
putative component `d'. Our component `e' was not detected by them]. The
galaxy lying 4 arcsec to the south is a foreground elliptical galaxy at $z
= 0.477$. Le F\`evre et~al \shortcite{fev88c} noted that 3C13 is optically
three magnitudes more luminous than a brightest cluster galaxy redshifted
to that distance, and suggested that this may be due to two factors:
firstly, the foreground elliptical galaxy may gravitationally amplify 3C13
by more than a magnitude; secondly, they suggested that component `a' may
be a foreground companion of this elliptical galaxy, and the surrounding
components `b' and `c' could be gravitationally lensed images of the host
radio galaxy at redshift $z=1.351$. Comparison of the morphology of these
regions as seen with the HST, together with the smooth, barely elongated,
emission in the infrared images makes the second possibility unlikely. In
addition, the fact that the K--magnitude of 3C13 is typical of the other 3CR
galaxies at that redshift suggests that the optical brightening is instead
due to a flat spectrum aligned component.  

Using Keck spectro-polarimetry, Cimatti \etal\ \shortcite{cim97} have
shown that the optical continuum of 3C13 is polarised, with the
orientation of the polarisation being roughly perpendicular to the axis of
the ultraviolet continuum. The fractional polarisation remains roughly
constant, at the 5 to 10\% level, from observed wavelengths of 4000 to
9000\AA. This polarised emission, associated with light scattered from an
obscured quasar nucleus, effective rules out a gravitational lensing
hypothesis for the morphology of this source.

\begin{figure*}
\centerline{
\psfig{figure=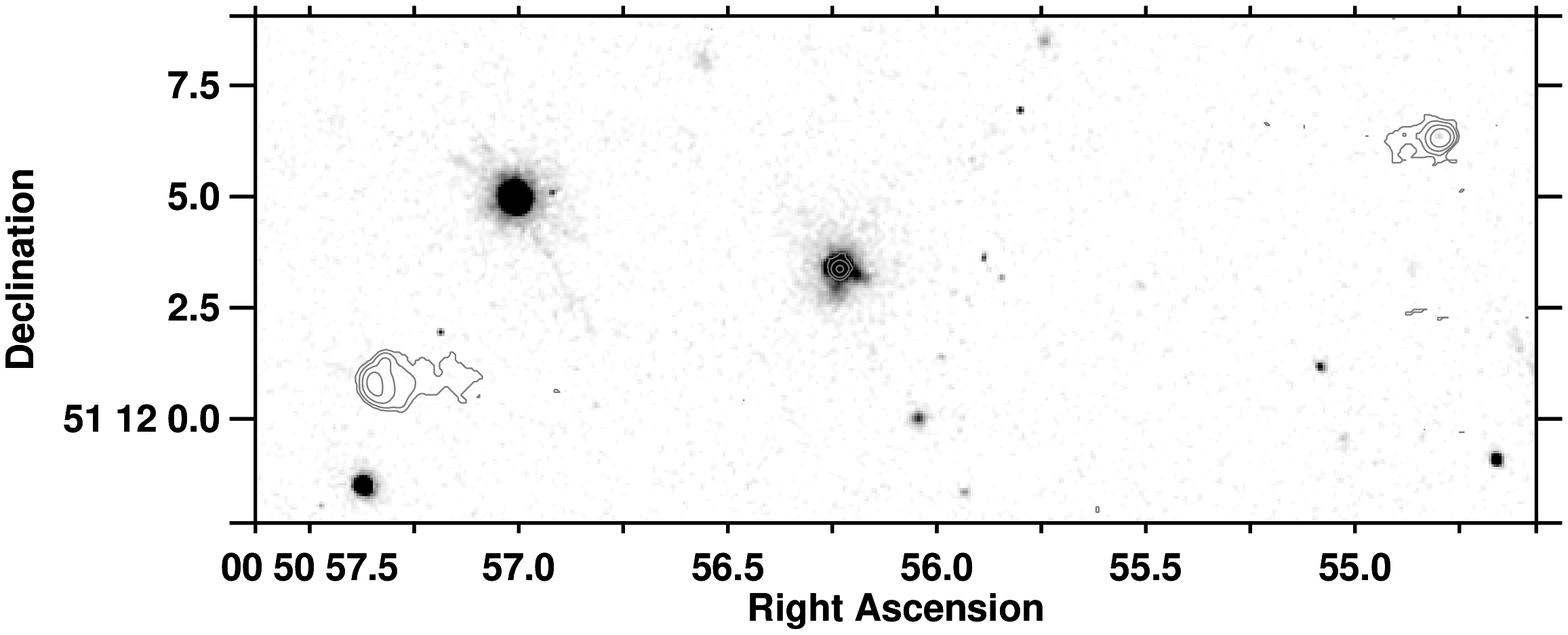,angle=90,clip=,width=13.5cm}
}
\smallskip
\centerline{
\psfig{figure=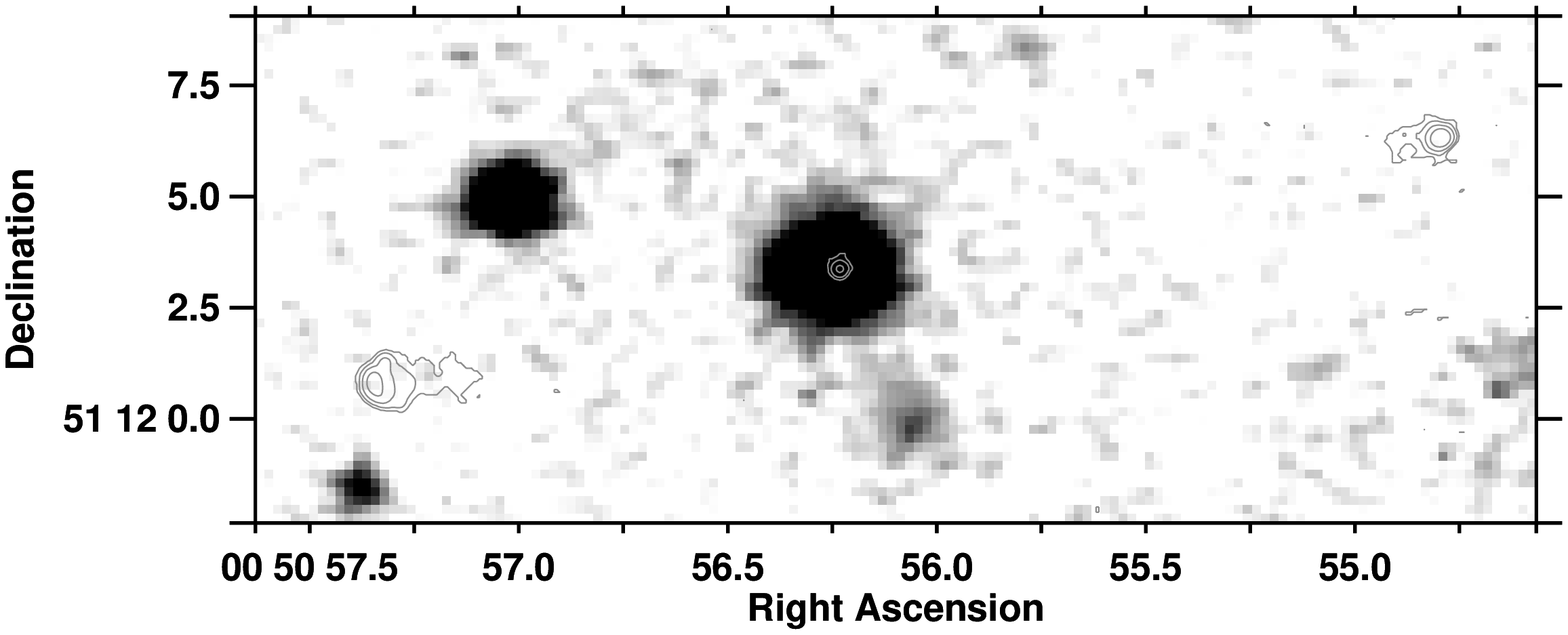,angle=90,clip=,width=13.5cm}} 
\caption{\label{fig3c22} (a) The sum of the HST images of 3C22 taken
through the f622W and f814W filters. Overlaid are contours of radio
emission from the A--array VLA observation. Contour levels are
320\,$\mu$Jy beam$^{-1}$\,$\times (1,4,16,64)$. (b) UKIRT K--band image of
3C22 with VLA radio contours overlaid as in (a).}
\end{figure*}

The UKIRT J--band image (Figure~\ref{fig3c13}c) shows a slight extension
along the radio axis and a faint emission region corresponding to the
southern component `b', although this is much less pronounced than in the
optical image, and is not detected at all in the K--band image
(Figure~\ref{fig3c13}d). It is interesting that the elongation of the
J--band image is more precisely aligned along the radio axis direction
than is the optical emission. Near--infrared spectrophotometry by Rawlings
\etal\ \shortcite{raw91a} has shown that the [OIII]~5007 line is strong in
this source ($\sim 4.5 \times 10^{-18}$ W m$^{-2}$), and contributes a
significant fraction ($\gta 10\%$) of the total J--band flux. This line
emission may be partially responsible for the J--band alignment.

\subsection*{3C22}

According to orientation--based unification schemes for 3CR radio sources
\cite{bar89}, radio galaxies and quasars could represent the same class of
object viewed at different angles to the line of sight. In these schemes,
quasars have their radio axis orientated within $45^{\circ}$ of the line
of sight, enabling the central active galactic nucleus and broad--line
regions to be seen, whilst radio galaxies are orientated within
$45^{\circ}$ of the plane of the sky, and have their central regions
obscured by a torus of material. Dust obscuration of these central regions
will be less severe at infrared wavelengths and so, assuming that the
unification model is correct, any broad emission lines or compact central
objects may be visible in the radio galaxies at these wavelengths.

Recently, broad \Ha\ emission has been detected from 3C22 with FWHM $\sim
5500 \pm 500$ km s$^{-1}$ \cite{eco95,raw95}. Coupled with the nucleated
appearance of the K--band image of this source, and the fact that its
K--band magnitude is $\approx 1.3$ magnitudes brighter than the mean
K--$z$ relation, this has led to suggestions that this source contains a
reddened quasar nucleus, perhaps observed close to the radio galaxy ---
quasar divide \cite{dun93,eco95,raw95}. The radio structure of the source
supports this hypothesis: the radio core is brighter than that of any
other source in our sample (see Table~\ref{vlatab}), and it is one of the
few radio galaxies at this redshift for which a radio jet has been
detected \cite{fer93}; detection of bright cores and jets is common in
quasars even at high redshift.

\begin{figure*}
\centerline{
\psfig{figure=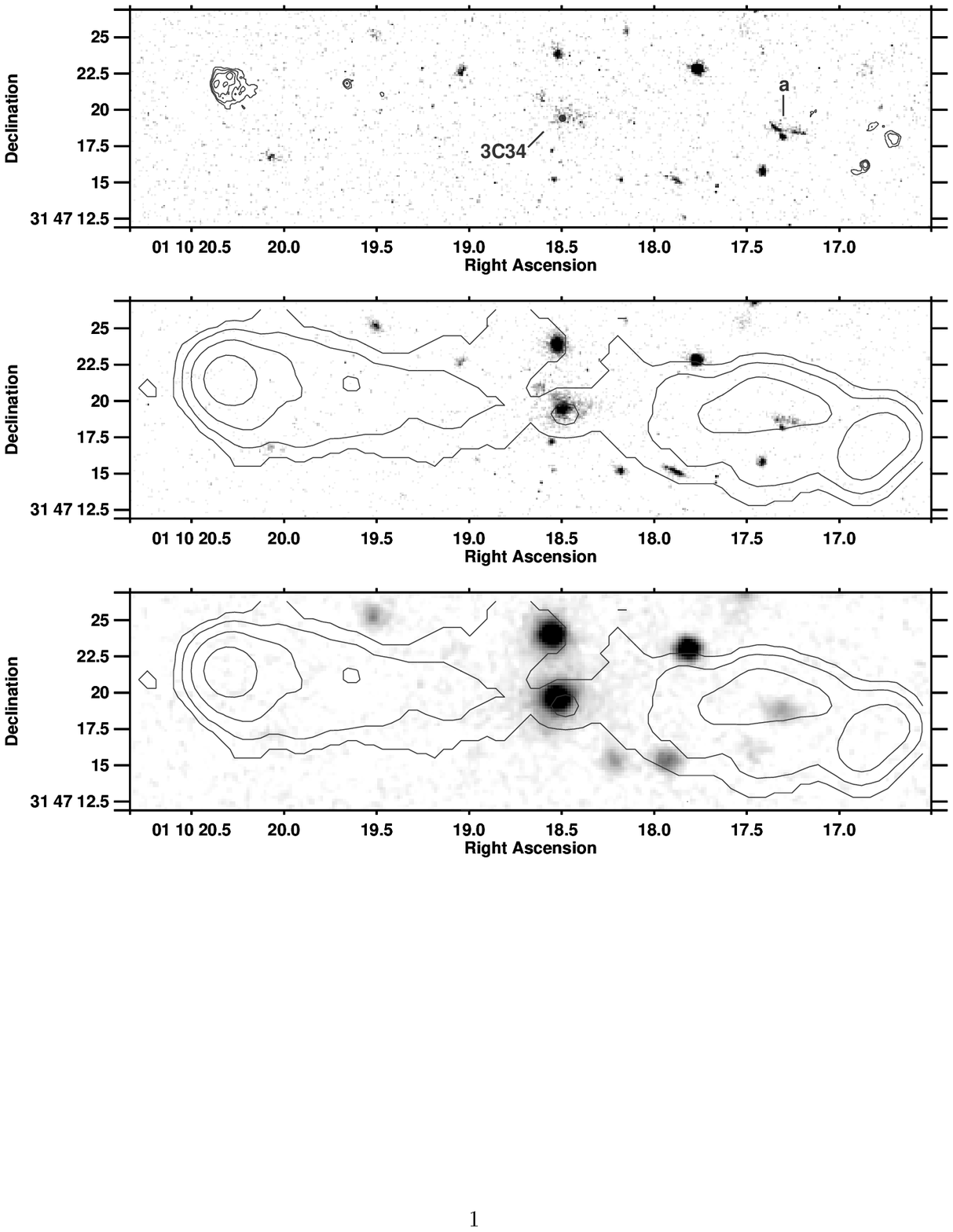,clip=,width=\textwidth}
}
\caption{\label{fig3c34} Images of the radio galaxy 3C34. (a) The HST
image taken using the f555W filter, with A--array VLA radio contours
overlaid; contour levels are $(1,2,4,8) \times 240 \mu$Jy beam$^{-1}$. The
object labelled `a' shows evidence for a strong interaction between the
radio jet and a galaxy in the cluster surrounding 3C34. (b) The HST image
taken using the f785LP filter, with radio contours from the C--array VLA
observation overlaid; contours levels are $(1,4,16,64) \times 260 \mu$Jy
beam$^{-1}$. (c) The UKIRT K--band image with C--array VLA radio contours
as in (b) overlaid.}
\end{figure*}

Our K--band image (Figure~\ref{fig3c22}b) also shows 3C22 to be compact,
although it is not totally unresolved. Unlike the other galaxies in the
sample, 3C22 is not well fitted by a de Vaucouleurs profile, , $I \propto
{\rm exp} \left [-7.67 \left (r/r_{\rm v}\right )^{1/4}\right ]$, but can
be fitted using the combination of a de Vaucouleurs profile and a point
source (see Paper II). Even in the HST image (Figure~\ref{fig3c22}a), the
galaxy is much more nucleated than other galaxies in the sample, but it
does show two deviations from circular symmetry. The first is a slight
extension of the central bright component, just to the south of west and
misaligned by about 30 degrees from radio emission extending from the core
towards the western hot--spot \cite{fer93}. The second is what appears to
be a small companion just to the south of the host galaxy and, again,
misaligned relative to the radio axis. If 3C22 is indeed a reddened
quasar, the nature of these two components is of great interest. The
f622W$-$f814W colours of the western and southern extensions are $1.08 \pm
0.15$ and $1.40 \pm 0.22$ respectively, as compared to the central bright
component which has f622W$-$f814W$= 1.26 \pm 0.04$. The western extension
therefore seems slightly bluer than the host galaxy/quasar, and the
southern extension slightly redder, although at low significance.

\subsection*{3C34}

This galaxy lies near the centre of a compact cluster \cite{mcc88} at
redshift $z = 0.689$. The galaxy is one of the reddest in the sample, and
is barely visible in the HST image obtained through the f555W filter
(Figure~\ref{fig3c34}a). In the f785LP filter image
(Figure~\ref{fig3c34}b) it appears as a bright central galaxy with
an extensive halo and faint companions.

Of particular interest in this field is the emission region at RA: 01 10
17.25, Dec: 31 47 18 (J2000), labelled object `a'. The f555W filter image
clearly shows the presence of two very long, narrow regions of optical
emission along the radio axis, with a `blob' of emission just to the south
of them. This image has been overlaid with the A--array VLA data so as not
to obscure this region, and to make it clear that these point towards the
northern component of a double hotspot in the western lobe of this
source. The other two images are overlaid with the C--array map to show
the more extended radio structure.

These knots are significantly bluer than the central radio galaxy:
their combined f555W$-$K colour is $4.08 \pm 0.17$, compared with
f555W$-$K$ = 5.61 \pm 0.10$ for 3C34. The cause of the elongation and
blue colour of this region has been discussed in detail by Best \etal\
\shortcite{bes97b}, who concluded that the most likely scenario is
that it is due to a massive burst of star formation induced by the
passage of the radio jet through a galaxy in the cluster surrounding
3C34.

\begin{figure}
\centerline{
\psfig{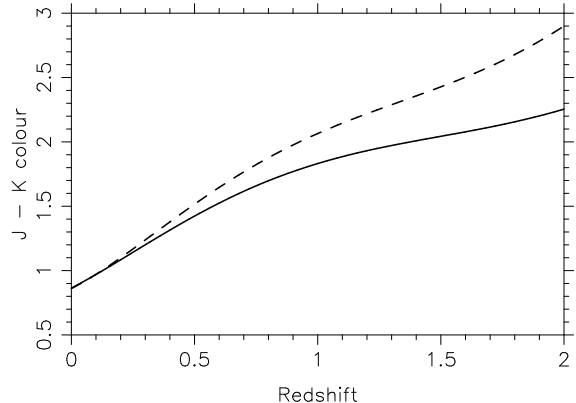}
}
\caption{\label{jkcols} The expected redshift dependence for the J$-$K
colour of elliptical galaxies. The dashed line corresponds simply to
redshifting the spectral energy distribution of a current epoch giant
elliptical galaxy with no evolution of its stellar populations. The solid
line provides the fit if the stellar populations evolve passively.}
\end{figure}

Support for this object being a cluster galaxy in which the radio jet has
induced activity of some kind comes from a number of observations.
Firstly, we can use the infrared J$-$K colour as an indicator of the
redshift of the galaxy, since this colour index is dominated by the old
stellar population (eg. Best \etal\ 1997a,b) and is a fairly strong
function of redshift out to $z \sim 1$. In Figure~\ref{jkcols} we plot the
redshift dependence of this colour, obtained by redshifting the spectral
energy distribution of a present--day elliptical galaxy with no evolution
(dashed line), and with passive evolution (solid line), by which we mean
that it is assumed that the elliptical galaxy formed at $z \sim 10$ and
that the stellar population has been passively evolving since then. Note
that although the radio galaxies themselves do not follow this plot beyond
a redshift $z \sim 0.8$, instead becoming bluer than the expected passive
evolution line, this is due to the contribution of the aligned ultraviolet
emission and the \Hb\ and [OIII]\,5007 emission lines to the J--band flux
density at these redshifts: for `ordinary' cluster galaxies, these effects
will be small, and so the J$-$K colour remains dominated by the old
stellar population. Object `a' has a J$-$K colour of $1.54 \pm 0.08$,
which is consistent with it lying at the same redshift as 3C34 (see
Figure~\ref{jkcols}).

Further evidence in support of this hypothesis comes from the radio
structure of the source. The enhanced region of radio emission lying to
the north of object `a' (Figure~\ref{fig3c34}b, Johnson \etal\ 1995) has a
radio spectral index less steep than that of the rest of the radio lobe,
and which increases away from the hotspot, indicating a region of rapid
backflow from the hotspot \cite{blu94}. This backflow loops around to the
north of object `a' rather than passing though it, consistent with object
`a' lying within the radio lobe, and the relativistic electrons flowing
out from the hotspot avoiding this region of higher gas density. A map of
the radio depolarisation properties of 3C34 between 6cm and 21cm shows
that galaxy `a' corresponds precisely to the position and morphology of a
`depolarisation silhouette', that is, a region in which the radio
depolarisation, due to Faraday depolarisation, is significantly higher
than in the surrounding lobe \cite{joh95}. Johnson \etal\ concluded that
the most likely cause of the depolarisation was that it was due to Faraday
depolarisation by the gas within a cluster galaxy lying in front of the
lobe, but it would also be consistent with the galaxy lying within the
lobe.

A combination of an old stellar population together with a short--lived
starburst induced by the radio jet provide an excellent fit to the
spectral energy distribution of this source, whilst other alignment
mechanisms can be shown to be of little importance. The reader is referred
to Best \etal\ \shortcite{bes97b} for a much fuller discussion of 3C34.

\begin{figure*}
\centerline{
\psfig{figure=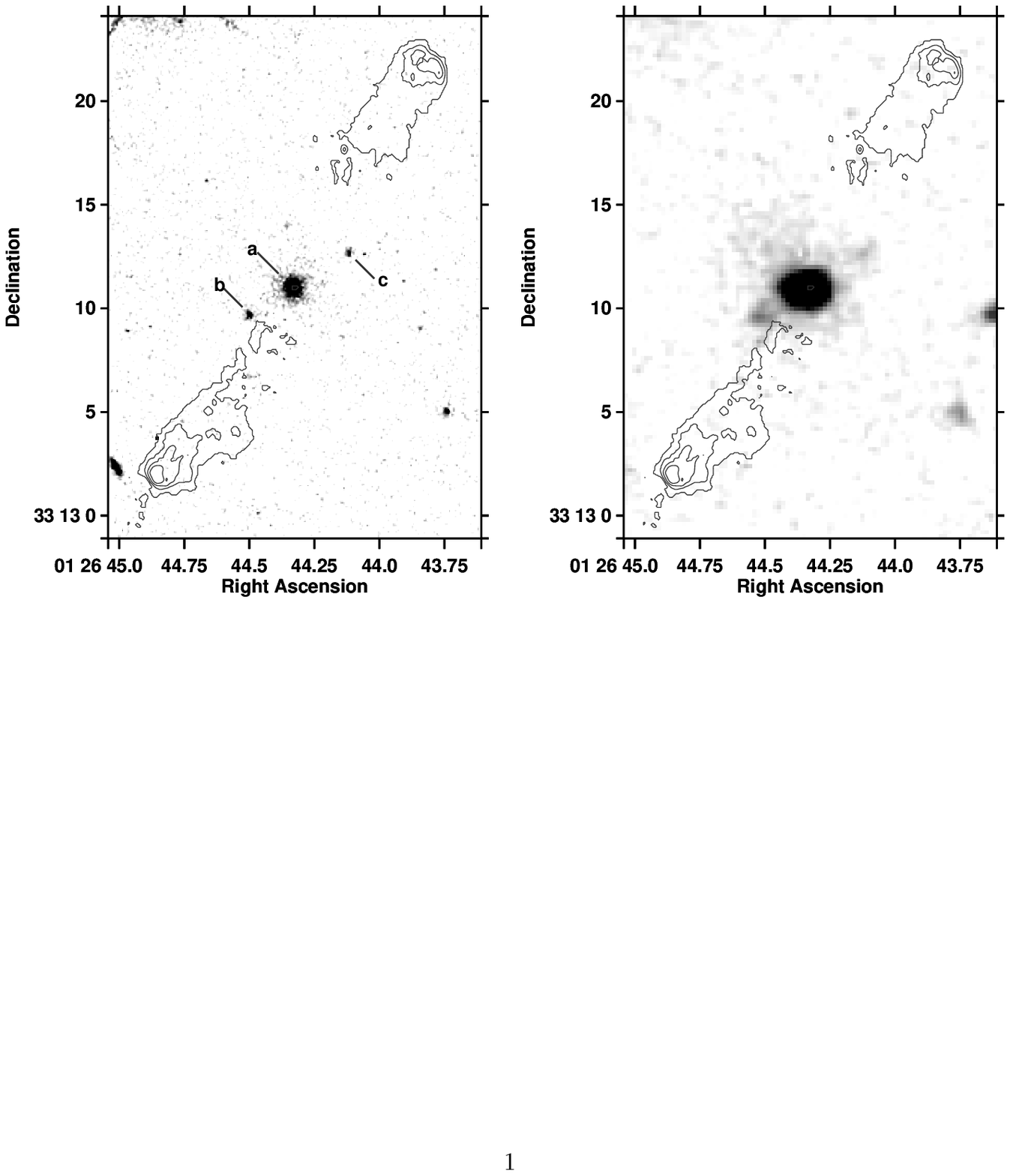,clip=,width=\textwidth}
}
\caption{\label{fig3c41} Images of the radio galaxy 3C41, overlaid with
contours of radio emission from the A and B array VLA observations;
contour levels are $(1,4,16,64) \times 160 \mu$Jy beam$^{-1}$. (a) The sum
of the HST images through the f555W and f785LP filters; (b) The UKIRT
K--band image. }
\end{figure*}

\begin{figure*}
\centerline{
\psfig{figure=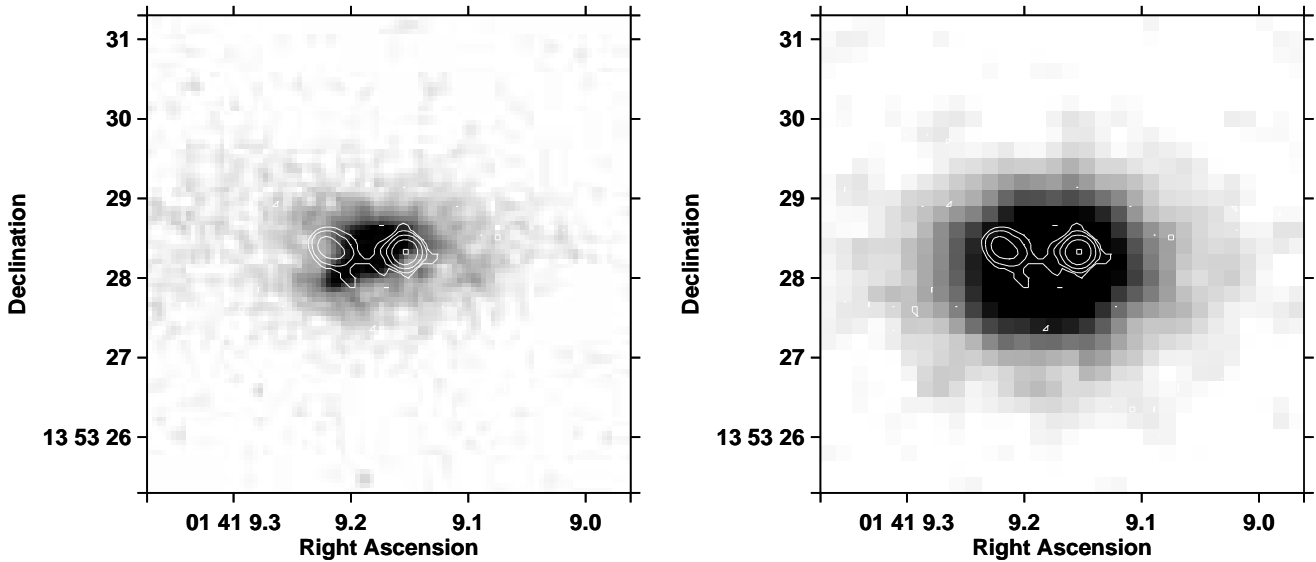,clip=,width=\textwidth}
}
\caption{\label{fig3c49} Images of the radio galaxy 3C49, with contours of
radio emission overlaid, as observed with the A array of the VLA. The
contour levels are $(1,4,16,64,256) \times 1200 \mu$Jy beam$^{-1}$. (a)
The sum of the HST images through the f555W and f814W filters. (b) The
UKIRT K--band image.}
\end{figure*}

\subsection*{3C41}

The galaxy 3C41, at redshift $z = 0.795$, appears to be almost totally
symmetrical in the HST image (Figure~\ref{fig3c41}a, galaxy `a').
Emission is also seen from two components, one on either side of the
galaxy to the ESE (`b') and the WNW (`c'), approximately 2.5 arcsec
away. Whether or not these are associated with the radio galaxy is
unclear. They do not appear in an image of the source centred on the
rest--frame [OII]~3727 line \cite{mcc88}, but they are faint and so this
does not exclude them from being at the same redshift. The two components
`b' and `c' are also visible in the K--band image (Figure~\ref{fig3c41}b),
and have f555W$-$K colours of $4.20 \pm 0.41$ and $4.25 \pm 0.25$
respectively, over a magnitude bluer than the host radio galaxy which has
f555W$-$K$= 5.85 \pm 0.08$. This suggests that, if they are associated
with the radio source, they may be optically active in some way, although
they are misaligned by about 25$^{\circ}$ from the radio axis. Apart from
these two companions, the optical emission is unremarkable.

The K--band image shows a bright central nucleus, and the K--magnitude of
3C41 is significantly brighter than the mean K$-z$ relation of the 3CR
galaxies. The infrared radial intensity profile suggests the presence of
an unresolved quasar component (see Paper II). There is also, however,
some evidence for a faint halo in the K--band, particularly to the
north--east. Eisenhardt and Chokshi \shortcite{eis90} also found
significant emission out to a radius of at least $6''$ in their K--band
image of this galaxy.

The radio contours indicate the presence of a weak radio jet pointing
towards the south--eastern hot--spot, a feature which is rare amongst
these high redshift radio galaxies.

\begin{figure*}
\centerline{
\psfig{figure=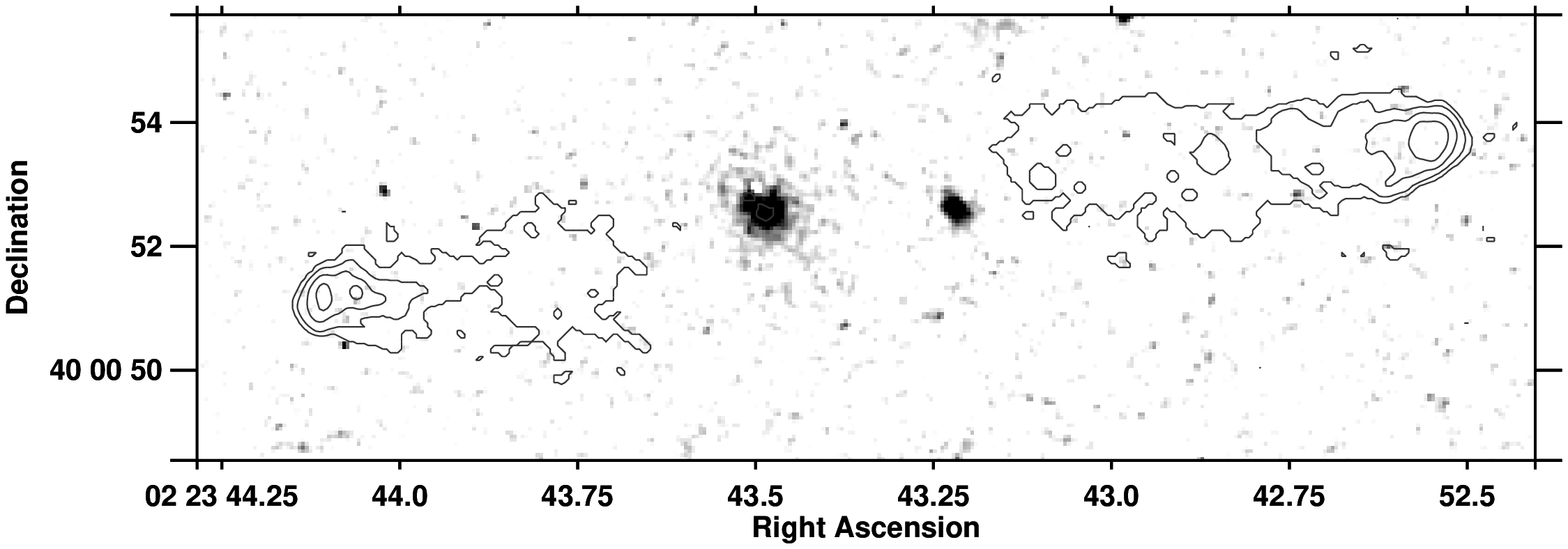,clip=,width=15cm,angle=90}
}
\centerline{
\psfig{figure=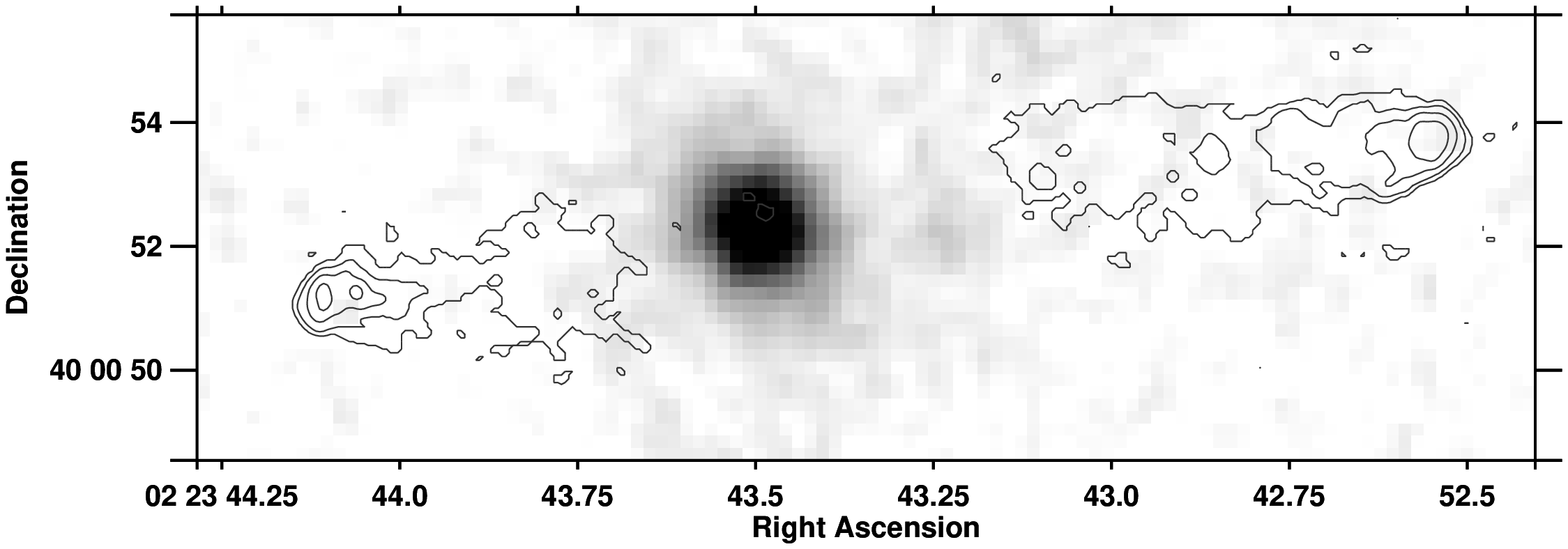,clip=,width=15cm,angle=90}
}
\caption{\label{fig3c65} (a) The sum of the two HST images of 3C65, taken
using the f675W and f814W filters. Overlaid are contour of the radio
emission from the A and B array VLA observations; contour levels are
$(1,4,16,64) \times 120 \mu$Jy beam$^{-1}$. (b) The UKIRT K--band image
with the the same radio contours overlaid as in (a).}
\end{figure*}

\subsection*{3C49}

3C49, at redshift $z = 0.621$, has a radio size of only 7~kpc, and is
one of two compact steep spectrum radio sources in the sample. The HST
image, shown in Figure~\ref{fig3c49}a, shows only a single bright
component, plus diffuse emission extending beyond the radio lobes. The
bright central emission is misaligned from the radio axis by about 20
degrees, but the more extended fainter emission is orientated within
five degrees of the axis of the radio emission. The infrared image
(Figure~\ref{fig3c49}b) shows that the radio emission lies well within
the envelope of the host galaxy. In the infrared waveband, the galaxy
shows a slight elongation along the radio axis.

This radio source has been mapped at high angular resolution by a number
of authors. Fanti \etal\ \shortcite{fan89} detected a weak central
component, positioned approximately two--thirds of the way from the
eastern to the western hotspot. This has a flux density of about 5\,mJy at
a wavelength of 6\,cm, and possesses a flat spectrum. Its identification
as the radio core was confirmed by later observations
\cite{bre92,san95}. Fanti \etal\ \shortcite{fan89} also derived an age for
the source of $10^5$ years, with a corresponding hotspot advance speed of
0.05 to $0.1c$, based on radio spectral ageing. For radio spectral ageing,
the radio spectrum is interpreted in terms of an ageing population of
electrons, and the break frequency of the synchrotron radiation can be
used to estimate the time elapsed since the electrons were last
accelerated \cite{pac70,liu92}. This time lapse is generally found to be
shortest at the hotspots, and increases along the lobe towards the central
regions of the source. The spectral age of the electrons in the oldest
regions of the lobe provides an estimate of the age of the radio
source. The age estimate for this source is consistent with the source
being young, rather than its small size being due to a slower advance
speed caused by a denser surrounding medium.

The western hotspot is brighter, more compact, and closer to the nucleus
than its eastern counterpart, which Fanti \etal\ \shortcite{fan89} ascribe
to the interaction of this component with denser material, requiring a
clumpy interstellar medium. Van Breugel \etal\ \shortcite{bre92} note that
both lobes are almost totally depolarised at 15~GHz, consistent with
Faraday depolarisation by the interstellar medium of the galaxy.

\subsection*{3C65}

3C65 ($z = 1.176$, RA: 02 23 43.47, Dec: 40 00 52.2, Figure~\ref{fig3c65})
is one of the most passive galaxies in the sample. The host galaxy is
redder than most of the 3CR sample ($V-K \approx 6$; f555W$-$f814W = $1.65
\pm 0.11$) and, apart from the presence of a bluer (f555W$-$f814W = $0.77
\pm 0.19$) companion galaxy 3 arcsec to the west along the radio axis, it
shows little evidence of any optical activity associated with the radio
phenomenon. This western companion appeared as two components, separated
north--south, in the ground--based image of Le F\`evre and Hammer
\shortcite{fev88b}; the HST image does not show multiple components, but
does show a NE--SW extension.

Recently, Lacy \etal\ \shortcite{lac95} have claimed the presence of a
point--like emission source at the centre of the infrared image, which
they have associated with an obscured quasar nucleus. Rigler and Lilly
\shortcite{rig94} compared the infrared profile of the galaxy with a de
Vaucouleurs law taking into account the effects of seeing, and
found a satisfactory fit without the need for an unresolved emission
source. They concluded that any unresolved component would make only a
small contribution to the total infrared flux from the galaxy.

\begin{figure}
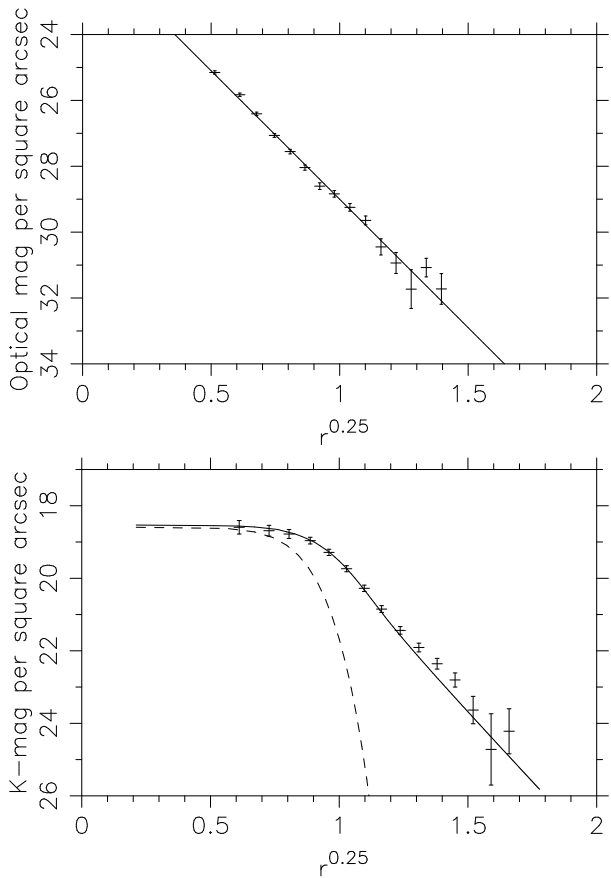

\centerline{
\psfig{figure=figure9a.ps,clip=,width=8cm,angle=-90}
}
\smallskip
\centerline{
\psfig{figure=figure9b.ps,clip=,width=8cm,angle=-90}
}
\caption{\label{devauc65} Plots of the radial surface brightness profile
of 3C65. (a) Comparison of the optical light profile (points marked with
error bars) with a de Vaucouleurs fit (solid line) of characteristic
radius 1.3 arcsec ($\approx 11$\,kpc). (b) Comparison of the infrared
light profile (points marked with error bars) with a de Vaucouleurs fit
(solid line) and a point source profile (dashed line), corrected for the
effects of seeing. The de Vaucouleurs profile has the characteristic
radius calculated by the fit to the optical profile. There is no evidence
for a nuclear contribution to this emission.}
\end{figure}

Since the HST image of this galaxy does not appear to be dominated by
active blue emission, and does not suffer from the effects of seeing, it
can be used to test whether or not the light profile of the underlying
galaxy follows a de Vaucouleurs profile. In Figure~\ref{devauc65}a, the
radial intensity profile of 3C65 as seen in the HST image is compared with
a de Vaucouleurs distribution. A very good match is obtained, consistent
both with the presence of a giant elliptical galaxy and with the
suggestion that there is little optical activity.

We can use this fit to provide a more accurate estimate of the
contribution any point source might make to the infrared image of 3C65. In
Figure~\ref{devauc65}b we plot the radial intensity profile of the K--band
image of the galaxy. The predicted de Vaucouleurs fit for a standard
spherically symmetrical giant elliptical galaxy with the characteristic
radius calculated above (solid line), and the profile of a point source
(dashed line), taking into account the effects of 1.3 arcsec seeing, are
also plotted. It can be seen that the galaxy profile is in reasonable
agreement with the de Vaucouleurs law; the excess of emission at large
radii may be associated with an extended halo around the galaxy, visible
in Figure~\ref{fig3c65}b, and characteristic of that seen around cD
galaxies (see Paper II for a further discussion of this). The addition of
a point source would not improve the fit. We agree with Rigler \& Lilly
(1994) that there is little evidence of a nuclear contribution to the
K--band emission of 3C65.

Detection of a 4000\ang\ break in the off-nuclear spectrum by Lacy \etal\
\shortcite{lac95} and by Stockton \etal\ \shortcite{sto95} indicates the
presence of an old (3--4 Gyr) stellar population in this galaxy,
corresponding to a formation redshift of $z_{\rm f} \gta 5$. This old
stellar population is seen relatively uncontaminated in the HST image.
Note that the fit of the de Vaucouleurs profile is generally not very good
for the HST images of the galaxies in our sample. The infrared images,
however, can be well fitted using a de Vaucouleurs profile (see Paper
II). This suggests that the old stellar population is present in most, if
not all, of the galaxies in the sample, but that in the HST images it is
generally swamped by emission induced by the radio activity. We discuss
these results in more detail in Paper II.

\subsection*{3C68.2}

\begin{figure*}
\centerline{
\psfig{figure=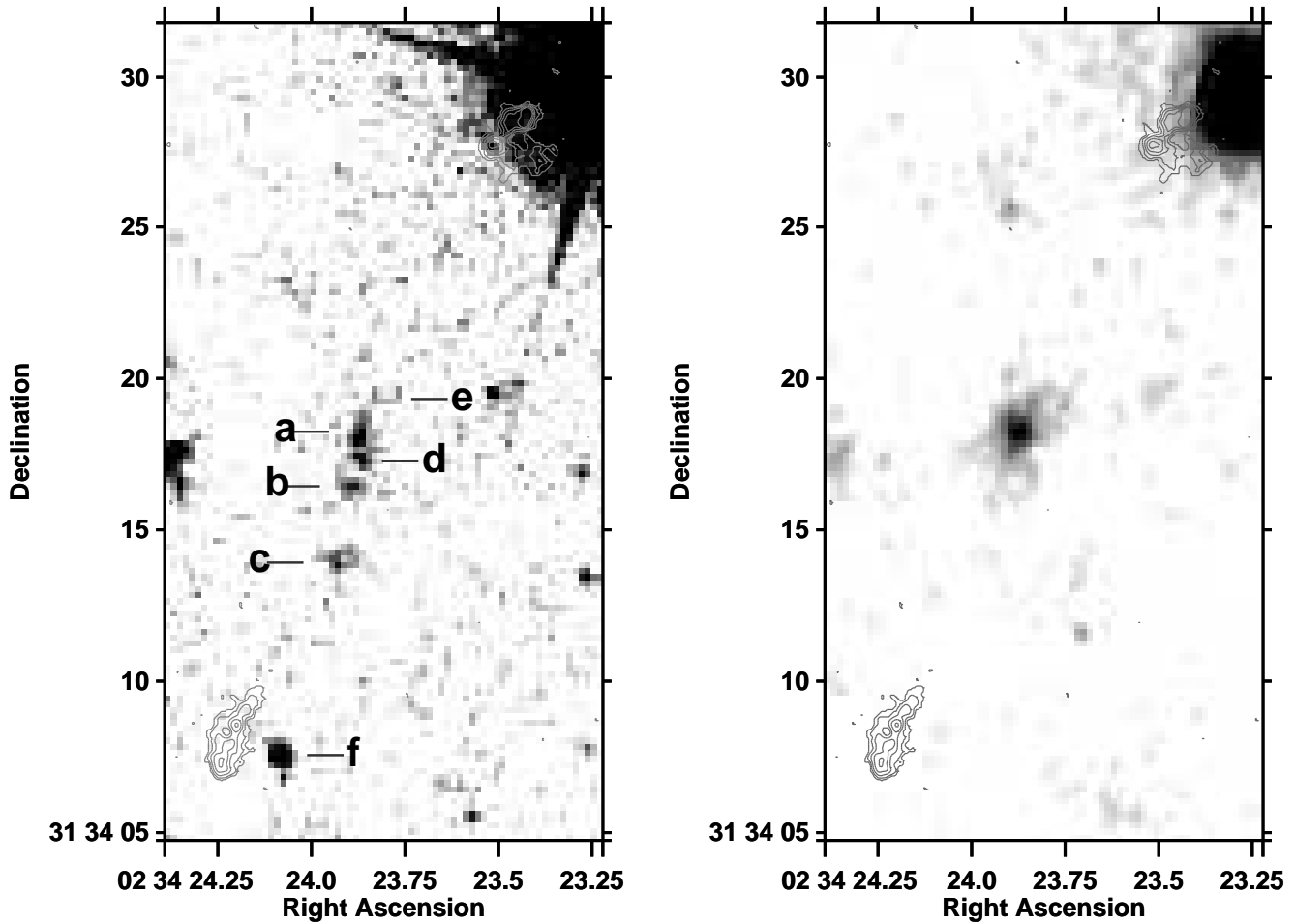,clip=,width=\textwidth}
}
\caption{\label{fig3c68} Images of the radio galaxy 3C68.2, with contours
of radio emission overlaid, from the A and B array VLA observations;
contour levels are $(1,2,4,8,16) \times 80 \mu$Jy beam$^{-1}$. (a) The HST
image of the galaxy, taken using the f785LP filter. (b) The UKIRT K--band
image. }
\end{figure*}

At redshift $z = 1.575$, 3C68.2 is one of the most distant galaxies in the
sample. To improve the signal--to--noise ratio of the HST data, the pixels
have been summed in $2 \times 2$ blocks and then smoothed with a 0.2
arcsec Gaussian; consequently the resolution of the image shown in
Figure~\ref{fig3c68}a, is slightly lower than that of most of the HST
images, but is still sufficient to show that 3C68.2 is composed of a
string of 4 or 5 bright components, misaligned with respect to the radio
axis by slightly over 10 degrees, and extending over 50~kpc.  Components
`a', `b' and `c' correspond to those detected by Le F\`evre \etal\
\shortcite{fev88b}; in addition to these, we note the presence of an
emission knot, `d', between `a' and `b', and a low signal--to--noise
detection of emission to the north of `a' --- component `e'. We see no
evidence of emission to the south of knot `c', as seen by Le F\`erve
\etal, although this may be due to the lower signal--to--noise in our
image.

The UKIRT images show none of this complexity. The K--band image
(Figure~\ref{fig3c68}b) is also somewhat extended and aligned along the
radio axis (even more closely than the optical image), but component `c',
in particular, is not detected in either the J or K--band. It is unclear
whether the brightest two emission regions in the HST image (`a' and `d'),
which lie towards the centre of the infrared emission, are separate
components or whether they are a single emission region with an obscuring
dust lane running across the centre.

Another component, `f', lies close to the southern radio hotspot. This
object appears bright on the HST image (f785LP$= 22.50 \pm 0.20$), but is
not detected at all in the K--band (K $\gta 21$). Its f785LP$-$K colour is
therefore $\lta 1.50$, which when compared to the central galaxy
(f785LP$-$K $= 3.66 \pm 0.16$) indicates that object `f' must be very
blue. It is quite conceivable that this object lies at the same redshift
as 3C68.2, its blue colour being due to scattering of light from an
obscured active nucleus, or to line emission, or to star formation induced
by shocks associated with the nearby radio hotspot. In the last case, an
interesting comparison could be made with the galaxy in the lobe of 3C34
discussed earlier, since component `f' would be expected to be a young,
on--going starburst, rather that an older decaying one.

\subsection*{3C217}

\begin{figure*}
\centerline{
\psfig{figure=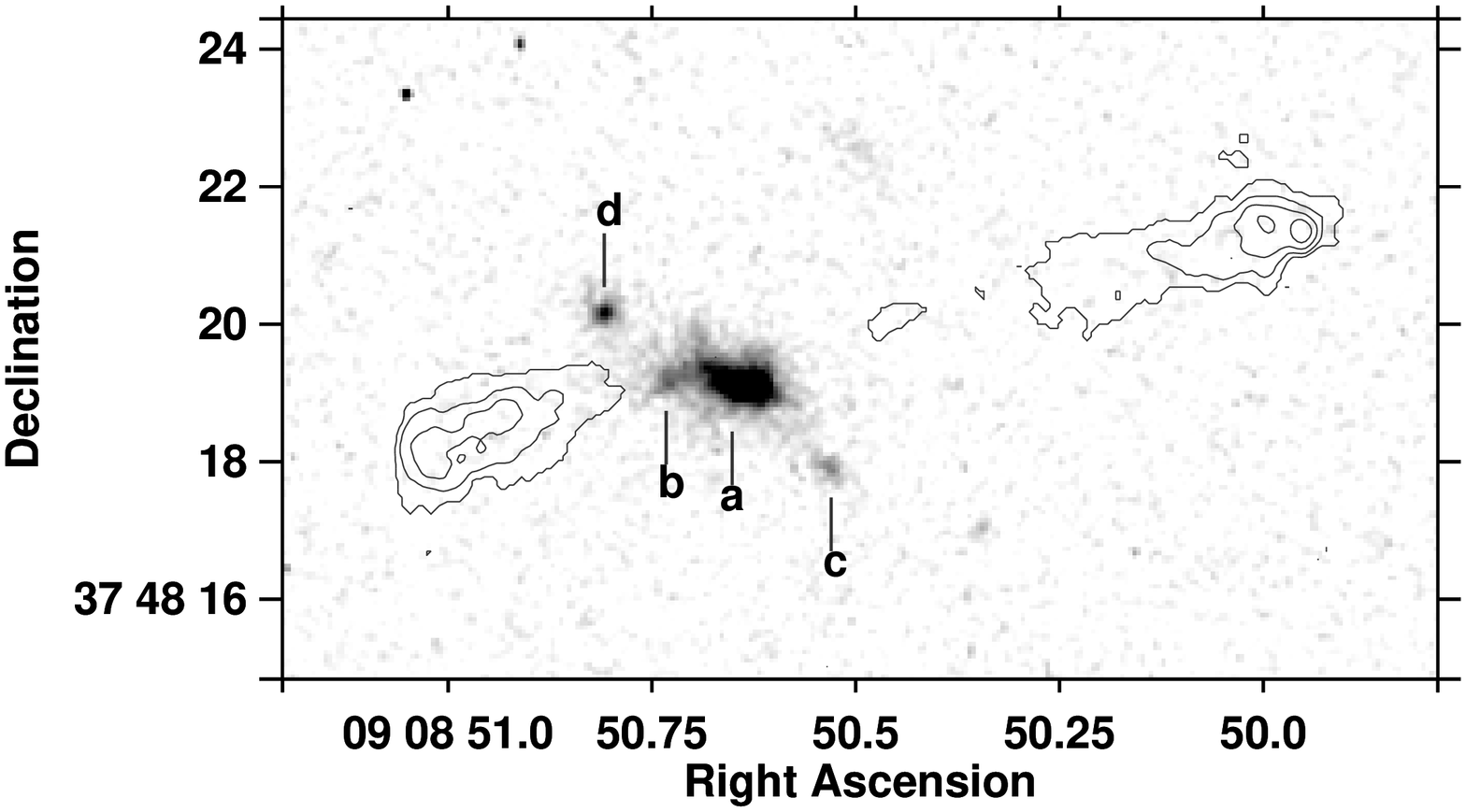,clip=,width=11cm,angle=90}
}
\centerline{
\psfig{figure=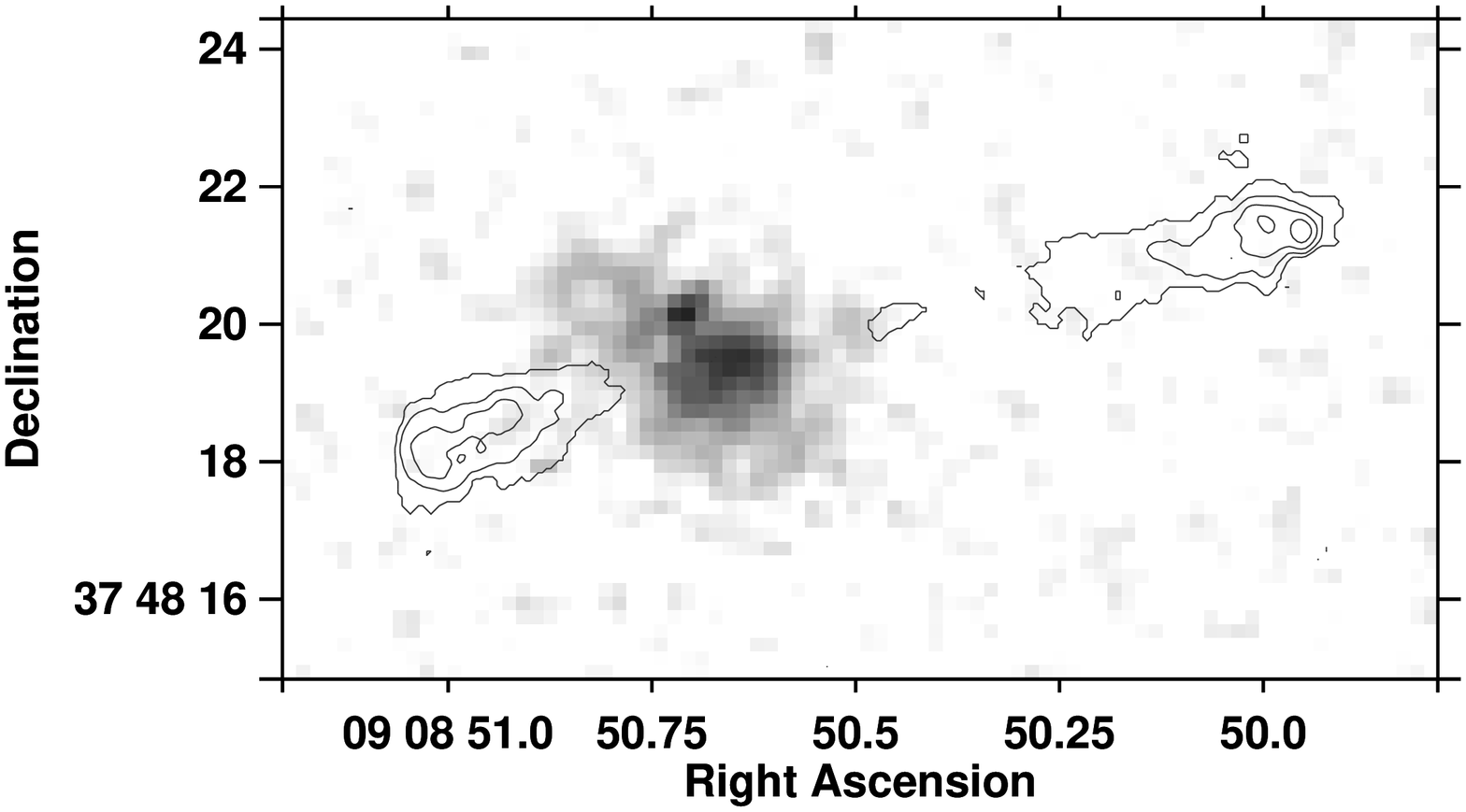,clip=,width=11cm,angle=90}
}
\caption{\label{fig3c217} Images of the radio galaxy 3C217. (a) The sum of
the HST images observed using the f622W and f814W filters, overlaid with
contours of radio emission as observed in the A and B array VLA
observations; contour levels are $(1,4,16,64) \times 120 \mu$Jy
beam$^{-1}$.  (b) The UKIRT K--band image, with radio contours as in (a).}
\end{figure*}

One of the bluest and most active galaxies in the sample, 3C217 ($z =
0.898$, Figure~\ref{fig3c217}), is also one of the most misaligned. The
bright central emission region (`a') is elongated and misaligned from the
radio axis by nearly $40^{\circ}$. Just to the east of this, a string of
knots (`b') lie in a curve, arcing round towards the eastern radio lobe,
and possibly tracking the path of the radio jet.  These knots possess an
f622W$-$f814W colour of $1.30 \pm 0.13$, which is significantly bluer than
that of the central emission region (f622W$-$f814W = $1.54 \pm
0.06$). Further components lie some 3 arcsec away to the south--west (`c')
and to the north--east (`d'). Component `d' possesses a similar colour to
the central regions, but `c' is much bluer with f622W$-$f814W = $0.85 \pm
0.32$.

This galaxy also possesses one of the most distorted infrared images,
showing the central galaxy, another bright component, and extended diffuse
emission to the north--east probably associated with component `d'
(Figure~\ref{fig3c217}b). As noted by Rigler \etal\ \shortcite{rig92} and
Dunlop and Peacock \shortcite{dun93}, the galaxy appears as extended in
the infrared as in the optical images. Their image through a narrow--band
filter centred on [OII]~3727 shows that the line emission is roughly
symmetrical and confined mainly to the central object, and so this cannot
be the cause of the observed elongation.

\subsection*{3C226}

\begin{figure*}
\centerline{
\psfig{figure=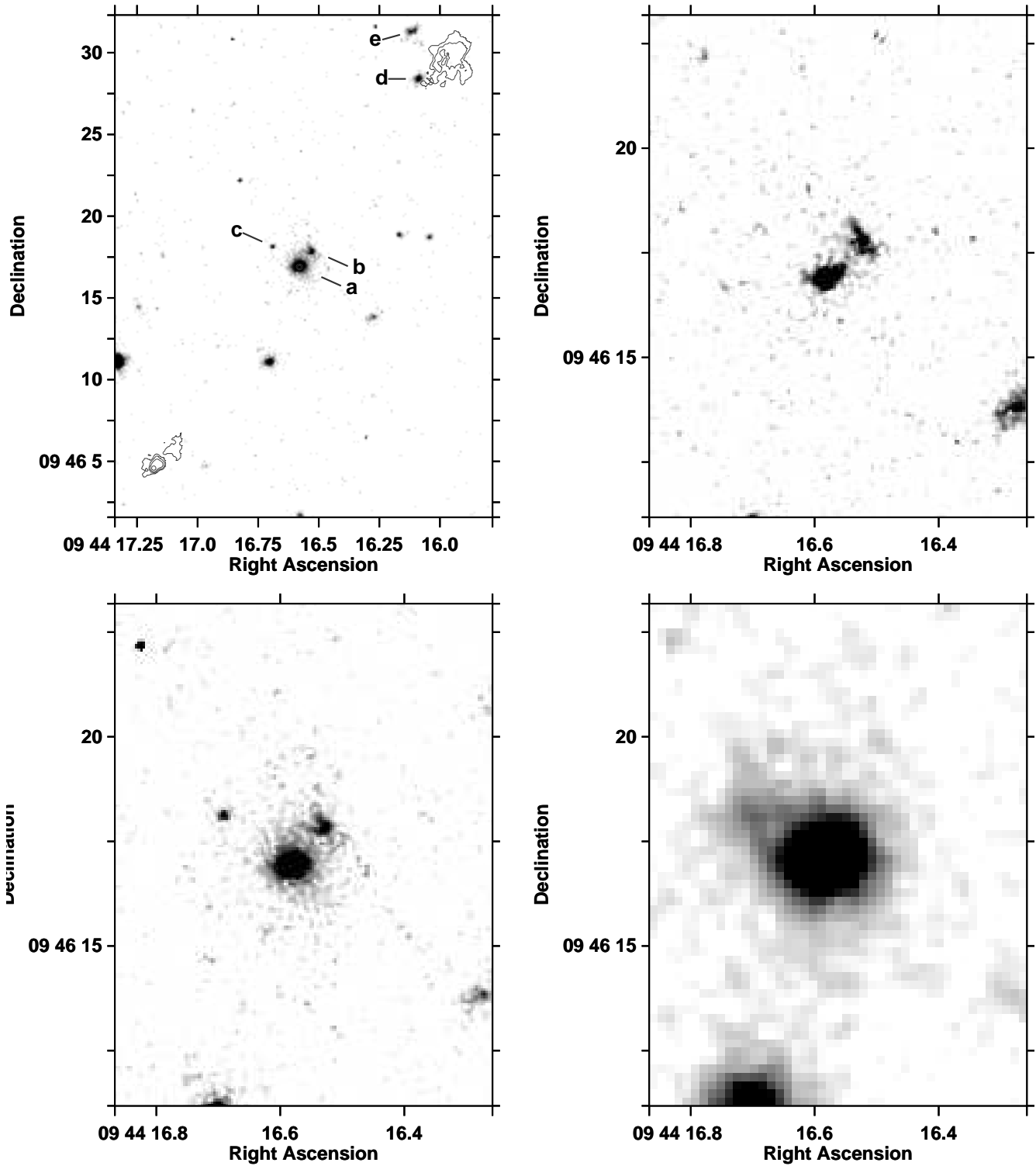,clip=,width=\textwidth}
}
\caption{\label{fig3c226} Images of the radio galaxy 3C226. (a -- top
left) The f785LP HST image, overlaid with contours of radio emission as
observed using the A and C arrays of the VLA. Contour levels are
$(1,4,16,64) \times 240 \mu$Jy beam$^{-1}$. (b -- top right) An
enlargement of the central regions of the f555W HST image. (c -- bottom
left) The central regions of the f785LP HST image to the same scale as
(b). (d -- bottom right) The UKIRT K--band image to the same scale as
(b).}
\end{figure*}

3C226, at redshift $z = 0.818$ (Figure~\ref{fig3c226}), displays a double
morphology, with a bright central emission region (`a') corresponding to
the position of the radio core, and a companion (`b') 10 to 15 kpc to the
NW along the radio axis. This companion is significantly bluer than the
central galaxy, with an f555W$-$f785LP colour of $1.96 \pm 0.14$ as
compared to $2.49 \pm 0.07$ for the latter. The f555W image (enlarged in
Figure~\ref{fig3c226}b) shows this companion itself to be extended
perpendicular to the radio axis. This component is not visible in our
K--band image (Figure~\ref{fig3c226}d), although it is marginally detected
in that of Rigler \etal\ \shortcite{rig92}; in both of these infrared
images, there is an extension to the north--east (perpendicular to the
radio axis), corresponding to a very red companion (`c'), visible in the
f785LP observation (Figure~\ref{fig3c226}a,c), but barely detected at
555\,nm. There are also two emission regions close to the north-western
radio lobe (`d' and `e'), the more northerly being particularly blue
(f555W$-$f785LP$ = 1.01 \pm 0.12$), although it is unclear whether these
are connected with the radio source.

If we use J$-$K colour as an indicator of the redshift of a source, since
at $z \lta 1$ this is relatively unaffected by any flat--spectrum UV
component, then 3C226 has a colour of J$-$K$= 1.71 \pm 0.15$, consistent
with that expected a standard giant elliptical galaxy at $z \sim 0.8$ (see
Figure~\ref{jkcols}), whilst components `d' and `e' have colours of
J$-$K$= 1.61 \pm 0.31$ and J$-$K$= 1.10 \pm 0.45$ respectively. This
indicates that, if these objects contain old stars, `d' may well be at the
same redshift at 3C226 , whilst the bluer object `e' is likely to be
foreground.

Di Serego Alighieri \etal\ \shortcite{dis94a} detected polarised continuum
emission from this source in three different wavebands. They measured
polarisation percentages of $P_B = 12.3 \pm 2.3\%$, $P_V = 13.3 \pm 4.2\%$
and $P_i = 2.5 \pm 1.4\%$, for the B, V and i bands respectively. To
account for the decrease in polarisation between the V and i bands, they
note that the i band lies at longer wavelengths than rest--frame 4000\ang
, and associate the decrease with dilution of the scattered component by
an underlying old stellar population. The amplitude of the 4000\ang\ break
required in the old stellar population to produce the observed decrease in
polarisation implies an age of $\sim 5$ Gyr, corresponding to a formation
redshift of $z_{\rm f} \gta 10$. These authors have also detected broad
MgII~2798 emission from the host galaxy of this source and, although the
narrow lines are all unpolarised, the broad MgII line is polarised at the
same percentage as the continuum. This is presumed to be due to scattered
light from the broad line region close to the active nucleus. If
scattering by hot electrons associated with the cluster gas ($T \sim
10^7$\,K) were responsible, spectral features such as this would be
smeared by Doppler effects, and so dust scattering or scattering by
electrons associated with the warm ($T \sim 10^4$\,K) emission line gas
are favoured.

\subsection*{3C239}

\begin{figure*}
\centerline{
\psfig{figure=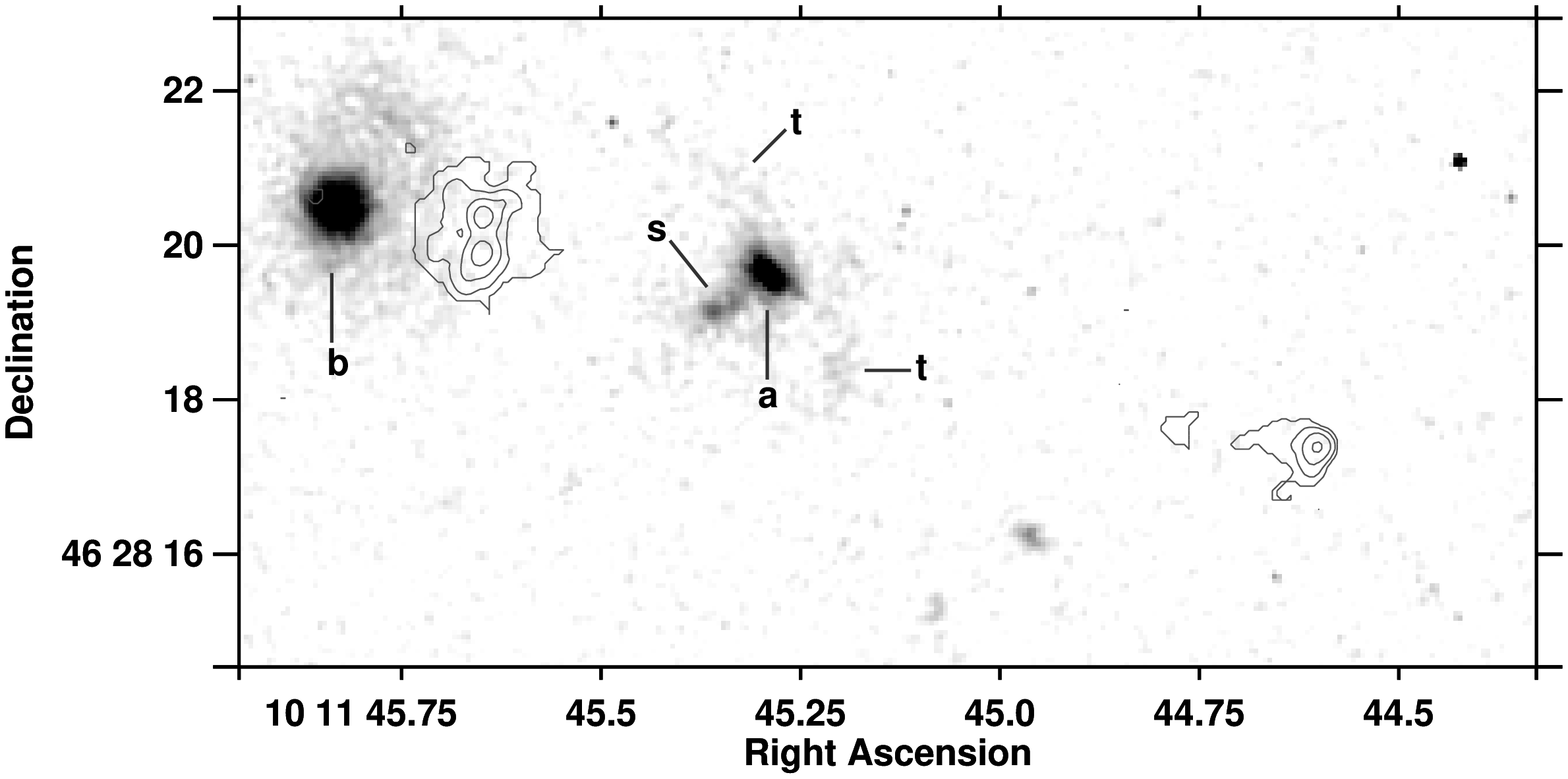,clip=,width=12cm,angle=90}
}
\centerline{
\psfig{figure=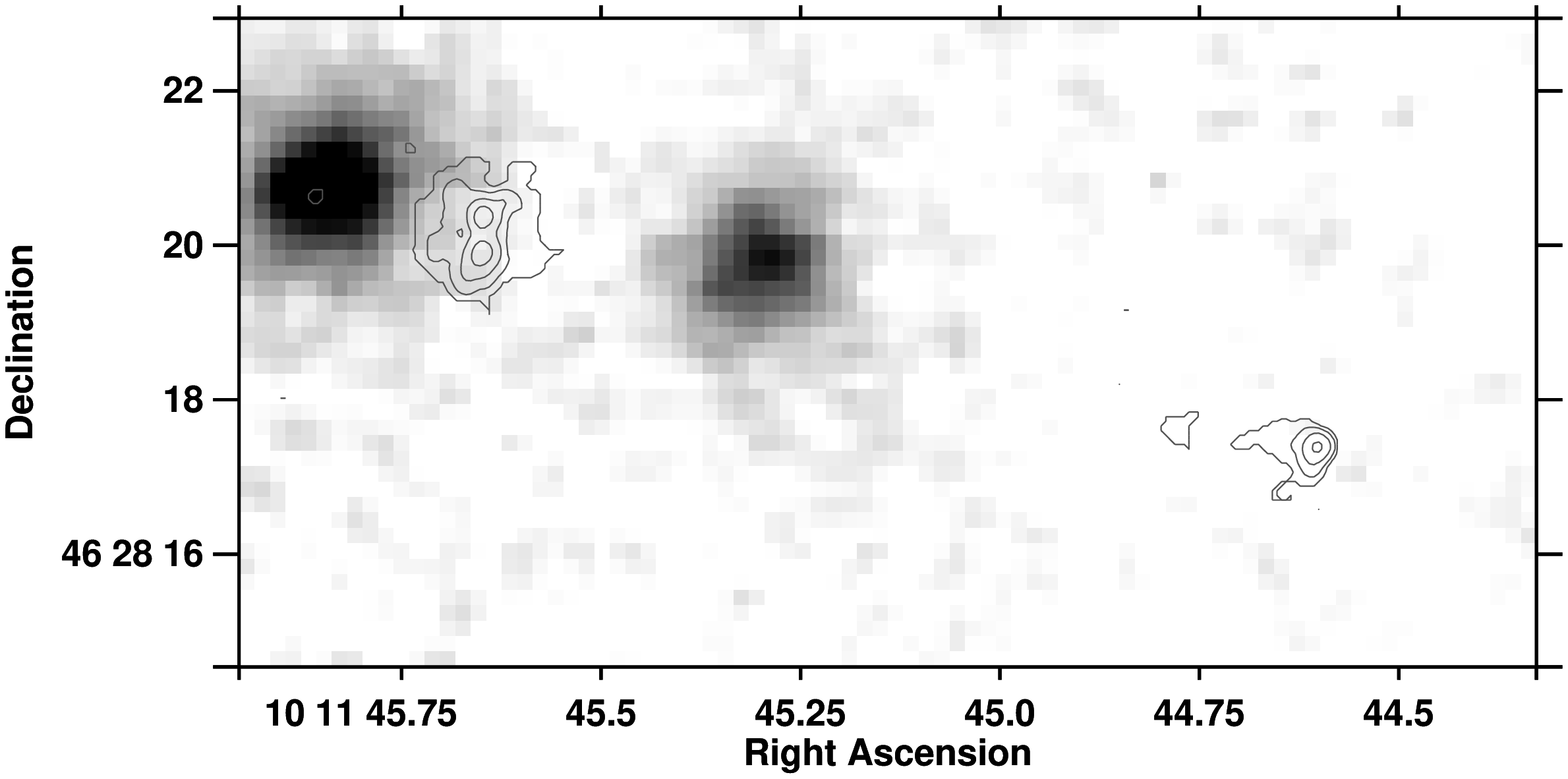,clip=,width=12cm,angle=90}
}
\centerline{
\psfig{figure=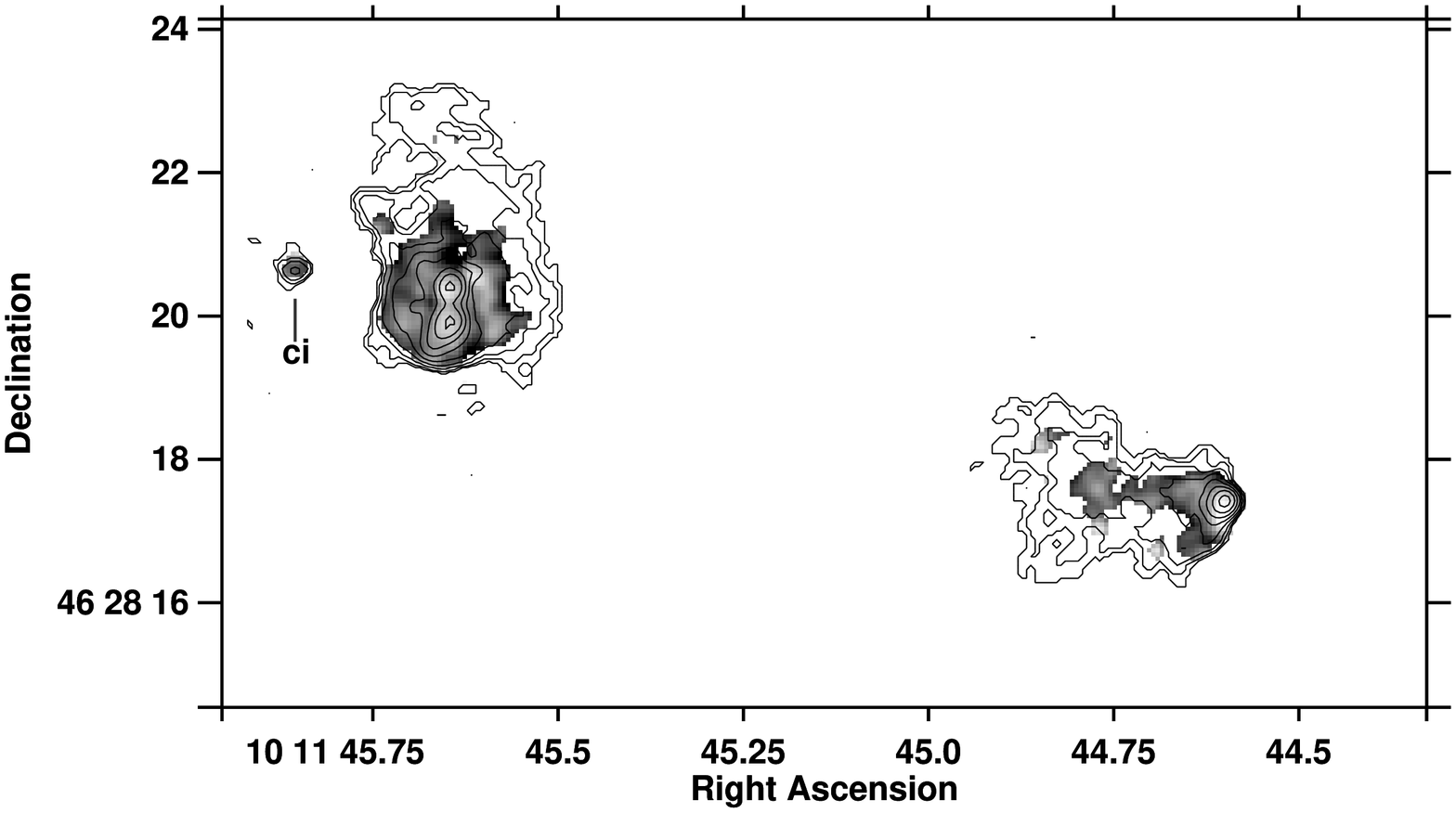,clip=,width=12cm,angle=90}
}
\caption{\label{fig3c239} Images of the radio galaxy 3C239. (a) The sum of
the two HST images taken using the f785LP and f814W filters, overlaid with
contours of radio emission, at levels of $(1,4,16,64) \times 160 \mu$Jy
beam$^{-1}$, from A and B array observations. (b) The UKIRT K--band image
with the same VLA radio contours overlaid. (c) The radio emission at
1.4~GHz, observed by MERLIN (courtesy of D. Law--Green), with contour
levels of $(1,2,4,8,16,32,64,128,256) \times 400 \mu$Jy beam$^{-1}$. The
grey--scale is the radio spectral index calculated between 1.4 and 8.4
GHz, and ranges from $\alpha = 0.8$ (light grey/white) to $\alpha = 2.0$
(black), plotted only in regions where the radio emission was detected
with signal--to--noise ratio greater than 3 in both images.}
\end{figure*}

\begin{figure}
\centerline{
\psfig{figure=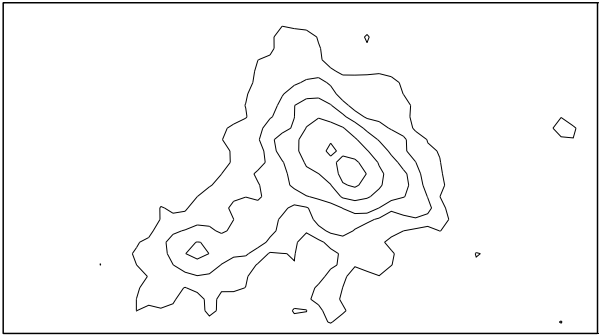,clip=,angle=90,width=8cm}
}
\caption{\label{3c239cont} A contour plot of the sum of the two HST images
of 3C239 showing the two distinct central components and the string of
emission to the south--east.}
\end{figure}

At redshift $z = 1.781$, 3C239 is the most distant galaxy in our sample,
and is correspondingly the most powerful radio source. The HST image
(Figure~\ref{fig3c239}a) shows a bright central galaxy (`a') which
consists of two distinct emission regions aligned within 25 degrees of the
radio axis. These are shown more clearly in the contour plot in
Figure~\ref{3c239cont}. In addition, a string of components (`s') extends
from the south--east of the galaxy, misaligned from the radio axis by some
45$^{\circ}$. The field also displays two large faint `tails' (`t')
stretching to the north and the west of the galaxy, extending 5 arcsec in
length ($\approx 40$\,kpc). The infrared image (Figure~\ref{fig3c239}b)
shows no evidence of these tails of emission, but has a marginal extension
corresponding to the string of components to the south--east.

These structures are reminiscent of the aftermath of galaxy collisions
observed in the local universe. In this picture, the two bright central
regions would represent the two merging galaxies, whilst tidal
interactions would induce the formation of structures `s' and `t'. Such a
collision would be a prime candidate for the fuelling of a quiescent black
hole, and inducing radio source activity in this source.

As reported by Hammer and Le F\`evre \shortcite{ham90}, 3C239 lies in a
crowded field in which the colours of the galaxies appear bimodal. They
suggest the presence of two populations of galaxies, one at the redshift
of 3C239, perhaps forming one of the most distant known clusters, and the
other at lower redshift. This second population includes galaxy `b', which
lies some 5 arcsec to the east of 3C239, close to the eastern radio lobe. Our
VLA image shows this lobe to contain a double hot--spot. More interesting
is the 1.4~GHz MERLIN image taken by Law--Green (Figure~\ref{fig3c239}c)
which, in addition to the double hot--spot, shows arc--like structures
within the radio lobe; both the eastern edge of the lobe and the more
diffuse northern emission are strikingly curved. There is also radio
emission from just to the east of galaxy `b' (labelled `ci'), which
possesses a similar spectral index ($\alpha \approx 1.4$) as the mean of
the radio lobe.

This may be an example of a gravitationally lensed system, with a
foreground galaxy `b' causing a distortion of the eastern radio lobe to
produce a double hot--spot, a set of radio arcs, and a counter--image
lying just to the east of the lensing galaxy. Lens system modelling
confirms that such a scenario is possible, with plausible parameters for
the lensing galaxy \cite{law97}. The alternatives are that the
`counter--image' is either radio emission associated with galaxy `b', or a
secondary hot--spot of 3C239, but neither of these possibilities seems
likely (see Law--Green \etal\ 1997). 

Spectral ageing analysis of the radio emission indicates that the lobe --
hotspot separation velocity of this source is $\sim 0.12c$, suggesting
that the age of the radio source is of order a few million years
\cite{liu92}.

\subsection*{3C241}

\begin{figure}
\centerline{
\psfig{figure=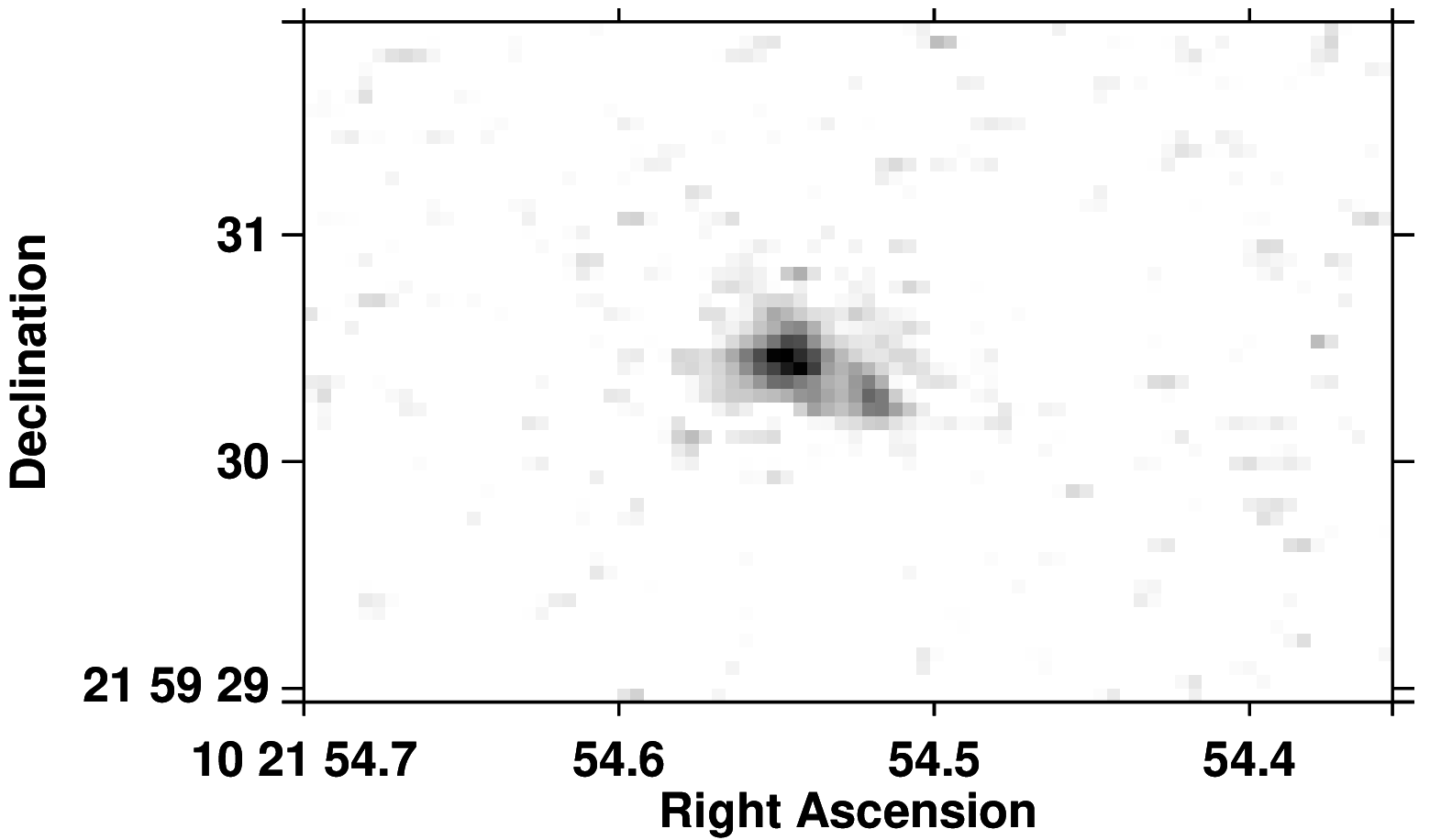,clip=,width=7.5cm,angle=90}
}
\centerline{
\psfig{figure=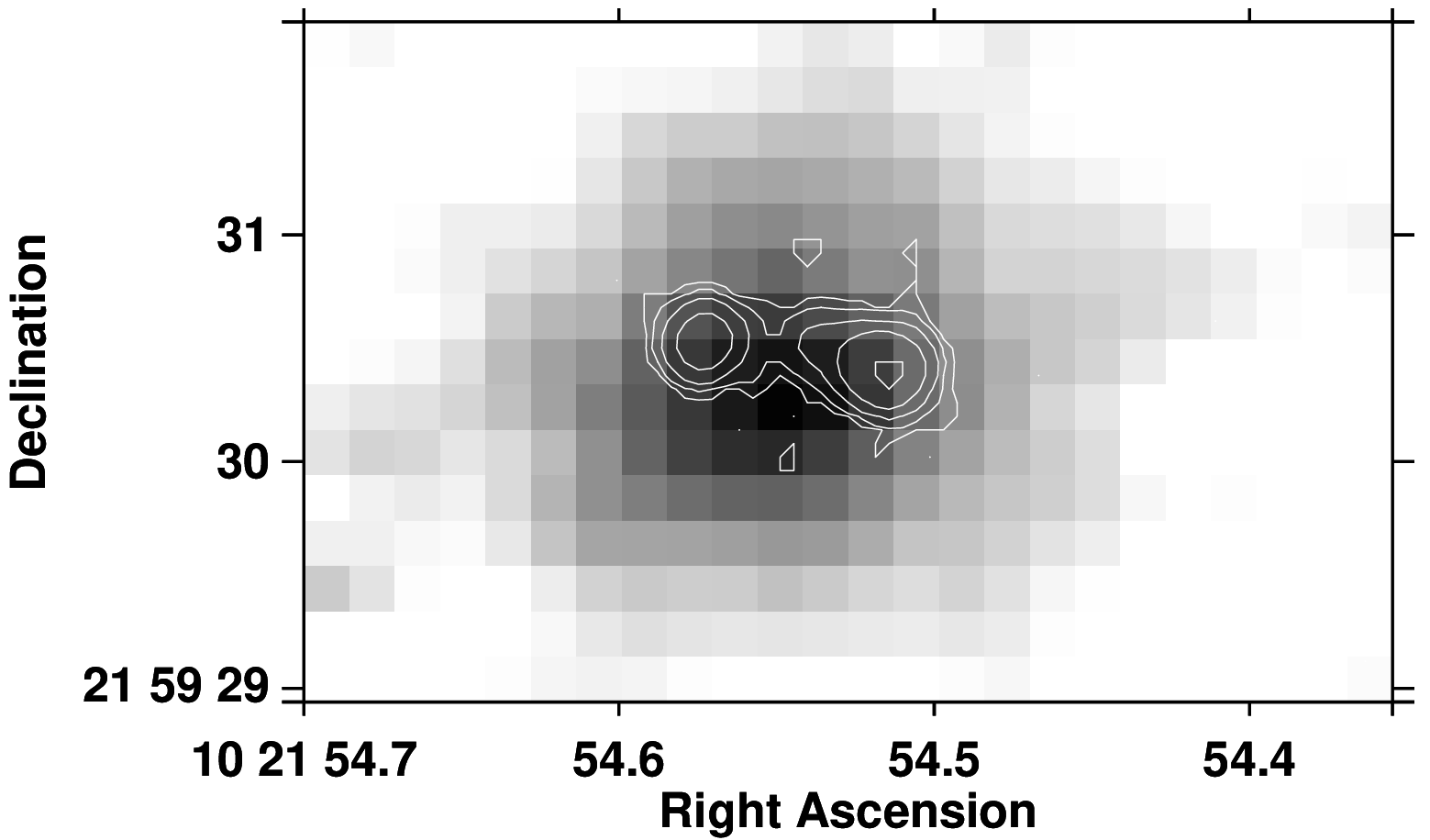,clip=,width=7.5cm,angle=90}
}
\caption{\label{fig3c241} Images of the radio galaxy 3C241. (a) The HST
image through the sum of the f785LP and F814W filters. (b) The UKIRT
K--band image, with the radio emission from the A array VLA observation
overlaid. Contour levels are $(1,4,16,64,256) \times 200\mu$Jy
beam$^{-1}$.}
\end{figure}

3C241, at redshift $z = 1.617$, is one of two compact steep spectrum
sources in our sample, having radio size just under 7~kpc. The HST image
(Figure~\ref{fig3c241}a) resolves the `elongation' seen in ground--based
imaging of this galaxy \cite{fev88a}, showing it to consist of two
separate emission regions aligned along the radio axis, on a scale similar
to the size of the radio source.  The radio emission is shown overlaid on
the K--band image (Figure~\ref{fig3c241}b) --- due to the small size of
this source, and the errors in the relative positioning of the optical and
radio frames, the precise position of the radio lobes with respect to the
optical emission is uncertain and the radio contours have therefore been
omitted from the HST plot to enable the image to be seen more clearly. The
infrared image shows that the radio source is completely contained within
the host galaxy.

A high resolution, deep MERLIN image at 5~GHz taken by Sanghera \etal\
\shortcite{san95}, and a combined EVN--MERLIN image at 1.6~GHz by Fanti
\etal\ \shortcite{fan85} give more detailed pictures of the radio source.
The western lobe is double, with two marginally separated components
connected by a bridge of radio emission, and a flat spectrum radio core
has been detected, positioned nearly symmetrically between the two lobes
seen in Figure~\ref{fig3c241}b. Both radio lobes are highly depolarised,
consistent with them lying within the host galaxy and Faraday
depolarisation having taken place within the interstellar medium.

\subsection*{3C247}

\begin{figure*}
\centerline{
\psfig{figure=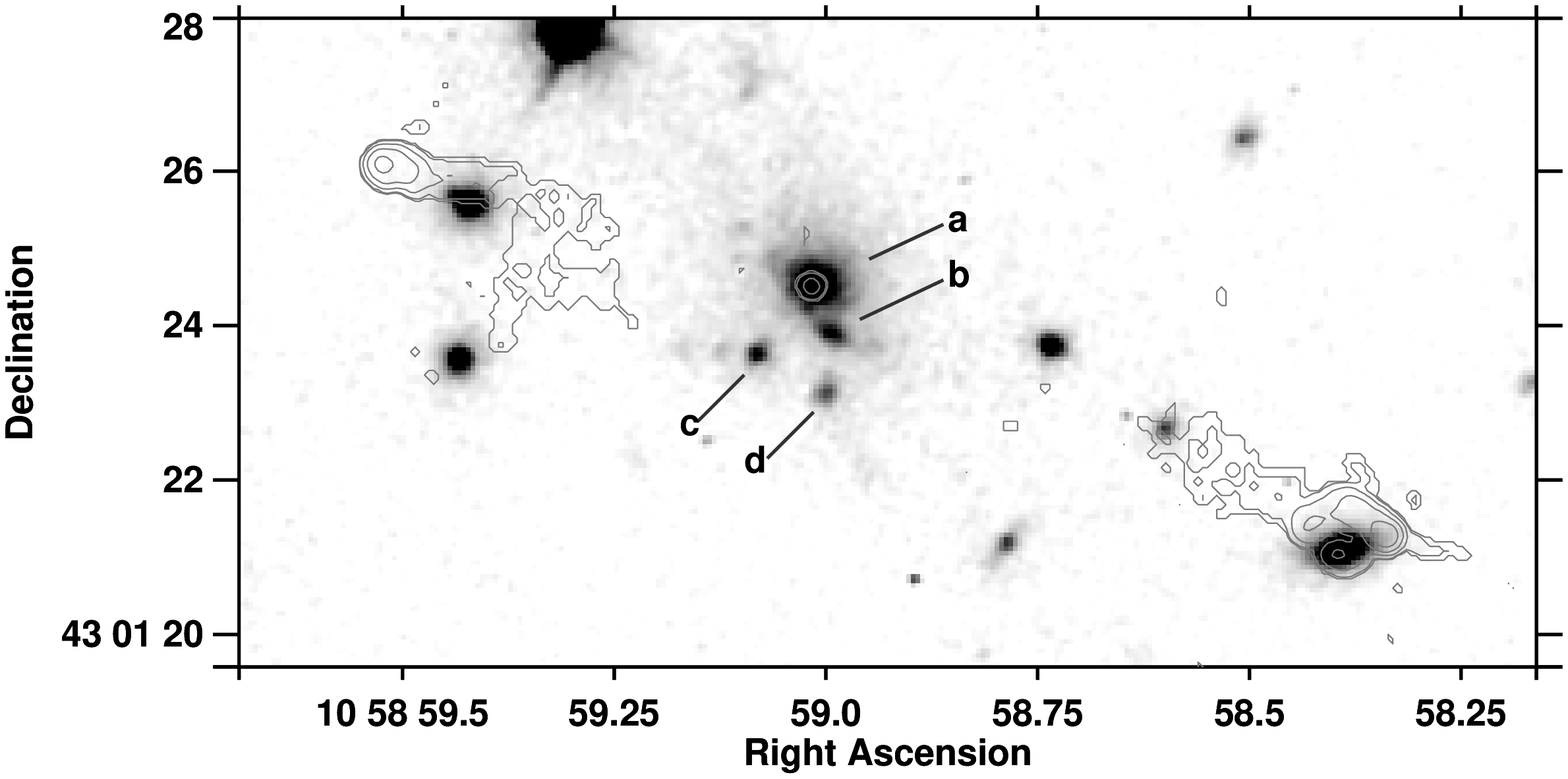,clip=,width=14cm,angle=90}
}
\centerline{
\psfig{figure=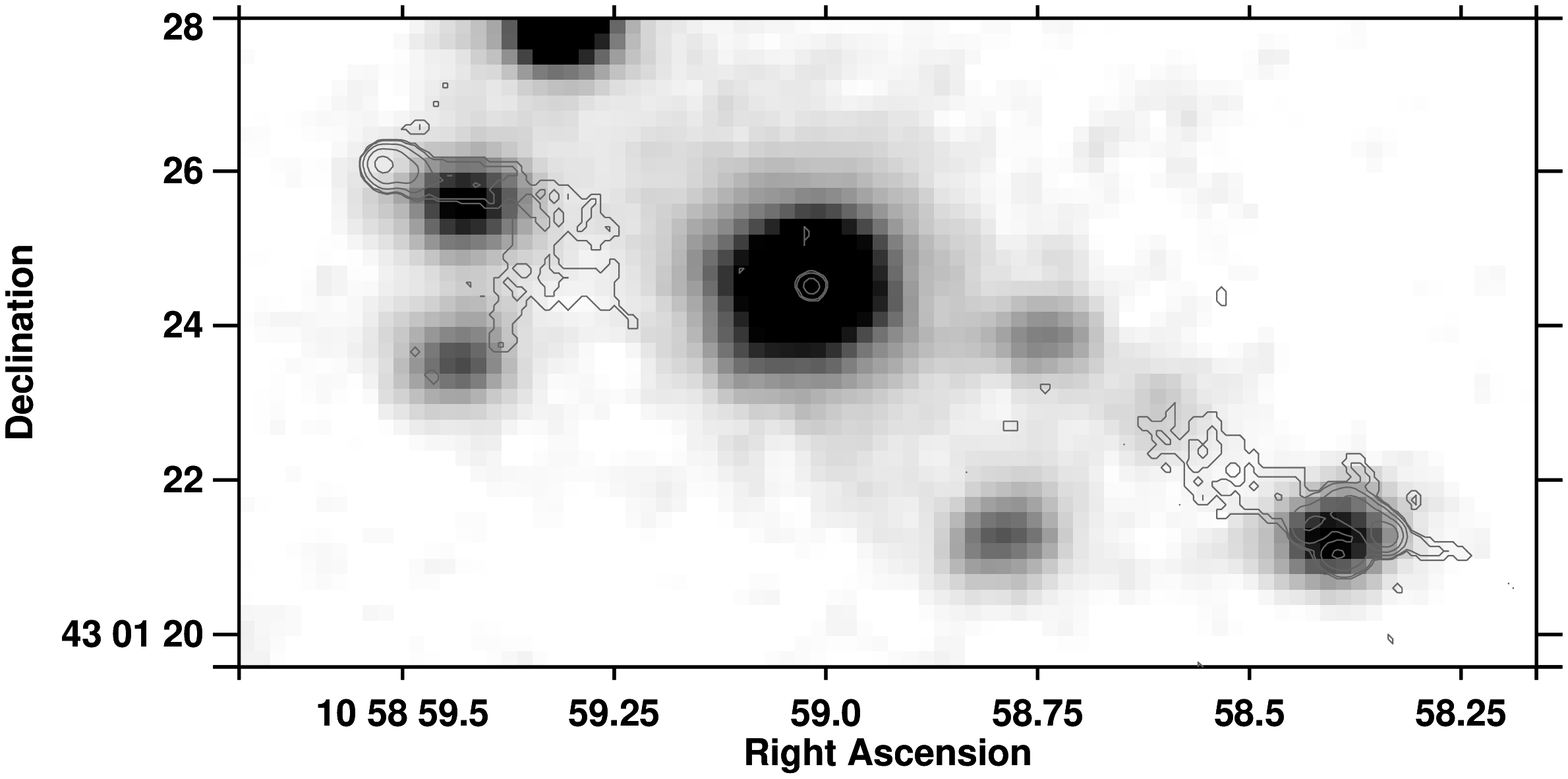,clip=,width=14cm,angle=90}
}
\caption{\label{fig3c247} Images of the radio galaxy 3C247. (a) The sum of
the two HST images, taken using the f555W and f814W filters, with contours
of radio emission overlaid. Radio contours are at $(1,2,8,32,128) \times
220 \mu$Jy beam$^{-1}$. (b) The UKIRT K--band image with radio contours as
in (a).}
\end{figure*}

The galaxy associated with the radio source 3C247, at redshift $z =
0.749$, lies in a very crowded field. The HST image of this galaxy
(Figure~\ref{fig3c247}a) shows a symmetrical central galaxy (`a'), with a
close companion, `b', lying about 0.8 arcsec to the south. Two other
galaxies, `c' and `d', lie within the envelope of the infrared emission,
which is displayed in Figure~\ref{fig3c247}b. The f555W$-$f814W colours of
these four galaxies are, respectively: $2.52 \pm 0.05$, $2.31 \pm 0.13$,
$2.43 \pm 0.16$ and $1.34 \pm 0.14$. Components `a', `b' and `c' are
therefore red, inactive objects, whilst component `d' is much bluer.

McCarthy \shortcite{mcc88} showed that the [OII]~3727 emission of this
source is extended towards the north--east. Our HST images clearly show
extended diffuse emission in this direction, which may be entirely due to
line emission, or may also contain continuum emission. McCarthy's
[OII]~3727 image also shows an extension to the south--west, with emission
from regions close to the galaxies seen near the radio lobe in
Figure~\ref{fig3c247}, indicating the possible presence of a cluster
surrounding 3C247. This extended emission line gas may be the cause of the
high radio depolarisation observed throughout the regions of the radio
lobes close to the nucleus \cite{liu91b}, and possibly also for the
distorted nature of the radio emission.

Radio spectral ageing analysis of this source has yielded an age of 3-5
million years, corresponding to a hotspot advance velocity of about $0.1c$
\cite{liu92}.

\subsection*{3C252}

\begin{figure*}
\centerline{
\psfig{figure=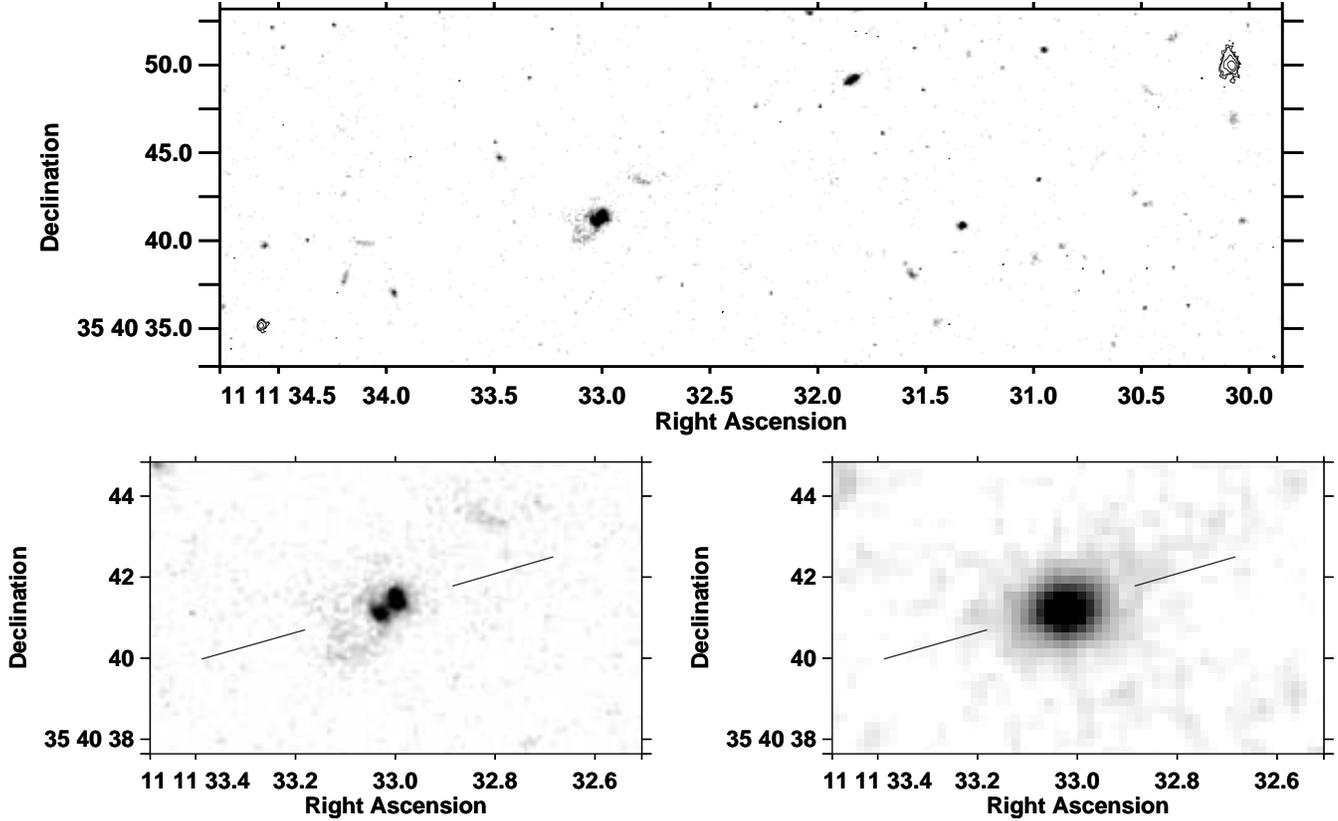,clip=,width=\textwidth}
}
\caption{\label{fig3c252} Images of the radio galaxy 3C252. (a) The sum of
the two HST images, taken using the f622W and f814W filters, with contours
of radio emission overlaid. Radio contours are $(1,2,8,32,128) \times 160
\mu$Jy beam$^{-1}$. (b) An enlargement of the combined HST image of the
central galaxy, (c) The UKIRT K--band image to the same scale as (b).}
\end{figure*}

The HST image of 3C252 (Figures~\ref{fig3c252}a,b), at redshift $z =
1.105$ shows two bright knots aligned along the radio axis. The close
separation of these knots makes it unclear whether they represent distinct
emission regions or whether a dust lane runs through the central regions
of the galaxy, as has been suggested for 3C324 \cite{lon95}. The f622W$-$f814W
colours of the two knots are $0.68 \pm 0.06$ and $0.88 \pm 0.09$ for the
western and eastern components respectively. Although these colours may
suggest that they are two distinct emission regions, this difference could
also be caused by differential reddening in a putative dust lane. In this
case the eastern component, being redder, would be seen through a greater
optical depth of dust and would therefore be the lobe orientated away from
us. This is consistent with this lobe being the shorter lobe in a
relativistic hotspot advance model (eg. Best \etal\ 1995). \nocite{bes95a}

The UKIRT K--band image (Figure~\ref{fig3c252}c) shows a giant elliptical
galaxy, with no evidence for the double optical components, although the
poorer resolution of this image would make these difficult to detect. It
does show a slight elongation along the radio axis, as does the image of
Rigler \etal\ \shortcite{rig92}, although that of Dunlop and Peacock
\shortcite{dun93} does not.

To the south--east of the central galaxy, the HST image shows a diffuse
emission region, of triangular morphology similar to that expected of an
ionisation cone of light scattered from an obscured nucleus. Surprisingly,
however, this region appears redder than the central emission regions
(f622W$-$f814W = $1.01 \pm 0.24$), although not at high
significance. Another region of diffuse emission lies 3 arcsec to the
north--west. Hammer and Le F\`evre \shortcite{ham90} suggested that this
object may have an arc--like structure characteristic of gravitational
lensing, but the HST image suggests instead that it is part of a more
diffuse structure.

\subsection*{3C265}

\begin{figure*}
\centerline{
\psfig{figure=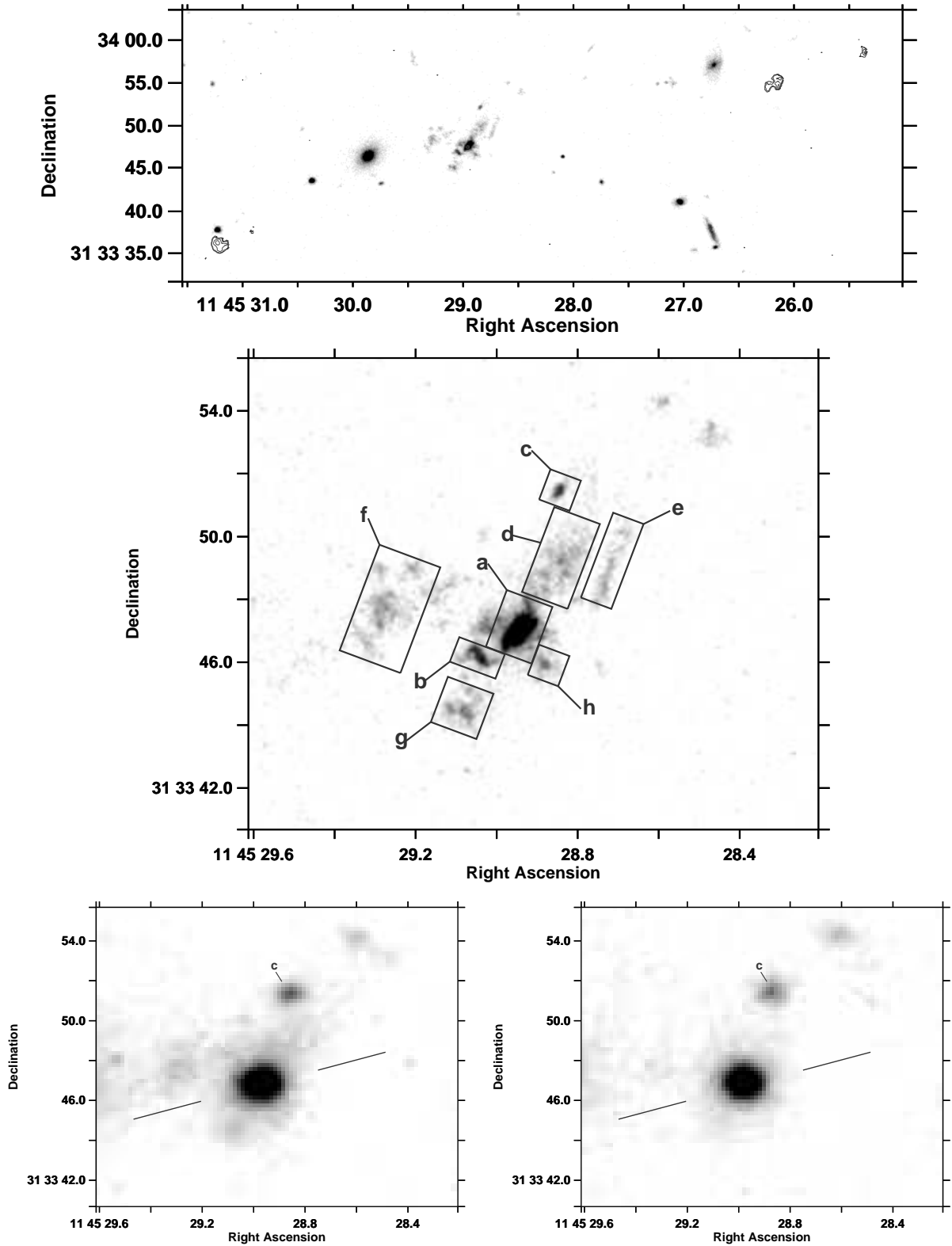,clip=,width=16cm}
}
\caption{\label{fig3c265} Images of the radio galaxy 3C265. (a) The sum of
the HST images overlaid upon contours of the radio emission as observed
with the A--array of the VLA; contour levels are $(1,2,8,32,128) \times
160 \mu$Jy beam$^{-1}$. The structure of the more diffuse radio lobes can
be found in the paper by Fernini \etal\ (1993). (b) An enlargement of the
central region of the HST image, showing the regions used for the colour
analysis. (c) The UKIRT J--band image, showing the same region of sky as
(b). (d) The UKIRT K--band image to the same scale as (c).}
\end{figure*}

The largest radio source in the sample, 3C265 ($z = 0.811$), possesses
perhaps the most bizarre optical morphology of all the galaxies in the
sample (Figures~\ref{fig3c265}a,b). The unusually bright central region is
completely surrounded by emission regions or companion galaxies, extending
nearly 50~kpc from the centre. This source was discussed by Longair \etal\
\shortcite{lon95} who considered that the optical structures may either be
the aftermath of a galaxy merger or alternatively be associated with large
clouds of gas, possibly cooling out of intracluster gas due to compression
by the passage of the radio beams. The second possibility is consistent
with the detection of X--ray gas surrounding a number of these powerful
radio galaxies (Crawford and Fabian 1996b, and references therein)
\nocite{cra96b} which suggests that cooling flows of up to $\sim
1000\,M_{\odot}\rm{yr}^{-1}$ may be infalling onto some central radio
galaxies.  This process would be particularly important in this, the
largest and hence one of the oldest radio sources.

Tadhunter \shortcite{tad91} reported the presence of high velocity gas in
the extended emission line regions of this source. To the SE of the
nucleus he detected three components of emission--line gas, at $-100$,
$+750$ and $+1550\,\rm{km}\,\rm{s}^{-1}$ with respect to the velocity at
the continuum centroid. These velocities are detectable in many lines, and
cannot be caused by gravitational effects from a single galaxy, nor
probably from a cluster. Tadhunter's preferred explanation is that these
emission line regions are associated with cocoons of material expanding
around the radio jets, perhaps remnants of the bow--shocks associated with
the passage of the radio jet. This supports the hypothesis of a cooling
flow stimulated by the radio jets. Although the contribution of line
emission may be high in the f785LP filter and the narrow--band [OII]~3727
image of Rigler \etal\ \shortcite{rig92} shows that the line emission
extends over the entire area of the source, the f555W image contains
little line emission and displays a similar morphology. Therefore the
extended structures cannot be solely due to line emission.

Jannuzi \etal\ \shortcite{jan91} have shown that this galaxy is polarised
at the 8 -- 9\% level, with the position angle of the electric field
vector orientated perpendicular to the extended UV emission of the host
galaxy rather than to the radio axis. Hammer \etal\ \shortcite{ham93}
presented a spectrum of the galaxy and argued that there is no evidence of
a 4000\ang\ break, although Cimatti \etal\ \shortcite{cim94} argued that
the presence of emission lines close to 4000\ang , and dilution by a UV
component make it difficult to quantify this result. Dey and Spinrad
\shortcite{dey96} have detected a weak stellar absorption feature (the
CaII~3933 K--line) in their spectrum, indicating that the system does
contain stars. They also observed broad MgII~2798 emission, both from the
nuclear region and from the spatially extended galaxian light, indicating
the presence of an obscured quasar nucleus towards the centre of this
source. Di Serego Alighieri \etal\ \shortcite{dis96} have observed the
source using spectropolarimetry. They confirm the level of continuum
polarisation measured by Jannuzi \etal\ \shortcite{jan91}, and also detect
broad MgII~2798 line emission in polarised light.

In contrast to the bizarre optical morphology, the K--band image
(Figure~\ref{fig3c265}d) shows only a bright central galaxy, together with
a red companion, galaxy `c', 4 arcsec to the NNW. In the J--band
(Figure~\ref{fig3c265}c) some of the diffuse emission is also visible. In
Table~\ref{265cols} we list the optical colours of the various components
of this source. The companion galaxy `c' is significantly redder than the
extended diffuse emission, whilst the two most southerly components are
both much bluer. There is no immediately obvious explanation for these
colour variations. In most models of the alignment effect the extended
aligned emission should be bluer than the central emission, but comparing,
for example, components `a' and `f', this is clearly not the case in
3C265. This might suggest that the optical morphology is influenced by
dust extinction; for example, this might be the cause of the dark strip
between components `d' and `e'. Reddening by this same dust might then
account for the colour differentials throughout the source. As a test, we
split object `d' into three strips of 3 arcsec by 0.5 arcsec and
calculated the colours in each of these regions; the values are given in
Table~\ref{265cols}. Again we see very strong colour gradients, in this
case even within a single emission region. This is characteristic of
differential reddening by dust but, surprisingly, the westernmost
regions of component `d' are the bluest, whereas if the gap between `d'
and `e' were due to dust extinction then this ought to be the reddest
region. Also, were dust extinction responsible for the morphology and
colours of this source, then the intrinsic `un--extincted' flux density
must be far higher than that observed, and already we observe this source
to be by far the brightest 3CR galaxy in the sample (eg. see
Table~\ref{hsttab}).

This source remains a mystery. Both its optical luminosity and its bizarre
morphology make it unique amongst the 3CR radio galaxies, whilst the fact
that it has a very high optical activity despite being the largest radio
source in the sample is also somewhat different from the general trend of
the other galaxies (eg. Best \etal\ 1996a). This might indicate that only
for the very largest sources has there been sufficient time for gas
compressed by the radio bow shocks to cool; the diffuse emission regions
near the centre of the other large radio sources such as 3C252, 3C277.2
and 3C356 may represent the start of this process in those sources.

\begin{table}
\begin{tabular}{ccc}
Component & f555W$-$f785LP & Error \\
    a     &   1.91         & 0.03  \\   
    b     &   1.89         & 0.09  \\
    c     &   2.57         & 0.20  \\
    d     &   2.15         & 0.07  \\
    e     &   2.00         & 0.13  \\
    f     &   2.22         & 0.09  \\
    g     &   1.59         & 0.11  \\
    h     &   1.05         & 0.19  \\
          &                &       \\
 d (east) &   2.45         & 0.15  \\
d (centre)&   2.29         & 0.14  \\
 d (west) &   1.79         & 0.13  \\
\end{tabular}
\caption{\label{265cols} The f555W$-$f785LP colours of the various
different components comprising 3C265. The different regions are
identified in Figure~\ref{fig3c265}b.}
\end{table}

\subsection*{3C266}

\begin{figure*}
\centerline{
\psfig{figure=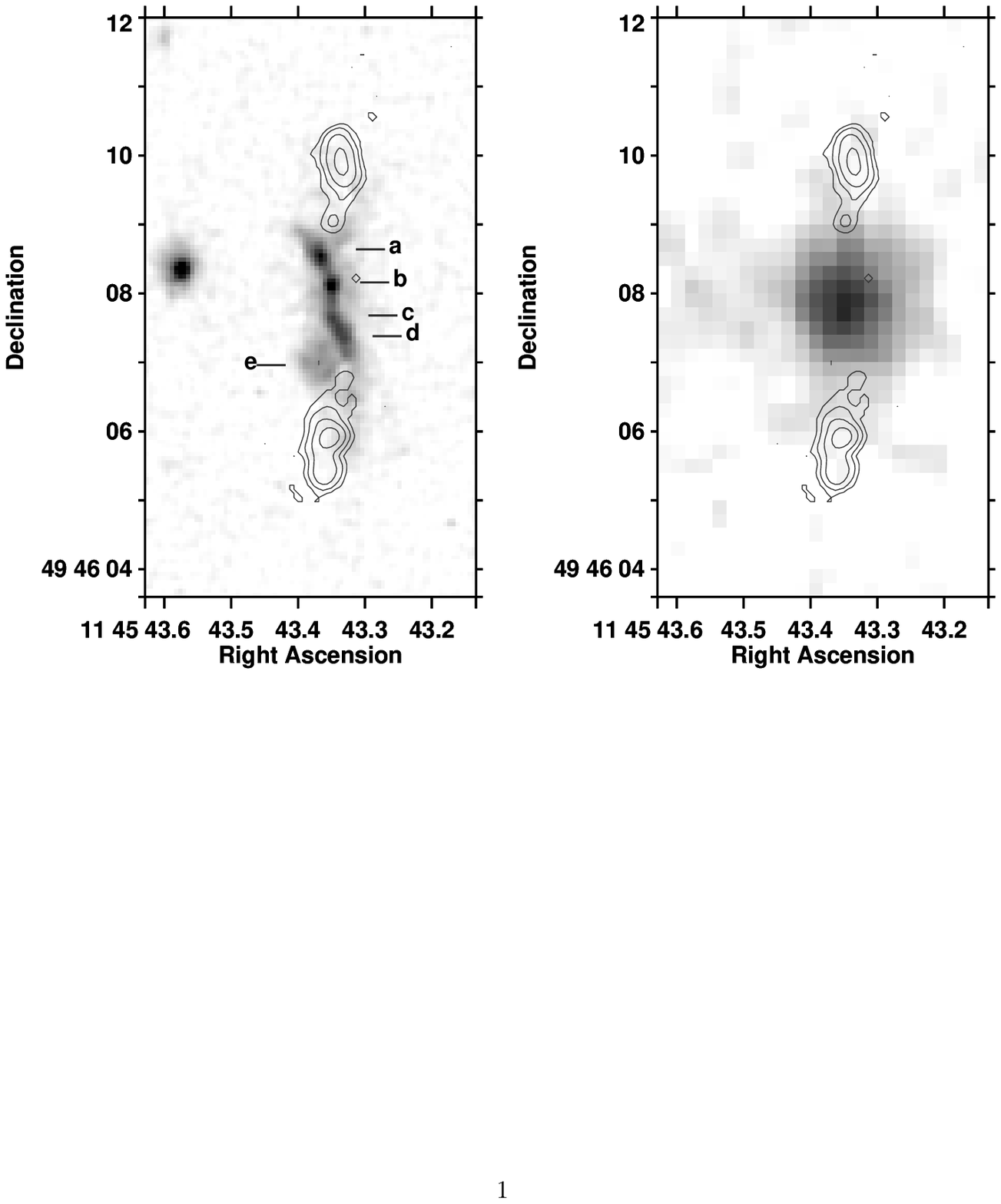,clip=,width=13.5cm}
}
\caption{\label{fig3c266} Images of the radio galaxy 3C266. Overlaid are
contour of radio emission as observed using the A array of the VLA. Radio
contour levels are $(1,4,16,64) \times 200 \mu$Jy beam$^{-1}$. (a) The sum
of the three HST images, taken using the f555W, f702W and f814W
filters. (b) The UKIRT K--band image. }
\end{figure*}

3C266, at redshift $z = 1.275$, is one of the smaller double radio sources
in the sample, extending just under 40~kpc. Misaligned only $10^{\circ}$
from the radio axis, a string of optically bright knots extends 15~kpc in
length from the inner edge of one radio lobe to the other
(Figure~\ref{fig3c266}a).  Underlying these, and even more closely aligned
with the radio structures is highly elongated diffuse emission. Relative
to these optical features, the K--band emission is much more compact and
circularly symmetric showing only a slight north--south extension
(Figure~\ref{fig3c266}b, Rigler \etal\ 1992, Dunlop and Peacock 1993).
Images taken through a narrow--band filter centred on [OII]~3727 show that
the line emission associated with 3C266 is as extended along the radio
axis as the continuum \cite{rig92,ham90}; Jackson and Rawlings
\shortcite{jac97} have recently detected extended [OIII]~5007 line
emission as well.

A companion galaxy lies 3 arcsec to the east, although this is not visible
in the narrow band [OII]~3727 images, and so could possibly be a foreground
object. Hammer \etal\ \shortcite{ham86} suggested that the extended
morphology of 3C266 could be a result of gravitational lensing either by
this companion, or by the rich cluster A1374 ($z = 0.209$), the centre of
which lies only 3 arcmins away from 3C266. The HST image does not support
a multiple--image lensing interpretation; whilst some brightening by
gravitational amplification cannot be ruled out, the fact that the
K--magnitude of this galaxy is fainter than the mean K$-z$ relationship
means that this is unlikely to be significant.

\begin{table}
\begin{tabular}{ccc}
Component & f555W$-$f814W & Error \\
    a     &      1.32     & 0.11  \\
    b     &      1.54     & 0.12  \\
    c     &      1.46     & 0.17  \\
    d     &      1.41     & 0.11  \\
    e     &      1.27     & 0.14  \\
\end{tabular}
\caption{\label{266cols} The f555W$-$f814W colours of the various
different components comprising 3C266. The various components are
identified on Figure~\ref{fig3c266}a.}
\end{table}

In Table~\ref{266cols} we list the colours of the various knots in
3C266. Although the results are not highly significant, it is interesting
that the further a knot is from the central regions of the galaxy, the
bluer it appears to be. There are a number of possible reasons why this
might be the case: (i) if the alignment is due to star--formation induced
by the passage of the radio jet, then the knots closest to the radio
hotspots will represent the most recent regions of star formation and
therefore will be bluest; (ii) the two filters contain different amounts
of line emission --- if the line emission is not evenly distributed
throughout the source, this could lead to apparent colour
differences; (iii) differential reddening by dust intrinsic to the source
can affect the observed colours --- in particular if the source contains a
central dust lane this could redden the two central knots; (iv) dilution
of the central regions of the active ultraviolet emission by a an
underlying red old stellar population. High resolution observations in the
ultraviolet and infrared wavebands would provide further colour
information and, together with mapping of the emission line structures of
these sources, would help to distinguish between these alternatives.

Spectral ageing of the radio structures suggests that the lobe--hotspot
separation velocity in this source is of order $0.18c$ \cite{liu92} which
would suggest an age for the radio source of 1--2 million years. Liu
\etal\ also report the possible detection of a radio core, of flux density
0.4~mJy at 5~GHz, located symmetrically between the radio hotspots.

\subsection*{3C267}

\begin{figure*}
\centerline{
\psfig{figure=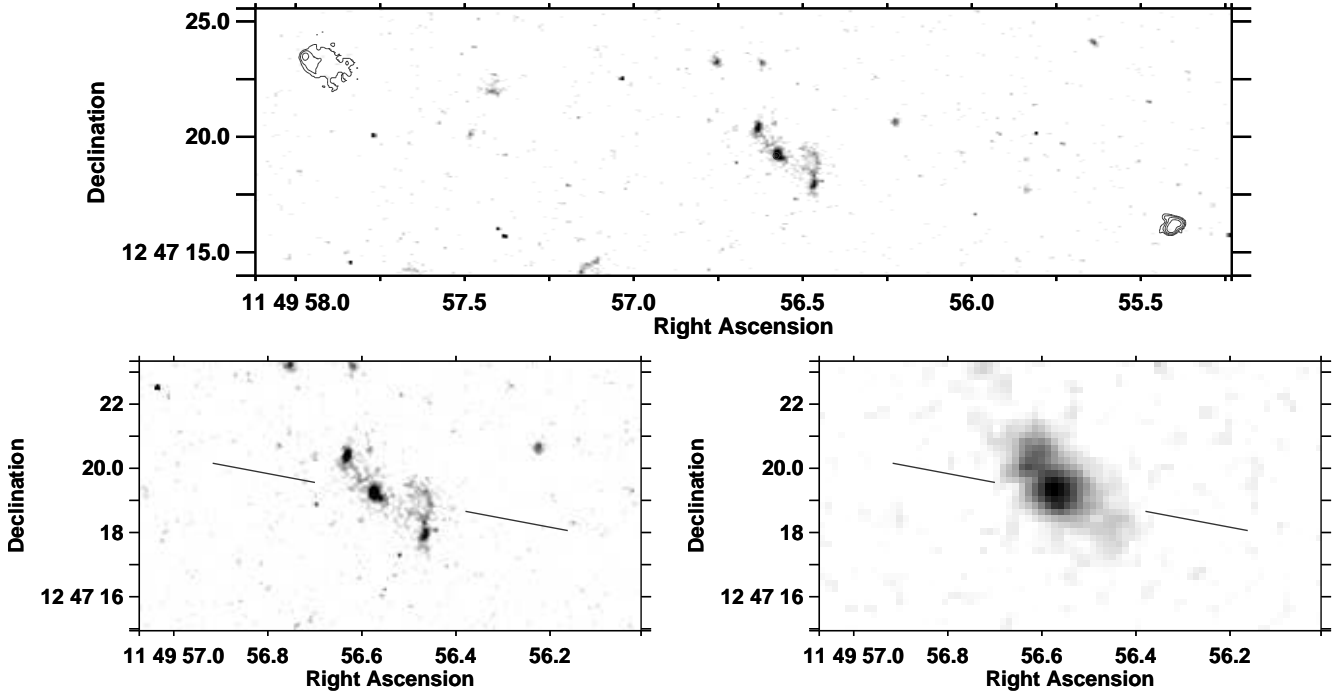,clip=,width=\textwidth}
}
\caption{\label{fig3c267} Images of the radio galaxy 3C267. (a) The sum of
the two HST images, taken using the filters f702W and f791W. Overlaid are
contours of radio emission from the A array observation, with contour
levels of $(1,4,16,64) \times 240 \mu$Jy beam$^{-1}$. (b) An enlargement
of the central regions of the HST images. (c) The UKIRT K--band image to
the same scale as (b).}
\end{figure*}

The HST image of 3C267, overlaid on the A--array VLA radio map, is
presented in Figure~\ref{fig3c267}a. An enlarged image of the central
galaxy is shown in Figure~\ref{fig3c267}b. This galaxy, at redshift $z =
1.114$, consists of three knots of emission separated by over 10~kpc, and
these are misaligned from the radio axis by about $27^{\circ}$.  Faint
diffuse emission stretches between the knots, and there is also an
arc--like structure, reaching from the north of the western knot back
towards the central knot. Best \etal\ \shortcite{bes96a} suggested that
this could plausibly be due to relaxation of a star--forming region within
the gravitational potential of the host galaxy. This arc structure and the
south--western emission region have an f555W$-$K colour of $2.36 \pm
0.25$, a magnitude bluer than the rest of the galaxy ($3.40 \pm 0.15$),
and are barely visible in the K--band image (Figure~\ref{fig3c267}c). The
K--band emission is more nucleated that the optical emission but, as in
the infrared images of Rigler \etal\ \shortcite{rig92} and Dunlop and
Peacock \shortcite{dun93}, does show an elongation in the same direction
as the optical image.

A narrow band image of this field taken by McCarthy \shortcite{mcc88}
through a filter centred on the [OII]~3727 emission line, showed that
three galaxies to the south--west of 3C267 have excess emission through
this filter. This suggests that they may be at the same redshift, meaning
that 3C267 may be a member of a distant cluster.

\subsection*{3C277.2}

\begin{figure*}
\centerline{
\psfig{figure=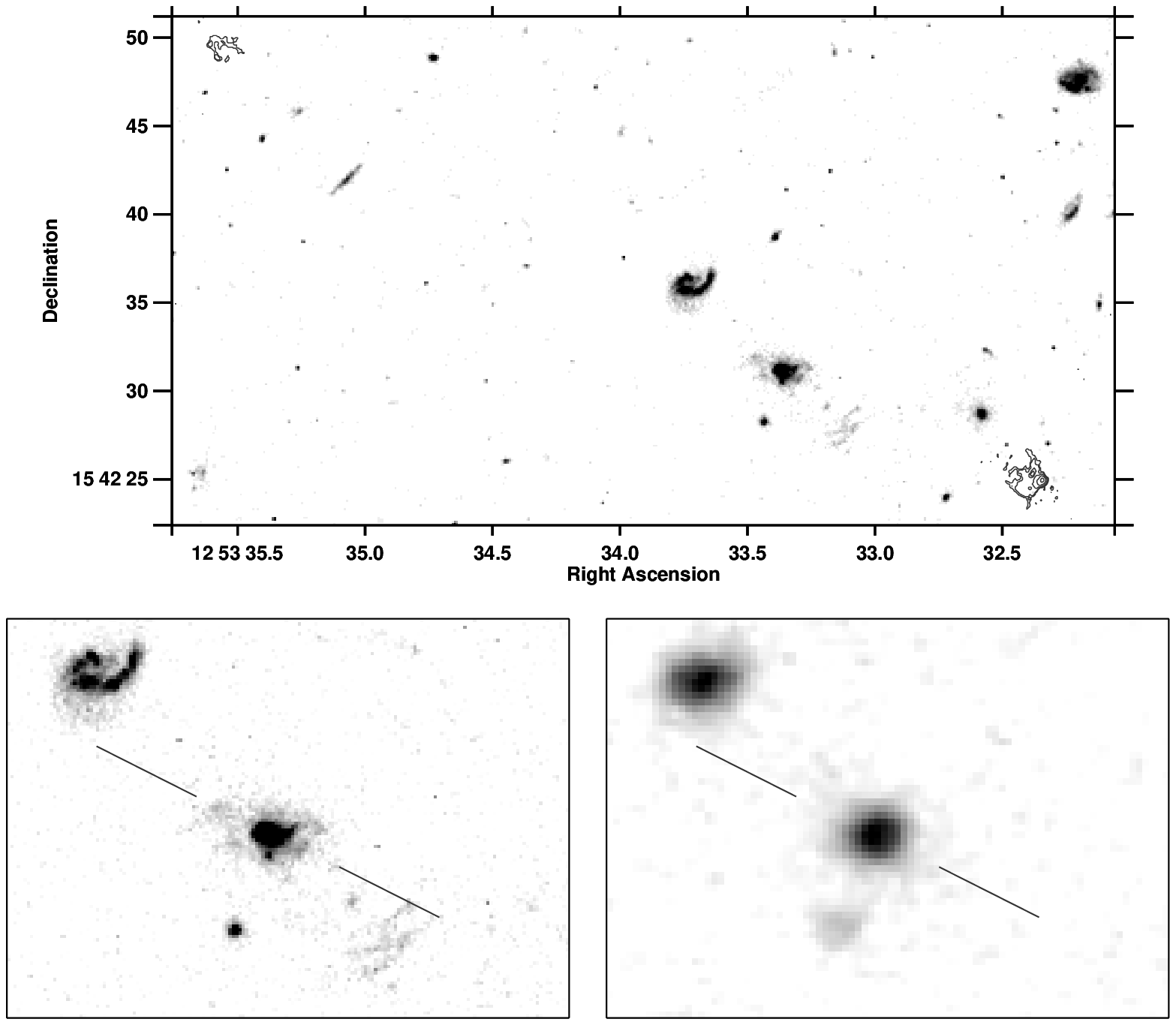,clip=,width=\textwidth}
}
\caption{\label{fig3c277.2} Images of the radio galaxy 3C277.2 (a) The sum
of the two HST images, taken using the filters f555W and f814W. Overlaid
are contours of radio emission from the A array observation, with contour
levels of $(1,2,8,32,128) \times 220 \mu$Jy beam$^{-1}$. (b) An
enlargement of the central regions of the HST images. (c) The UKIRT
K--band image to the same scale as (b).}
\end{figure*}

The HST images of 3C277.2 ($z = 0.766$, RA: 12 53 33.4, Dec: 15 42 31),
with VLA contours overlaid, is shown in Figure~\ref{fig3c277.2}a, and an
enlargement of the central regions in Figure~\ref{fig3c277.2}b.  The radio
galaxy shows a small elongation along the radio axis, and a region of
diffuse emission to the south--west. This emission coincides with the peak
flux density from a region of extended [OII]~3727 emission \cite{mcc88},
stretching over 30 arcseconds from the central radio galaxy. This emission
line gas causes strong depolarisation of the radio emission from the
western lobe, whilst the eastern lobe has no significant depolarisation
\cite{ped89a}. The western hotspot lies much closer to the radio core than
that on the other side, perhaps due to strong interaction of the radio jet
with the emission line gas on this side of the source. The emission from
this region in the HST images lies roughly along the inner edge of the
radio lobe, as defined by lower resolution radio maps, and is somewhat
extended along the lobe boundary, providing further evidence for such an
interaction.

Emission from these regions is barely visible in the J--band image, and
not seen at all in the K--band image (Figure~\ref{fig3c277.2}c). Instead
the infrared images show a fairly symmetrical galaxy, together with a
companion to the south--east. This companion is redder than the central
radio galaxy, with an f814W$-$K colour of $3.02 \pm 0.22$ as compared to
$2.43 \pm 0.10$. It is clearly present on McCarthy's \shortcite{mcc88}
narrow band [OII] image, and so is likely to lie at the same
redshift. Other objects in the field, predominantly to the south--west,
also show excess [OII] emission, suggesting the presence of a cluster
associated with 3C277.2. The strange galaxy to the north--east of the
radio galaxy is not seen in the [OII] image, and so is likely to be
a foreground object.

The optical emission from the radio galaxy is quite highly polarised in
the B--band, at about the 20\% level, suggesting that a significant
proportion of the ultraviolet emission from this source is non--stellar in
origin \cite{dis89}. The position angle of the polarisation is $164 \pm 6$
degrees, nearly perpendicular to the radio axis. A dust mass of about
$10^8 M_{\odot}$ is required to produce sufficient scattered light. At
longer wavelengths, however, the polarisation falls sharply, being $7.3
\pm 1.4$\% in the i band \cite{dis94a}. This band lies longward of
4000\ang\ in the rest--frame of the source, and so this result can be
explained by dilution of a flat spectrum scattered component by emission
from an old stellar population of age at least 2~Gyr. This is consistent
with the smooth morphology of the infrared images.

In an X--ray observation of this field, the source was detected with a
signal--to--noise ratio of only 3, and an X--ray luminosity of $\sim 4
\times 10^{43}$ erg\,s$^{-1}$ found in the ROSAT X--ray band
\cite{cra95}. This emission is likely to be predominately due to the hot
intracluster gas.

\subsection*{3C280}

\begin{figure*}
\centerline{
\psfig{figure=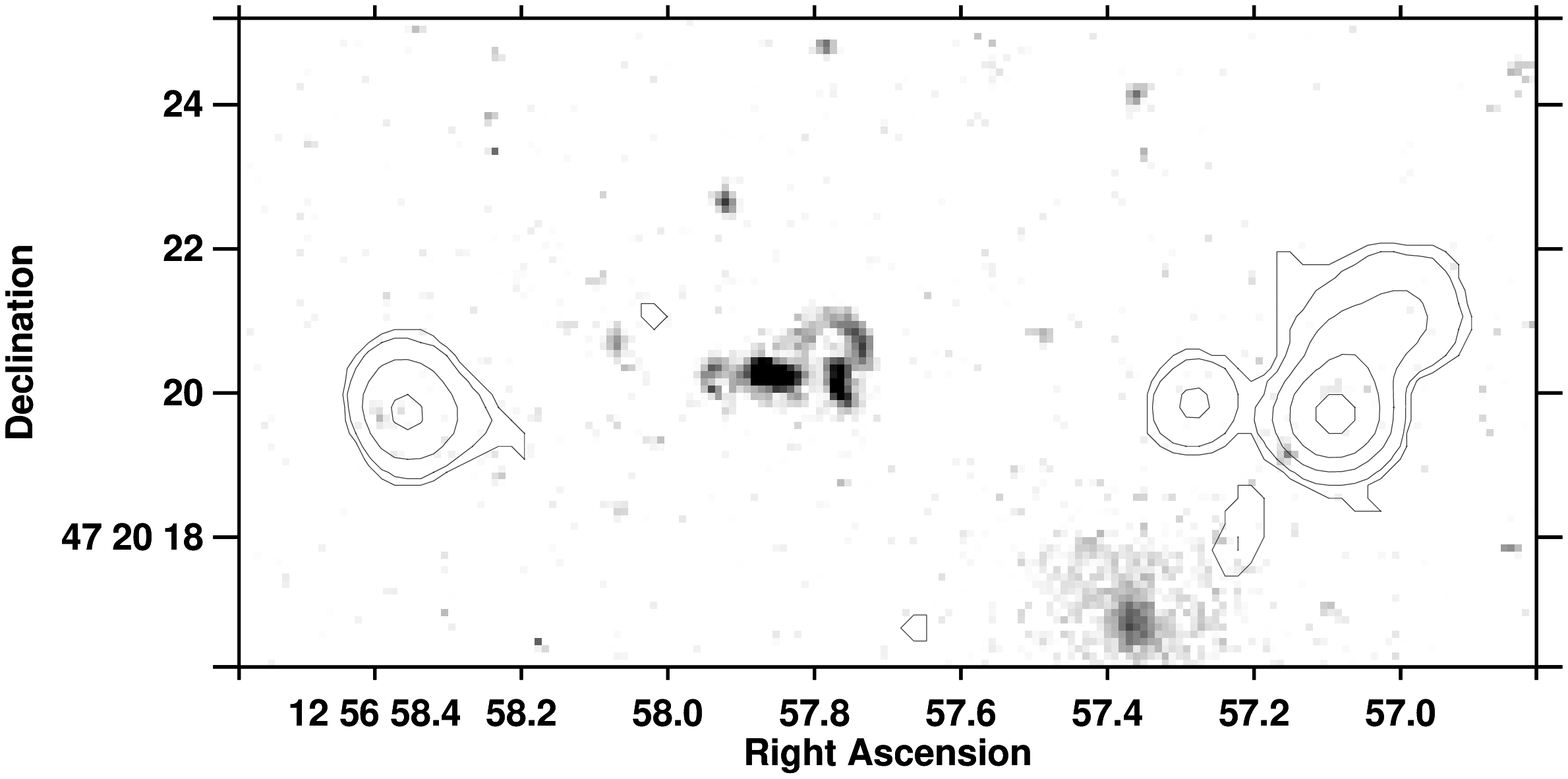,clip=,angle=90,width=13cm}
}
\centerline{
\psfig{figure=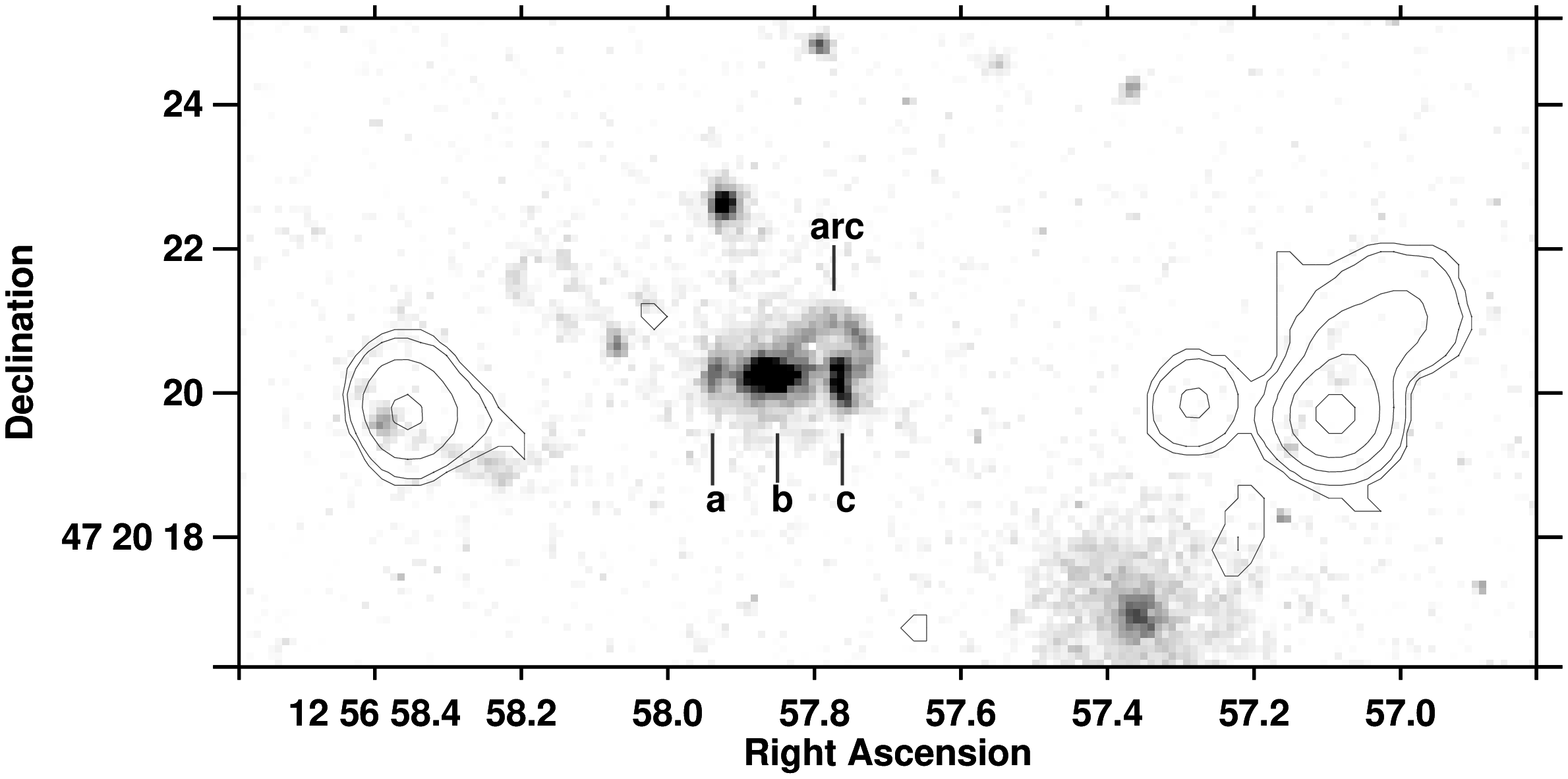,clip=,angle=90,width=13cm}
}
\centerline{
\psfig{figure=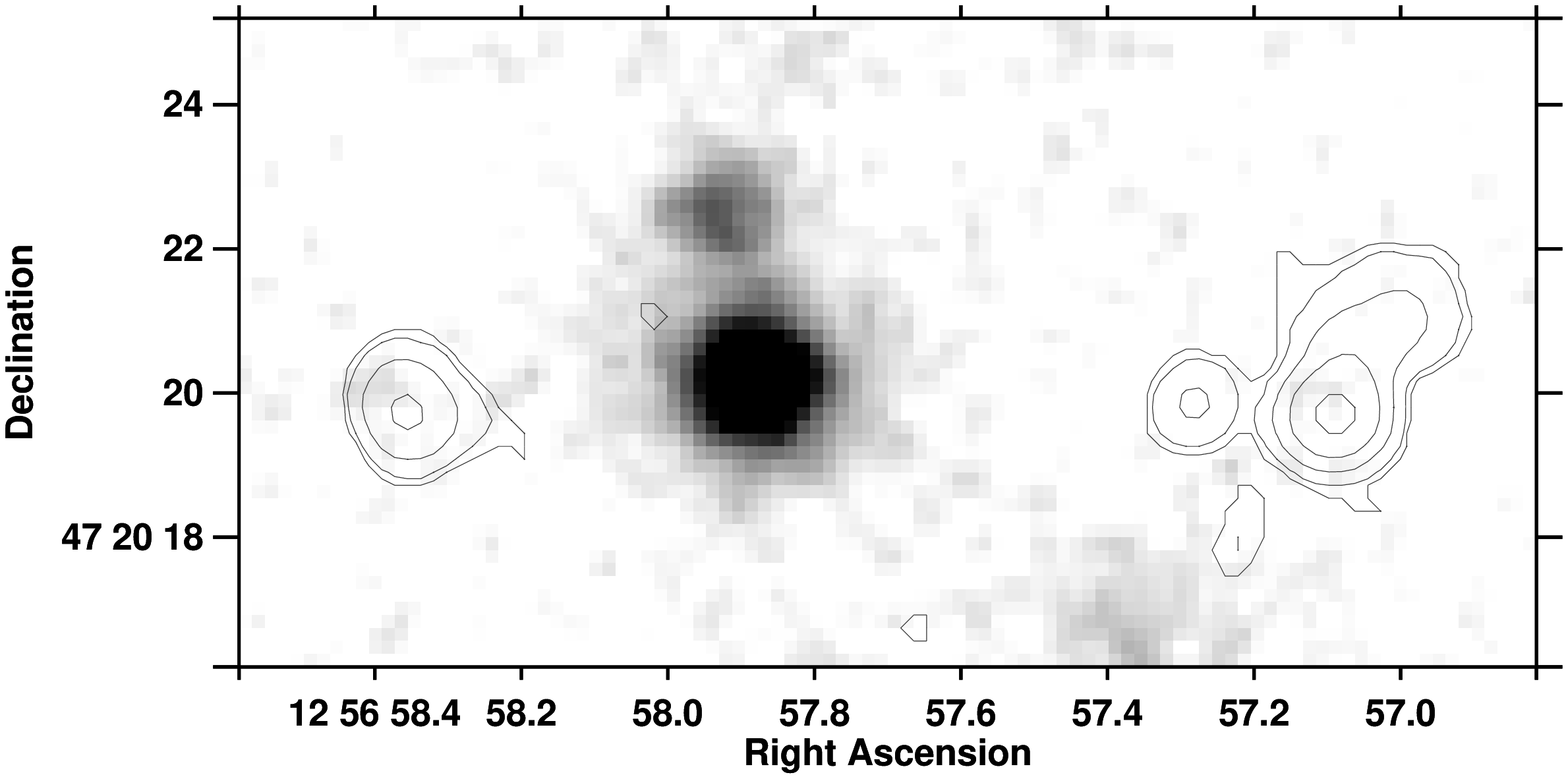,clip=,angle=90,width=13cm}
}
\caption{\label{fig3c280} Images of the radio galaxy 3C280. Overlaid are
contours from the VLA B array radio observation, with contour levels at
$(1,2,8,32,128) \times 2400 \mu$Jy beam$^{-1}$. (a) The HST image through
the f622W filter. (b) The HST image through the f814W filter. (c) The
UKIRT K--band image.}
\end{figure*}

The most striking feature of the HST image of 3C280 ($z = 0.996$,
Figures~\ref{fig3c280}a,b) is the bright arc of emission stretching back
from the western--most emission region towards the bright central emission
of the galaxy. As in the case of 3C267, Best \etal\ \shortcite{bes96a}
suggested that this could plausibly be relaxation of a star--forming
region within the gravitational potential of the host galaxy. This
arc--like structure lies at the extremity of the K--band emission from the
host galaxy (Figure~\ref{fig3c280}c), not corresponding to any structure
in that image, and so must represent very blue material. 

In Table~\ref{280cols} we present the f622W$-$f814W colours of the various
regions comprising 3C280. The central region, `b', is the reddest, whilst
the western component, `c', and the arc produce the bluest emission. We
divided the arc into two regions, the half nearest the western emission
and the half nearest the centre of the radio galaxy; the western region of
the arc is significantly bluer. This proves that the emission from the arc
cannot be due to gravitational lensing of component `b' by a foreground
component `c' or indeed by any other object. It is consistent with
relaxation of a star forming region, with the youngest stars having been
formed close to the western knot, and the oldest stars having fallen back
towards the central galaxy under gravity. It would also be consistent with
some reddening of the inner regions of the arc by dust extinction, or by
`dilution' of the ultraviolet emission by the underlying red old stellar
population.

The K--band image itself, shows a single galaxy, elongated slightly in the
east--west direction, plus a second galaxy to the north (see also Dunlop
and Peacock 1993). This red companion can also be seen on the HST
image. It is not certain that this galaxy lies at the same redshift at
3C280, but its J$-$K colour is $1.58 \pm 0.22$, consistent with it being
at that redshift (see Figure~\ref{jkcols}).

\begin{table}
\begin{tabular}{ccc}
Component & f622W$-$f814W & Error \\
    a     &      0.86     & 0.15  \\
    b     &      1.03     & 0.05  \\
    c     &      0.72     & 0.10  \\
    arc   &      0.62     & 0.13  \\
arc (west)&      0.44     & 0.16  \\
arc (cent.)&     0.79     & 0.19  \\
\end{tabular}
\caption{\label{280cols} The f622W$-$f814W colours of the various
different components comprising 3C280. These components are identified in Figure~\ref{fig3c280}b.}
\end{table}

3C280 possess strong aligned [OII]~3727 emission, which extends 5 arcsec
to the east of the host galaxy and surrounds the eastern radio lobe
\cite{rig92}. The [OII] emission from this region has a systemic velocity
of over 500\,kms$^{-1}$ with respect to that of the host galaxy. Emission
from this region of the source is detected in the f814W HST image
(Figure~\ref{fig3c280}b); this filter contains the [OII]~3727 emission
line at the redshift of 3C280, and so the emission seen from this region
is almost certainly dominated by line emission. The radio emission from
the eastern lobe is highly depolarised, probably due to Faraday
depolarisation by the emission line gas, whilst the western lobe is less
depolarised \cite{liu91b}.  A lobe -- hotspot separation velocity of $\sim
0.13c$ is derived by Liu \etal\ (1992) from radio spectral ageing fits,
indicating an age of a few million years.
 
Worrall \etal\ \shortcite{wor94} have detected 3C280 in the X--ray
waveband using ROSAT observations. They detected both unresolved and
extended emission, and argue that the extended emission is produced by hot
plasma of insufficient density for a cooling flow to have begun. Crawford
and Fabian \shortcite{cra95} argue that the unresolved X--ray emission
could be associated with a cooling flow in the central regions.

\subsection*{3C289}

\begin{figure}
\centerline{
\psfig{figure=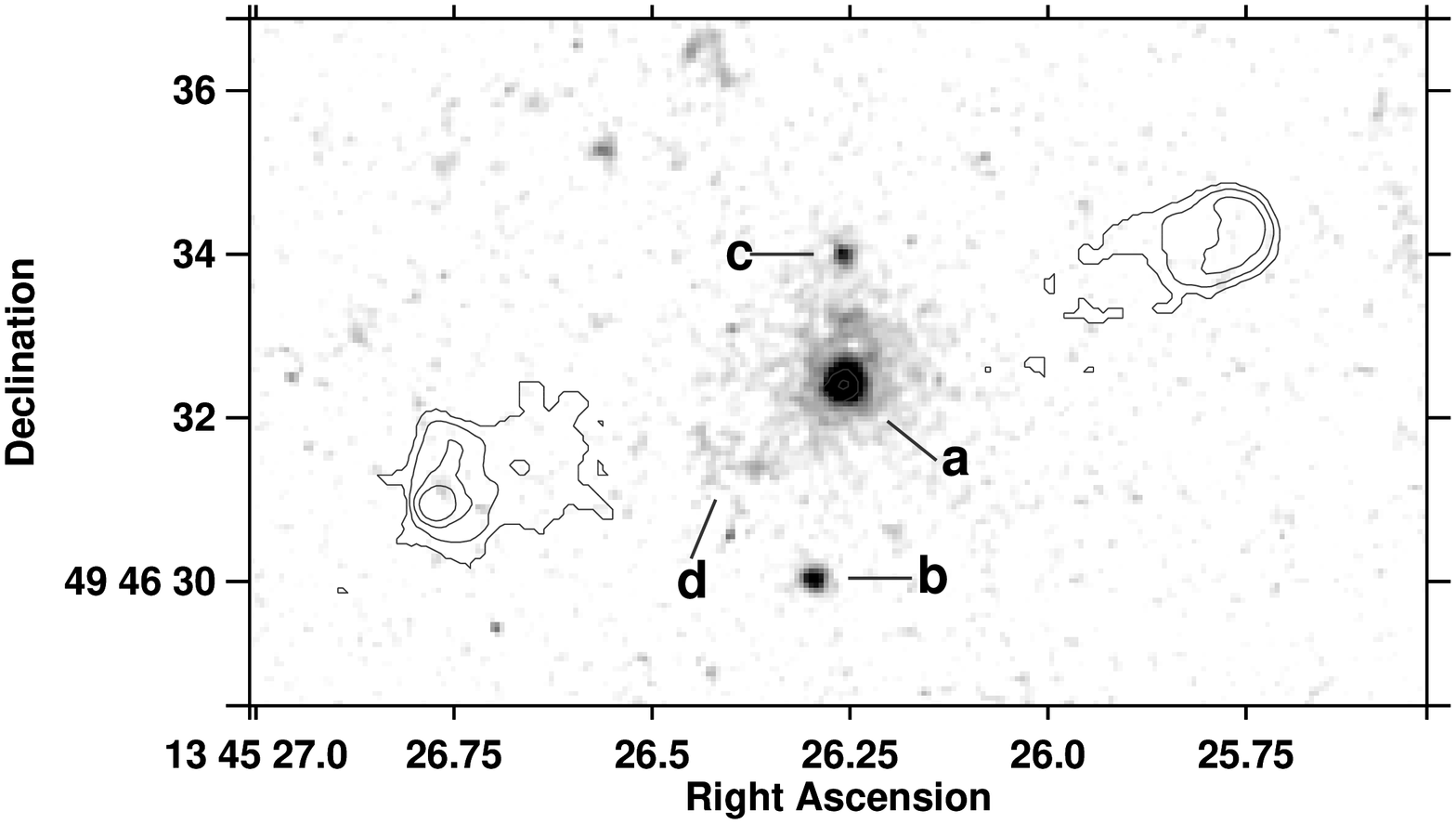,clip=,angle=90,width=8cm}
}
\centerline{
\psfig{figure=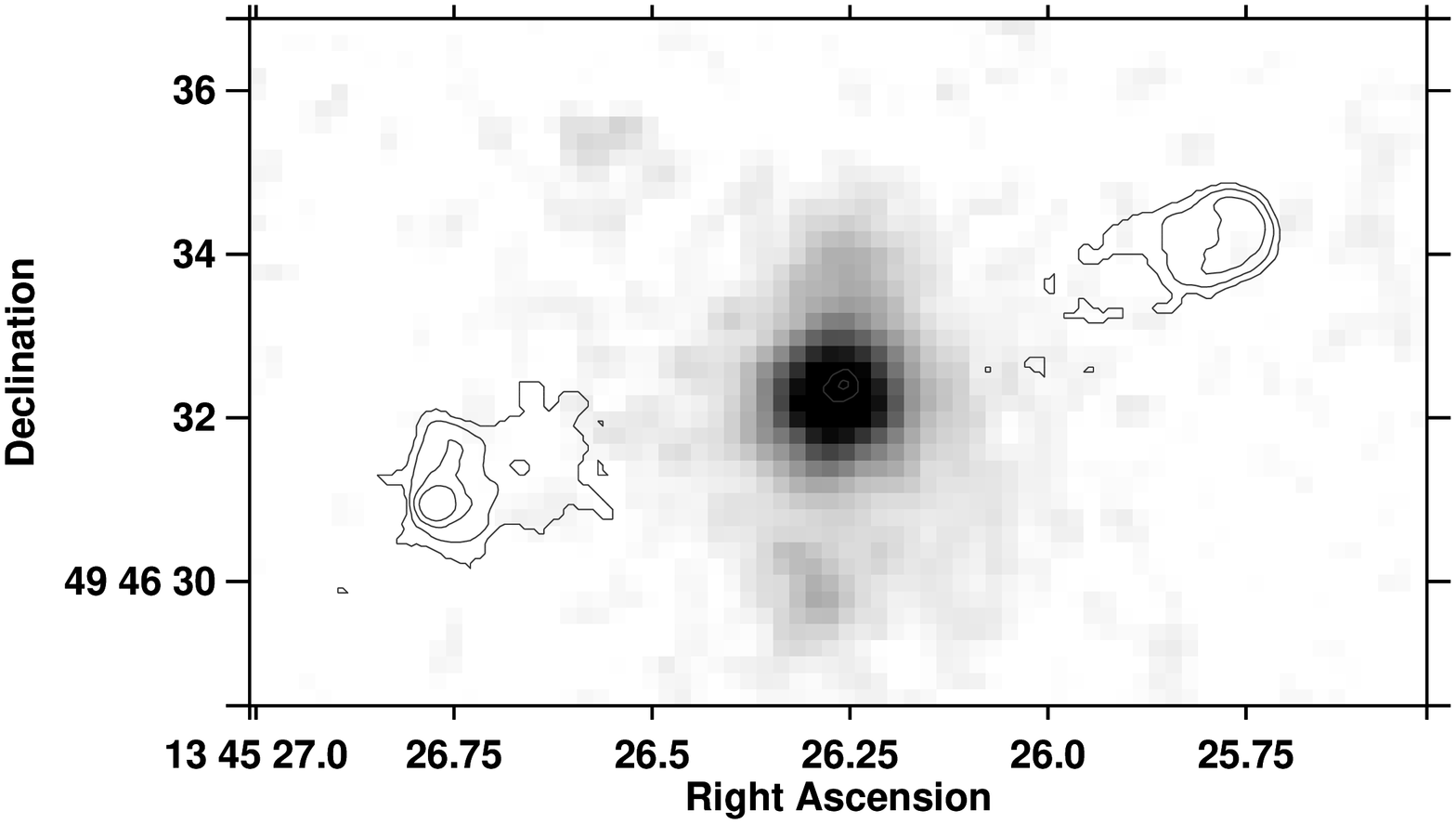,clip=,angle=90,width=8cm}
}
\caption{\label{fig3c289} Images of the galaxy 3C289, overlaid with
contours of the radio emission from the A and B array VLA observations;
contour levels are $(1,4,16,64) \times 200 \mu$Jy beam$^{-1}$. (a) The sum
of the two HST images taken using the f622W and f814W filters. (b) The
UKIRT K--band image.}
\end{figure}

3C289 ($z = 0.967$) does not show much optical activity, although it is
not entirely passive. The HST image, displayed in Figure~\ref{fig3c289}a,
consists of a central bright region (`a') surrounded by diffuse
emission. Two compact components lie 2.5 arcsec to the south (`b') and 1.5
arcsec to the north (`c'), and there is also a region of diffuse aligned
emission (`d') to the south--east of the galaxy, close to the radio
axis. The K--band image is presented in Figure~\ref{fig3c289}b; component
`b' can be seen as a southern extension of this image (demonstrating that
the north--south extension of Dunlop and Peacock's \shortcite{dun93} image
is no artefact) but component `c' is barely visible and there is little
evidence for the emission from `d'. The f622W$-$f814W colours of these
objects are: `a' --- $1.34 \pm 0.07$; `b' --- $0.66 \pm 0.37$; `c' ---
$1.98 \pm 0.36$; and `d' --- $1.48 \pm 0.39$.

The radio emission is highly symmetrical, with the hot--spots lying at the
extremities of two equally bright radio lobes, at the same distance from
the nucleus. Liu \etal\ \shortcite{liu92} have derived a lobe--hotspot
separation velocity of $\sim 0.1c$ from a radio spectral ageing analysis.

\subsection*{3C324}

\begin{figure*}
\centerline{
\psfig{figure=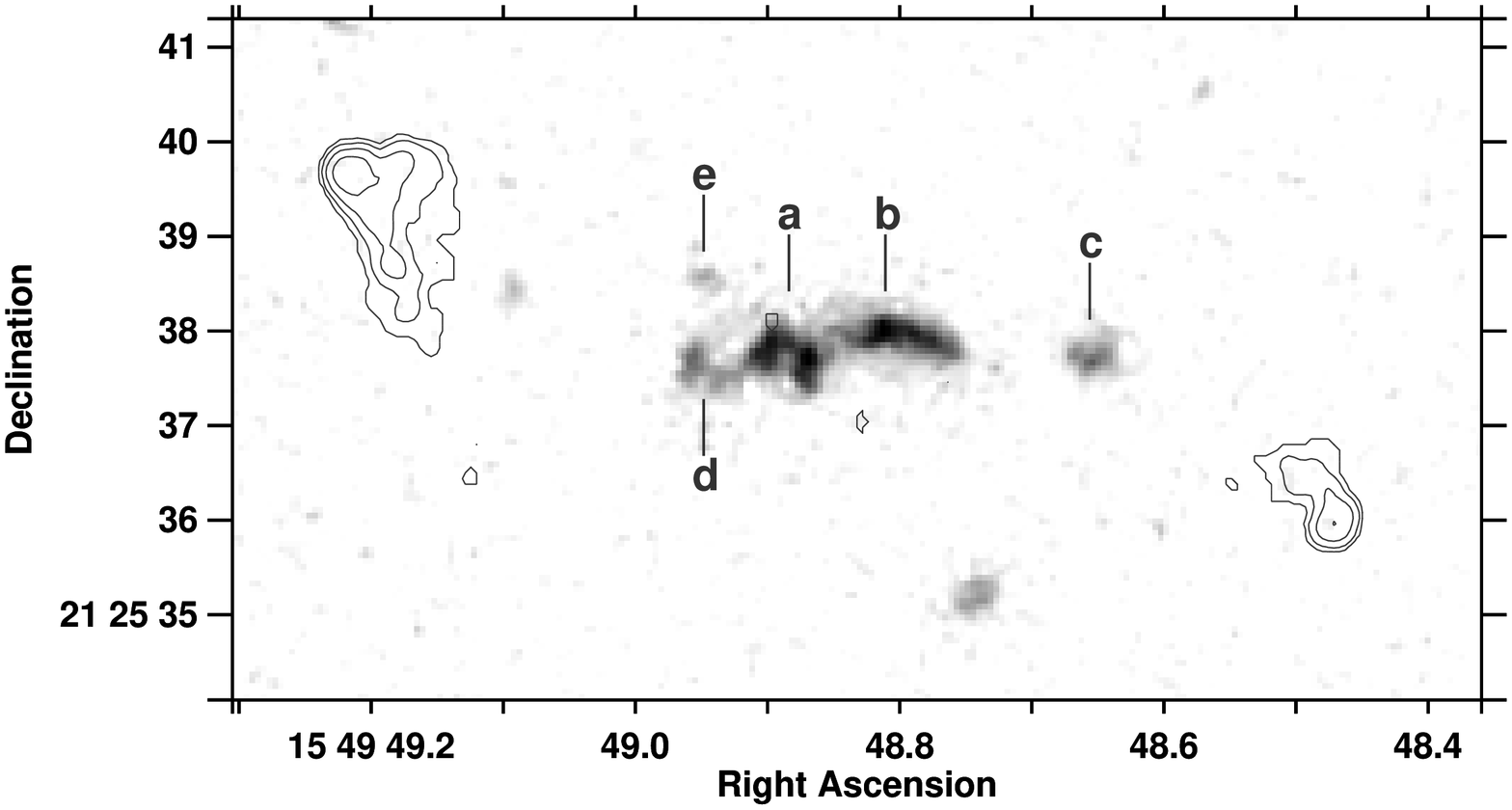,clip=,angle=90,width=13cm}
}
\centerline{
\psfig{figure=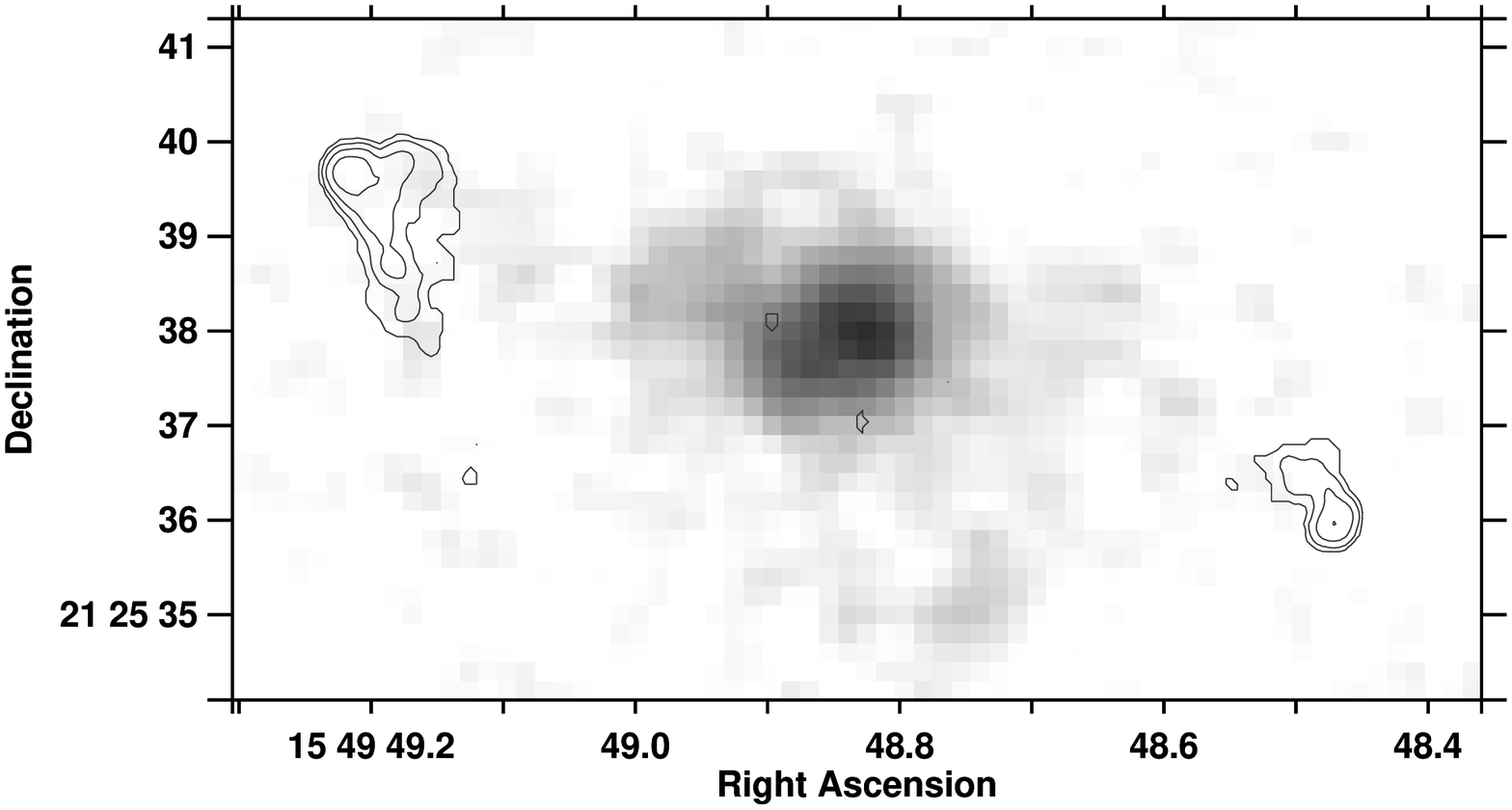,clip=,angle=90,width=13cm}
}
\caption{\label{fig3c324} Images of the galaxy 3C324, with contours of the
radio emission from the VLA A array observation overlaid. Contour levels
are $(1,4,16,64) \times 240 \mu$Jy beam$^{-1}$. (a) The sum of the two HST
images observed using the f702W and f791W filters. (b) The UKIRT K--band
image.}
\end{figure*}

3C324 lies towards the centre of a fairly rich cluster of galaxies at
redshift $z = 1.206$ \cite{dic97a}. The HST image in
Figure~\ref{fig3c324}a shows a number of components, misaligned with
respect to the axis defined by the radio hotspots by 30$^{\circ}$ (see
also the ground--based images of Hammer and Le F\`evre 1990 and Rigler
\etal\ 1992). Lower resolution VLA images by Fernini \etal\
\shortcite{fer93} show radio emission extending towards the radio galaxy
from the southern end of the eastern lobe and the northern end of the
western lobe and, if this is the axis along which the relativistic
material is being transported, the optical structures follow it reasonably
well. Dickinson \etal\ \shortcite{dic96} have suggested that the apparent
anti--clockwise rotation of the radio and ultraviolet structures with
decreasing distance from the active nucleus may represent precession of
the AGN emission axis.

The HST image of this galaxy was discussed by Longair \etal\
\shortcite{lon95}, who showed that the morphology was consistent with a
dust lane passing through the central regions of the galaxy between the
emission peaks labelled `a' and `b', obscuring the optical emission from
this region. Dickinson \etal\ \shortcite{dic96} redshifted a relatively
nearby brightest cluster galaxy to a redshift $z = 1.2$, and calculated
that extinction with E(B$-$V)$\approx 0.3$ is required to obscure the
central regions of the galaxy. No significant obscuration occurs at
infrared wavelengths, and the UKIRT K--band image is sharply peaked at the
centre of the galaxy (Figure~\ref{fig3c324}b). It is slightly elongated to
the north--east, due to the presence of a faint red companion, just
visible in the HST image (`e').  Dunlop and Peacock \shortcite{dun93} used
\clean ing and maximum entropy methods to deconvolve a deep K--band image,
and suggested that the result was not smooth, but clumpy. Their predicted
structure bears a vague resemblance to the morphology seen in our optical
HST image.

Di Serego Alighieri \etal\ \shortcite{dis93} and Cimatti \etal\
\shortcite{cim96} have measured an optical polarisation of between $11$
and $18\%$ from this galaxy, the percentage polarisation being roughly
constant with wavelength. The polarisation is extended and increases in
the outer regions of the source, suggesting the presence of diluting
ultraviolet emission which decreases with distance from the nucleus. Using
spectropolarimetry these authors have detected a broad component to the
MgII~2798 line, and showed this line to be polarised, although not at the
same level as the continuum emission. The narrow emission lines have
little or no polarisation. Although the ultraviolet emission of 3C324 is
relatively red ($f_{\nu} \propto v^{-2}$; Dickinson \etal\ 1996), Cimatti
\etal\ (1996) suggest that this can be due to scattering by dust
particles, the scattered light being reddened by passing through further
dust clouds outside of the scattering regions.

Only small colour differences are observed between the knots in this
source (see Table~\ref{324cols}), but the two filters used were not widely
separated in wavelength and so any differences would be expected to be
small. In addition, both filters are strongly influenced by line emission
and, if distributed non-uniformly, this could swamp any colour gradients
in the source. Cimatti \etal\ \shortcite{cim96} suggest that the eastern
emission (`a' and `d') is redder than the western (`b'), but that the line
luminosities are higher in the west; this second effect is likely to be
the more important of the two in our observations, leading to the western
component appearing to be redder.

\begin{table}
\begin{tabular}{ccc}
Component & f702W$-$f791W & Error \\
    a     &   0.64        &  0.08 \\
    b     &   0.86        &  0.08 \\
    c     &   0.56        &  0.20 \\
    d     &   0.77        &  0.15 \\
    e     &   0.94        &  0.38 \\
\end{tabular}
\caption{\label{324cols} The f702W$-$f791W colours of the various
different components comprising 3C324. These components are indicated in Figure~\ref{fig3c324}a.}
\end{table}

Interestingly, Cimatti \etal\ \shortcite{cim96} find that the companion 2
arcsec to the west of 3C324 (galaxy `c') is significantly bluer than the
host radio galaxy (our HST colours also show it to be bluer, although at
low significance). This companion galaxy is unpolarised. They have
modelled the stellar population within this galaxy and find that it is
consistent with a star--formation rate of up to $70 M_{\odot}$\,yr$^{-1}$.
If star formation is indeed occurring in galaxy `c', and has been induced
by the passage of the radio jet through a galaxy in the cluster
surrounding 3C324, then it would seem natural to assume that such star
formation would also have been induced by the radio jets as they passed
through the host galaxy of 3C324 itself. Therefore, the dilution of the
scattered light by ultraviolet emission could well be due to young
stars. These stars would also provide a natural explanation for the origin
of some or all of the dust involved in the scattering process. It is
interesting to compare this result for galaxy `c' with that of galaxy `a'
in 3C34 \cite{bes97b}.

Di Serego Alighieri \etal\ \shortcite{dis93} find no evidence in their
spectra for a system at redshift $z = 0.84$, which had been previously
suggested by Le F\`evre \etal\ \shortcite{fev87}. They note that all the
features quoted by Le F\`evre \etal\ coincide with strong sky
features. This, together with the high resolution image, makes it unlikely
that the morphology of this source is determined by gravitational lensing.

\subsection*{3C337}

\begin{figure*}
\centerline{
\psfig{figure=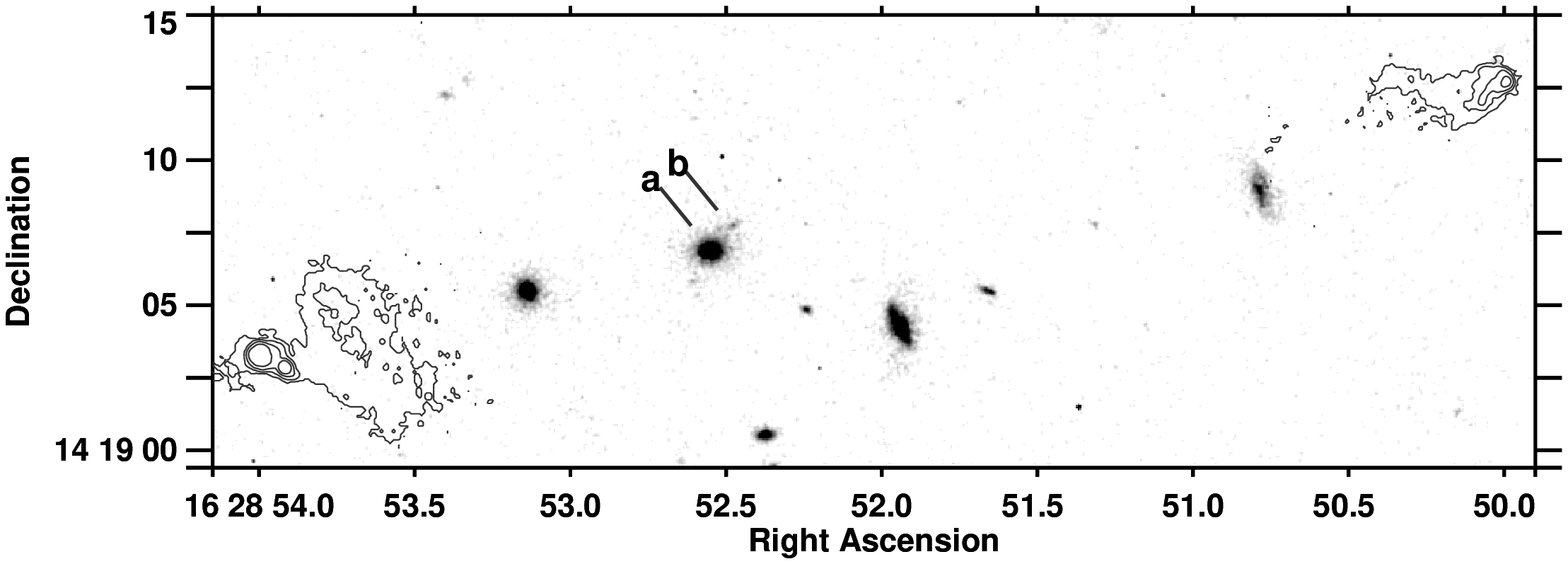,clip=,angle=90,width=15cm}
}
\centerline{
\psfig{figure=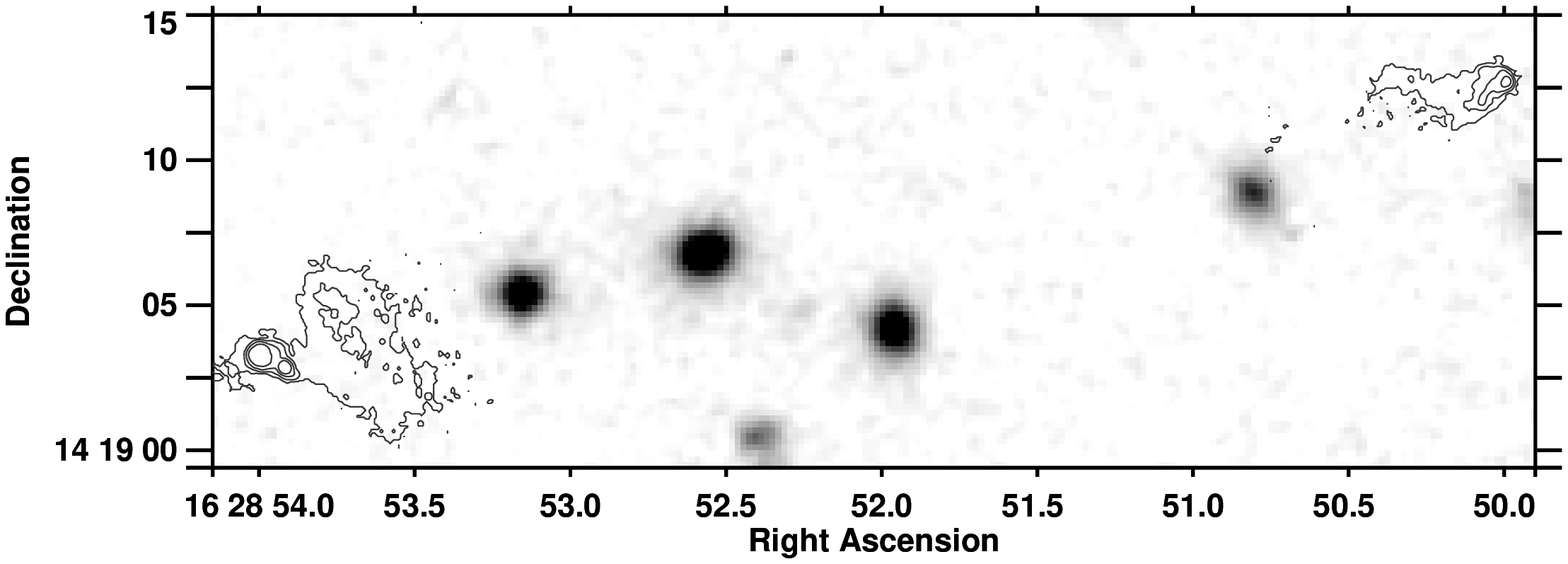,clip=,angle=90,width=15cm}
}
\caption{\label{fig3c337} (a) The sum of the two HST images of the radio
galaxy 3C337, observed through the f555W and f814W filters. Overlaid are
contours of radio emission from the A, B and C array observations using
the VLA. Contour levels are $(1,4,16,64) \times 100 \mu$Jy
beam$^{-1}$. (b) The UKIRT K--band image, with radio contours as in (a).}
\end{figure*}

The host galaxy associated with the radio source 3C337 lies towards the
centre of a cluster of galaxies \cite{mcc88} at a redshift of $z =
0.635$. The HST image of the galaxy (`a'), shown in
Figure~\ref{fig3c337}a, shows very little structure with just a single
small companion (`b') lying 1 to 2 arcsec to the NW of the host
galaxy. This knot has an f555W$-$f814W colour of $2.00 \pm 0.35$, bluer
than the $2.42 \pm 0.08$ colour of the central region. In the infrared
image, the galaxy shows no extensions, resembling a standard giant
elliptical galaxy (Figure~\ref{fig3c337}b).

This source shows strong nuclear and extended [OII]~3727 line emission,
the latter stretching over 100\,kpc from the central galaxy and extending
around the south of the eastern lobe \cite{mcc88}. The radio structure of
3C337 is asymmetric, with the eastern hotspots being only half as far
from the nucleus as the western hotspot, and the eastern lobe being much
more depolarised \cite{ped89b}. Each of these two effects is likely to be
due to the presence of this emission line gas near the eastern lobe.

\begin{figure*}
\centerline{
\psfig{figure=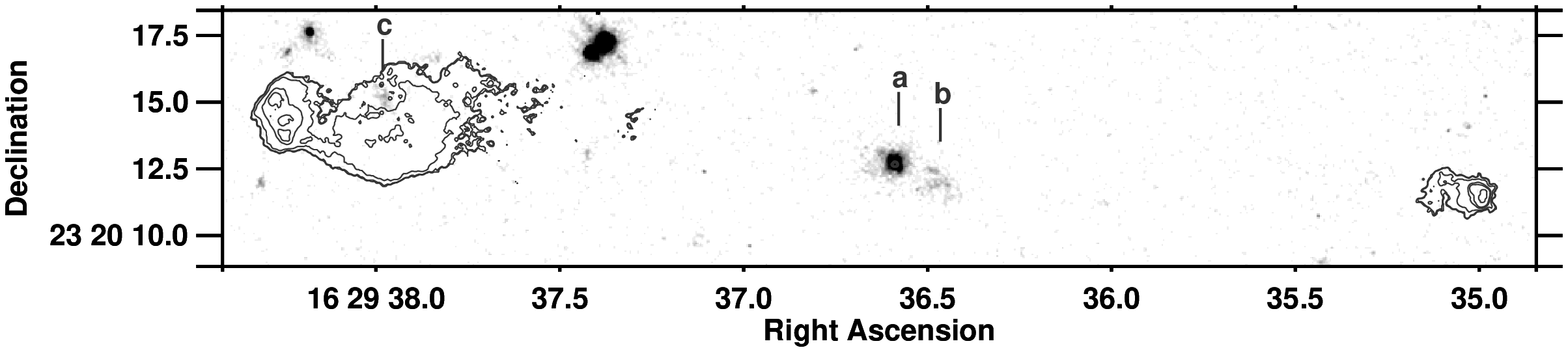,clip=,angle=90,width=\textwidth}
}
\centerline{
\psfig{figure=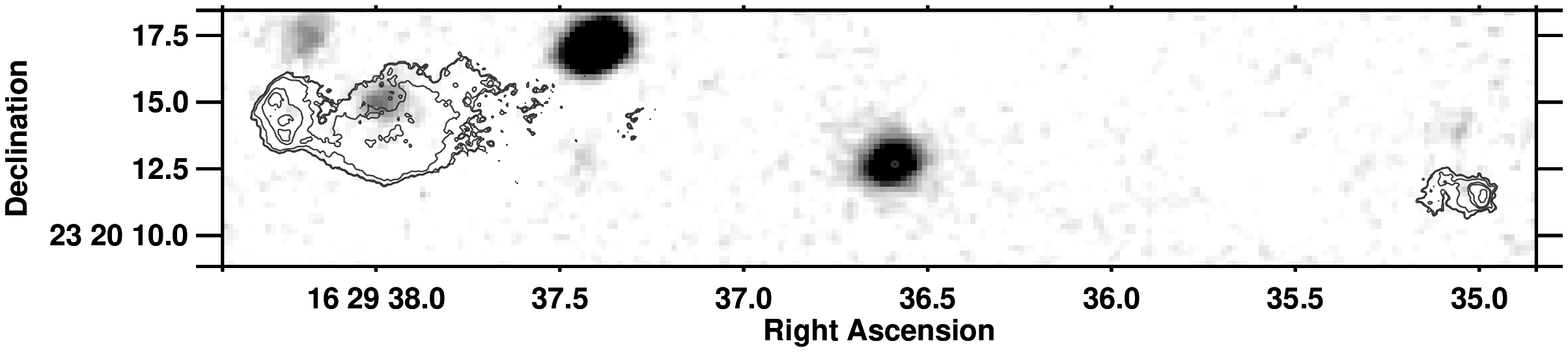,clip=,angle=90,width=\textwidth}
}
\caption{\label{fig3c340} (a) The sum of the two HST images of the radio
galaxy 3C340, observed through the f555W and f785LP filters. Overlaid are
contours of radio emission from the A, B and C array observations using
the VLA. Contour levels are $(1,2,8,32,64,128) \times 20 \mu$Jy
beam$^{-1}$. (b) The UKIRT K--band image, with radio contours as in (a).}
\end{figure*}

The eastern hotspot is double, the more southerly of the pair being more
compact and having polarisation properties consistent with it being the
current primary hotspot. The two hotspots protrude from a curved boundary
at the end of the eastern lobe. Radio spectral ageing estimates suggest
that the hotspots in this source are advancing at a speed of $\sim 0.07c$,
giving an age for the radio source of about $10^7$ years \cite{ped89b}.

\subsection*{3C340}

\begin{figure*}
\centerline{
\psfig{figure=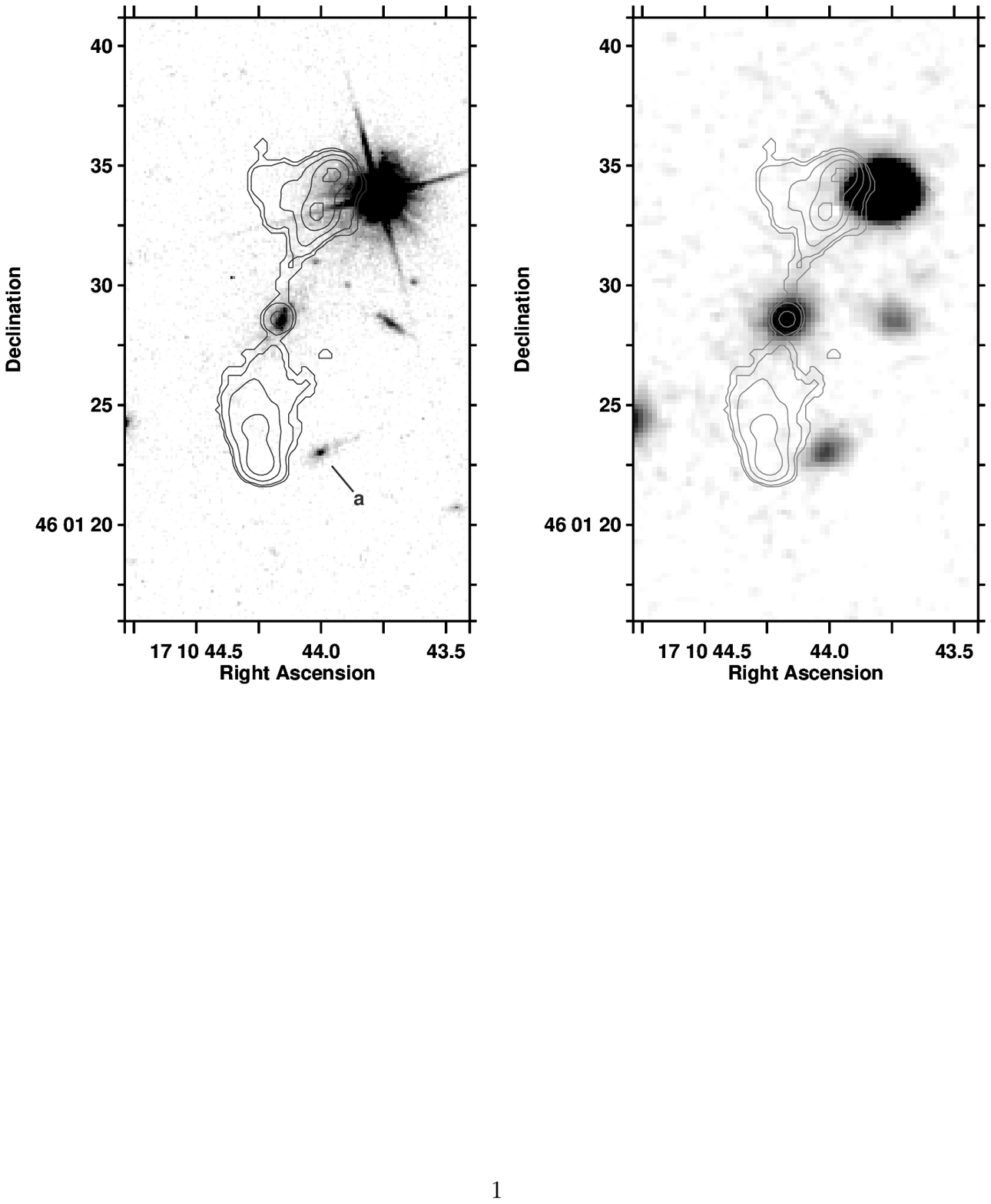,clip=,width=\textwidth}
}
\caption{\label{fig3c352} Images of the radio galaxy 3C352, with contours
of radio emission from the B array VLA observation overlaid. Contour
levels are $(1,2,8,32,128) \times 110 \mu$Jy beam$^{-1}$. (a) The sum of
the two HST images taken using the f555W and f814W filters. (b) The UKIRT
K--band image. }
\end{figure*}

The HST image of 3C340, galaxy `a', at redshift $z = 0.775$, is displayed
in Figure~\ref{fig3c340}a and shows a marginally aligned central galaxy
together with fainter emission (marked as `b') about 2 arcsec to the
WSW. The central galaxy has an f555W$-$f785LP colour of $2.51 \pm 0.06$,
whilst knot `b' is bluer with f555W$-$f785LP$= 2.14 \pm 0.18$. The knot
can also be seen quite strongly in the narrow--band [OII]~3727 image taken
by McCarthy \shortcite{mcc88}, but it is not visible in our infrared image
(Figure~\ref{fig3c340}b). Johnson \etal\ \shortcite{joh95} have detected a
weak radio jet candidate extending from the core towards the western lobe,
coincident with this emission line region.

The bright radio emission in the eastern lobe seems to pass around the
emission region at RA: 16 29 37.97, Dec: 23 20 15 (J2000), visible on both
the infrared and HST images and labelled `c'. This corresponds to the
region in which Johnson \etal\ \shortcite{joh95} have detected a
`depolarisation silhouette' in their deeper radio maps; that is, a point
where the depolarisation of the radio emission between 21\,cm and 6\,cm is
significantly higher than the rest of the lobe. This is attributed to
Faraday depolarisation by gas either within the lobe or in front of it.
They also note that this galaxy appears to be interacting with the larger
galaxy 3 arcsec to the north--east, with evidence (from ground
based--images) of a tidal tail of material from the latter, suggesting
that the gas involved in this interaction may be responsible for the
depolarisation.

If galaxy `c' is at the same redshift as 3C340, then backflowing material
within the radio lobe may be channelled round it by the higher density of
material. If this is the case then the host galaxy of 3C340 could be a
member of a group or cluster of galaxies. However, the depolarisation
throughout the rest of the lobes of 3C340 is remarkably low, indicating
that if this is the case then it must be a relatively poor cluster with
little cluster gas. Also, there is no evidence for emission from either of
the two galaxies in a narrow band [OII]~3727 image \cite{mcc88}.

\subsection*{3C352}

With a relatively bright star lying only 7 arcsec to the north--west,
3C352 at redshift $z=0.806$ is one of the most awkward sources to observe
from the ground, but the high angular resolution of the HST distinguishes
the galaxy clearly. The optical emission, shown in Figure~\ref{fig3c352}a,
is elongated, although not nearly as much as the [OII]~3727 emission line
region which extends over 8 arcsec \cite{mcc88,rig92}. The most extended
diffuse emission in the HST image is misaligned from the radio axis by
about 20 degrees, which is comparable to the misalignment of the [OII]
line emission. However, the bright central regions of the HST image,
although much less extended, are aligned almost exactly along the radio
axis.  The infrared K--band image is more compact and symmetrical, but
does show a slight extension along the radio axis (see
Figure~\ref{fig3c352}b, and also Dunlop and Peacock 1993).

Using Fabry--Perot interferometry, Hippelein and Meisenheimer
\shortcite{hip92} mapped the spatial and spectral distribution of the line
emission from this source, and attempted to remove it from an R--band
image of the galaxy. They suggested that the residual continuum emission
showed no significant elongation along the radio axis. A continuum image
taken just longward of the 4000\ang\ break shows marginal elongation, but
significantly less than that of the line emission \cite{rig92}.

Hippelein and Meisenheimer \shortcite{hip92} explained the velocity
profiles of the extended emission line region in terms of a bow--shock
associated with the radio jet. The bow--shock alone cannot supply
sufficient energy to excite the emission line regions, and so much of the
excitation is likely to be due to violent interactions of the gas with the
supersonic flow in the jet. In the central regions a proportion of the
excitation may be associated with ultraviolet emission from an AGN.

Several other objects in the field of 3C352 show [OII]~3727 emission,
indicating that the galaxy is likely to be a member of a cluster. One such
cluster member is the galaxy 5 arcsec SSW of 3C352, labelled galaxy `a' in
Figure~\ref{fig3c352}a. The [OII] emission from this object extends for
about 7 by 11 arcsec indicating a large gas mass. It is blue--shifted with
respect to 3C352 by 500\,km\,s$^{-1}$, with only small velocity
fluctuations and velocity widths within the cloud itself
\cite{hip92}. These authors suggested that this cloud may be excited by
an interaction of the gaseous halo surrounding the central galaxy with
accreting intracluster material. The HST image shows that this galaxy
possesses a peculiar morphology. The f555W$-$f814W colour of galaxy `a'
($2.55 \pm 0.26$) is redder than that of the central galaxy ($2.07 \pm
0.07$), indicating that the central galaxy may be more active than galaxy
`a', although this effect may be due to differing levels of line emission
in the two objects.

\subsection*{3C356}

\begin{figure}
\centerline{
\psfig{figure=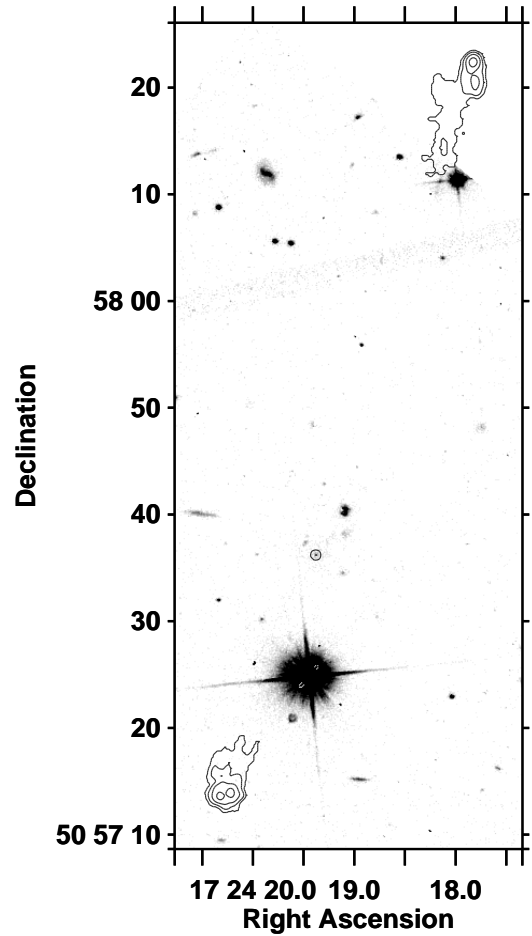,clip=,width=7cm,angle=90}
}
\caption{\label{3c356rad} The sum of the two HST images of 3C356, taken
using the f622W and f814W filters, with the B and C--array observations of
the 8.4~GHz VLA radio emission overlaid. The HST field does not extend
quite as far as the northern hotspot of this large radio source. The
`stripe' running across from Dec: 50 58 00 on the left--hand axis to Dec:
50 58 07 on the right--hand axis is the track of an object moving across
the field of view during the observation. Radio contour levels are
$(1,4,16,64) \times 220 \mu$Jy beam$^{-1}$.}
\end{figure}

\begin{figure*}
\centerline{
\psfig{figure=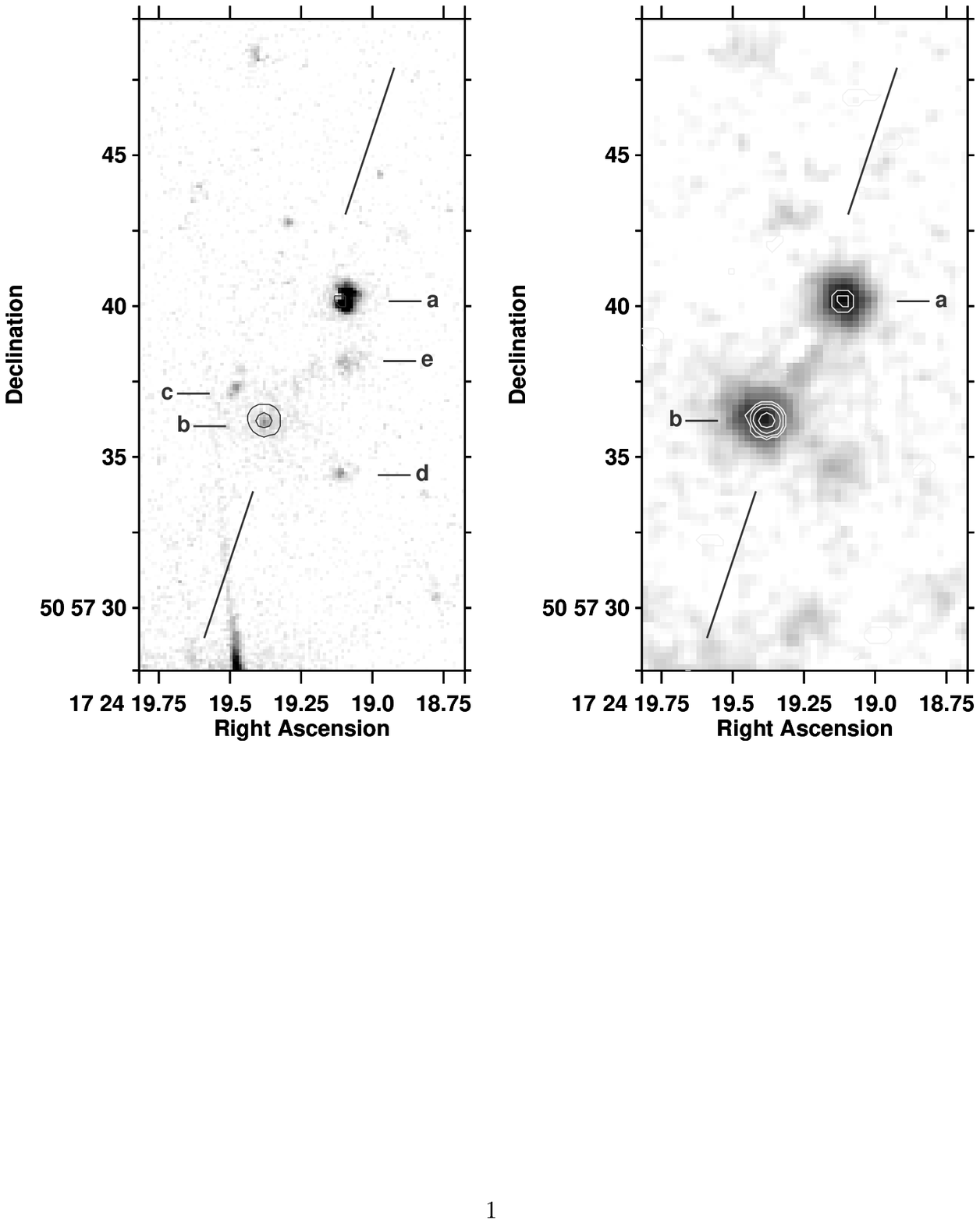,clip=,width=14cm}
}
\caption{\label{fig3c356} (a) An enlargement of the central regions of
3C356 as observed using the f622W and f814W filters of the HST. The
central radio emission is overlaid, with contour levels of $(1,4) \times
160 \mu$Jy beam$^{-1}$. (b) The UKIRT K--band image to the same scale as
(a), and with the central radio emission overlaid; contours are $(1,2,4,8)
\times 80 \mu$Jy beam$^{-1}$.}
\end{figure*}

Figure~\ref{3c356rad} shows the HST image of 3C356, with radio contours
overlaid. An enlarged picture of the central regions of this radio source
(Figure~\ref{fig3c356}), shows two unresolved radio `cores', separated by
only 5 arcsec and aligned approximately along the radio axis. Each of
these cores is associated with equally bright infrared galaxies at
redshift $z = 1.079$, and this has led to a dispute about which of the two
galaxies is the host galaxy of the radio source.

At 8.4~GHz (see Figure~\ref{fig3c356}), the radio emission associated with
the southern galaxy, `b', is brighter, with a flux density of 0.95~mJy,
whilst that from the northern galaxy, `a', is only 0.22~mJy (both of our
values are about 10\% smaller than the values quoted by Fernini \etal\
1993). The emission from `b' has a flat radio spectrum between 8.4~GHz and
5~GHz, $\alpha \approx 0.1$, typical of the compact cores of extended
radio sources. The spectral index of `a' is steeper, $\alpha \approx 1.1$,
more in line with that of a compact steep spectrum source.

Ground--based optical observations have shown galaxy `a' to be extremely
compact, whilst `b' is much larger with a very flat surface brightness
distribution \cite{fev88a,rig92}. Intrinsically, galaxy `b' is extended
perpendicular to the radio axis (Figure~\ref{fig3c356}, Rigler \etal\
1992, Dunlop and Peacock 1993). Both objects show strong [OII]~3727
emission, with the line emission from `a' again being much more centrally
condensed \cite{mcc88,eis90,rig92}.  Jackson and Rawlings
\shortcite{jac97} detect [OIII]~5007 emission from object `a', and
marginally from `b'. The detection of a 4000\ang\ break in both components
\cite{lac94} indicates that they both contain stars and must be at least
$10^7$ years old.

Rigler \etal\ \shortcite{rig92} suggest that the southern galaxy may have
wandered into the path of the radio beam from galaxy `a'. In contrast,
Lacy and Rawlings \shortcite{lac94} presented a model in which galaxy `b'
is the nucleus, and galaxy `a', which possesses ultra--high ionisation
narrow lines, is photoionised by hot gas generated in shocks driven into
the galaxy by interaction with the radio jet. They suggest that the
population of stars in galaxy `a' is only a few times $10^7$ years old,
and was formed when the jet originally pointed at the galaxy, but then
precessed away to point at the current hot--spot position, and has now
precessed back again. They cite the low carbon--to--oxygen ratio and the low
levels of dust in the galaxy as evidence of its young age.

Two aspects of the HST image suggest that this is not the case. Firstly,
galaxy `a' has been resolved by the HST, showing it to display a
`dumbbell'--shaped optical morphology. This double structure is
characteristic of a number of the galaxies in the sample, eg. 3C226, 3C241
and 3C252. Secondly, the optical emission from the southern galaxy `a' is
barely visible on the HST image, consisting of a set of three unresolved
emission regions orientated perpendicular to the radio axis: one lies at
the position of the infrared emission (`b'), a second 1 arcsec to the NE
(`c'), and the third nearly 3 arcsec to the WSW (`d'). Further emission
(`e') lies 2 arcsec to the south of the northern galaxy. This collection
of isolated emission regions bears no resemblance to any other galaxy in
our sample, and most of the emission lies outside the envelope of the
infrared emission from the southern component. It seems likely that `a' is
the host galaxy, and the disturbed structure seen around the southern
galaxy is associated with the aftermath of the passage of the radio jet
close to or through this galaxy.

Galaxy `a' shows a high continuum polarisation, rising from about 3\% at
restframe 4200\AA\ up to 15\% at 2000\AA\ \cite{cim97}. Broad MgII~2798
line emission is observed, and is polarised at the $6.8 \pm 1.2$\% level.
The narrow emission lines show little of no polarisation. The polarisation
of the southern galaxy `b' is much lower, reaching only $4.0 \pm 1.2$\% at
2000\AA. Based upon the assumption that the broad MgII~2798 line emission
and the scattered continuum emission should have the same fractional
polarisation, Cimatti \etal\ \shortcite{cim97} estimate that $50 \pm 15\%$
of the total emission at 2800\AA\ is associated with scattered light in
galaxy `a'. Following the prescription of Dickson \etal\
\shortcite{dic95}, they also calculate that as much as 25\% of the
ultraviolet emission may be associated with nebular continuum emission. On
the basis of energetics, these authors agree that the northern galaxy `a'
is more likely to harbour the hidden quasar nucleus.

In Table~\ref{356cols} we present the f622W$-$f814W colours of the five
different components labelled in Figure~\ref{fig3c356}a. These data show
that galaxy `b' is significantly redder than the northern galaxy `a', but
also shows that the other diffuse emission regions around this galaxy are
as blue as galaxy `a'. This would be consistent with these representing
regions which have been disrupted by the passage of the radio beam, and
are currently optically active. Alternatively these may be associated with
gas cooling after being compressed by the expanding radio lobes, as has
been suggested for 3C265.

\begin{table}
\begin{tabular}{ccc}
Component & f622W$-$f814W & Error \\
    a     &      1.06     & 0.05  \\
    b     &      1.92     & 0.24  \\
    c     &      0.95     & 0.23  \\
    d     &      1.43     & 0.29  \\
    e     &      1.03     & 0.21  \\
\end{tabular}
\caption{\label{356cols} The f622W$-$f814W colours of the various
different components comprising 3C356. These components are indicated in Figure~\ref{fig3c356}a.}
\end{table}

\begin{figure*}
\centerline{
\psfig{figure=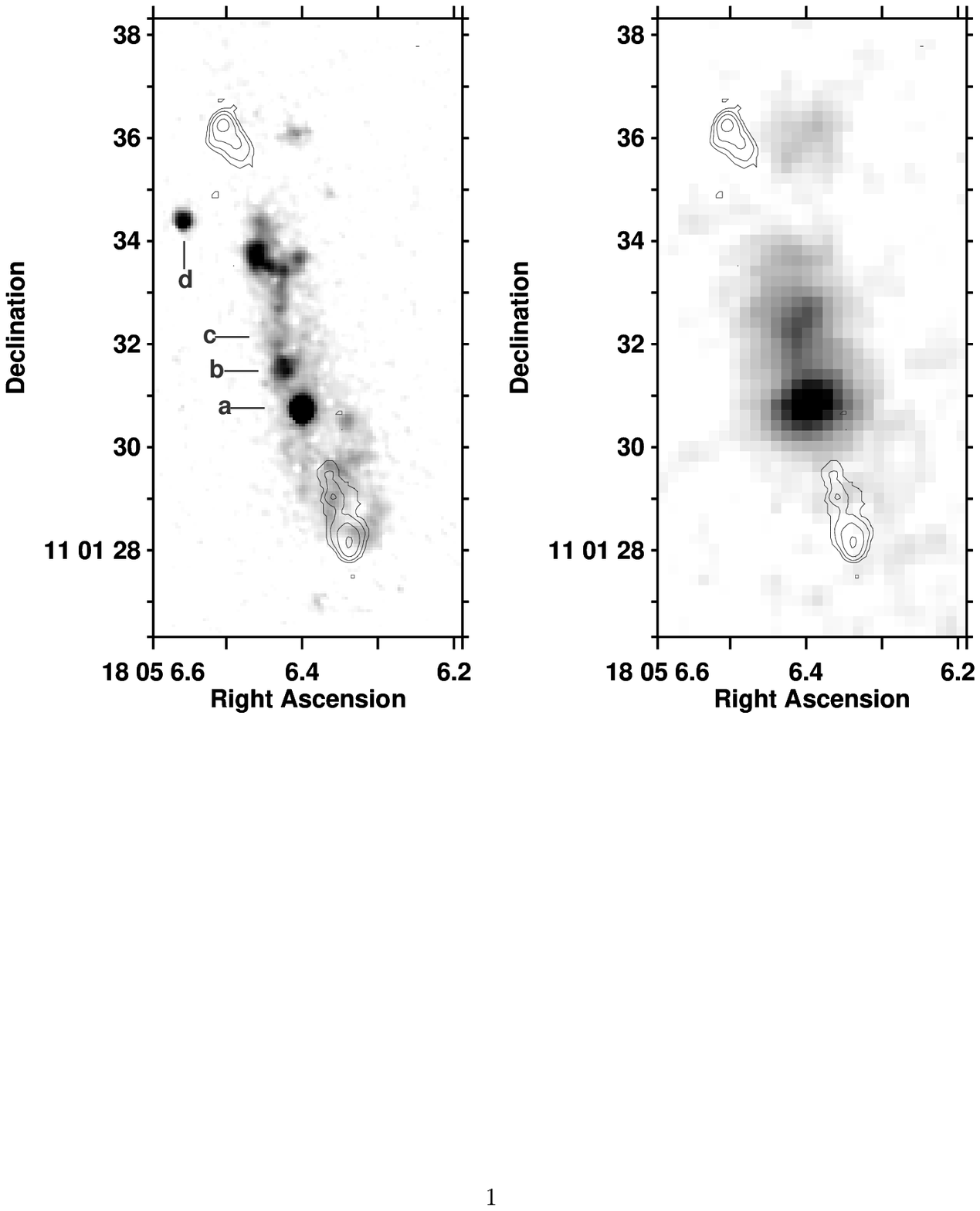,clip=,width=14cm}
}
\caption{\label{fig3c368} Images of the radio galaxy 3C368, overlaid with
contours of $(1,4,16,64) \times 120 \mu$Jy beam$^{-1}$ of the radio
emission from the 8.4~GHz A--array VLA observation. (a) The sum of the
images observed using the f702W and f791W filters of the HST. (b) The
UKIRT K--band image. }
\end{figure*}

X--ray emission has been detected from the vicinity of this source by
Crawford and Fabian \shortcite{cra93}, with an X--ray luminosity of $\sim
2.5 \times 10^{44} \rm{erg\,s}^{-1}$. A further observation using the
ROSAT HRI has shown that this X--ray emission is extended \cite{cra96a},
and not associated with a point source, as would be expected in the
jet--cloud interaction model \cite{lac94}. Crawford and Fabian
\shortcite{cra96a} associate these X--rays with emission from a dense halo
of hot gas cooling at a rate of 500 to 1500 $M_{\odot}$\,yr$^{-1}$; this
is sufficiently high that it may account for the origin of the emission
regions `c', `d' and `e'. There is marginal evidence that the X--ray
emission is elongated in the direction of the radio axis.

Rawlings \etal\ \shortcite{raw91a} have detected the emission line
[SIII]~9532, with a flux of $3.2 \pm 0.5 \times 10^{-18}\rm{W\,m}^{-2}$,
in the northern galaxy using near infrared spectroscopy. This will
contribute about 15\% of the K--band flux, but line emission alone is not
sufficient to account for the large difference in colour between the two
galaxies.

\subsection*{3C368}

3C368, at redshift $z=1.132$, is the best studied radio galaxy in this
sample, and perhaps of all galaxies at high redshift. The HST image of
this source, presented in Figure~\ref{fig3c368}a, has been discussed in
detail in a previous paper \cite{lon95}. The main points of note were the
following: (i) the bright object towards the centre of the galaxy
(labelled `a') is a galactic M--dwarf star \cite{ham91} --- a spectrum of
this object taken by Stockton \etal\ \shortcite{sto96a} confirms this
identification; (ii) the most promising candidate for the nucleus of the
galaxy was the emission region 1 arcsec to the north of the M--star (`b');
(iii) a bright string of knots stretches from this emission region towards
the northern hot--spot; (iv) the southern radio lobe lies within an
edge--brightened elliptical emission region, possibly associated with a
bow--shock phenomenon (see also Meisenheimer and Hippelein
1992)\nocite{mei92}.

Interestingly, there are no significant `colour' differences between the
central emission region `b' (f702W$-$f791W$= 0.71 \pm 0.08$), the northern
knots (f702W$-$f791W$= 0.77 \pm 0.06$) and the southern diffuse emission
(f702W$-$f791W$= 0.72 \pm 0.07$). As the two filters include significant
line emission, any non-uniformity in the line emission would produce
noticeable `colour' gradients, and so this result suggests that the line
emission has very similar morphology to the continuum in this source (see
also the [OII]~3727 image of Meisenheimer and Hippelein
1992).\nocite{mei92} Any variation in line emission must be matched by
corresponding variations in the intrinsic colours.

The infrared emission from 3C368 shows none of this complexity. The
M--star is clearly visible in the infrared K--band image
(Figure~\ref{fig3c368}b), and to the north of this the emission from the
galaxy itself can be seen. The lack of complexity of the infrared image is
not simply due to lower angular resolution: a 0.3 arcsec resolution
K--band image has been obtained by Stockton \etal\ \shortcite{sto96a}, and
it also shows only these simple structures. In their image, the centre of
the infrared galaxy lies fractionally north of the emission region `b'
which we previously suggested as a candidate for the nucleus, and is
coincident instead with the dip in the optical emission (`c'). We agree
with their interpretation that this is the position of the galaxy nucleus;
this suggests that, as in the case of 3C324, there may be dust obscuration
of the nucleus.

There is only a slight hint of emission from the southern region in the
K--band image. Also only faintly detected in the infrared images is
emission from object `d' at RA: 18 05 6.55, Dec: 11 01 34.5 in the HST
image. A colour--colour plot shows this object to lie in a region occupied
by galactic stars \cite{rig92}, and this hypothesis is supported by its
unresolved appearance on the HST image. 3C368 lies at low galactic
latitude, and many stars lie within an arcminute of the source.

Di Serego Alighieri \etal\ \shortcite{dis89} measured the polarisation of
the optical emission to be $P_V = 7.6 \pm 0.9\%$ and $P_R = 2.6 \pm 1.2\%$
in the V and R wavebands respectively. Scarrott \etal\ \shortcite{sca90}
carried out imaging polarimetry, showing that that the extended regions
were polarised, with the electric field vector perpendicular to the radio
axis, and that the scattering percentage decreased with increasing
wavelength. They attributed the latter observation to dilution of the
scattered component, possibly by an underlying old stellar population.

Van Breugel (private communication) has recently made HST and Keck
observations of this source and found no extended polarisation with either
telescope. The contradiction between these and the previous observations
has yet to be resolved; the time-scale between observations is clearly too
short for it to be due to variability. Van Breugel's Keck spectrum of
3C368 shows little evidence for stellar absorption lines.

The contribution to the spectrum of this source by nebular emission
mechanisms, that is, the free--free, free--bound and 2--photon continua
and Balmer continuum emission, has recently been calculated using
photoionisation models, based upon the line flux of the [OIII]~5007 line
\cite{dic95}. They show that up to 60\% of the ultraviolet emission from
3C368 may be associated with thermal emission from the nebular
gas. Stockton \etal\ \shortcite{sto96a} suggest that this may rise to as
much as 80\% in the northern emission knots. Rawlings \etal\
\shortcite{raw91a} have cast doubt on whether photoionisation models are
appropriate for this source, since they detect a low [SIII] / [OII] ratio,
suggesting low ionisation, but also strong high ionisation NeIII lines in
the optical spectrum. The use of different ionisation models would not,
however, affect the result that nebular continuum emission from this
source is highly significant.

X--ray emission from the vicinity of the radio source has been detected
with an X--ray luminosity $\sim 1.5 \times 10^{44}\rm{erg\,s}^{-1}$
\cite{cra95}.

\subsection*{3C437}

\begin{figure}
\centerline{
\psfig{figure=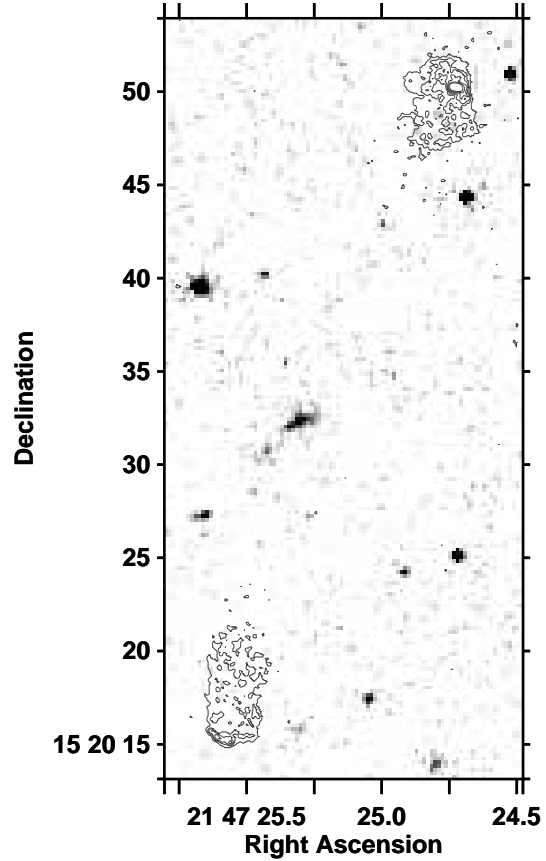,clip=,width=7cm,angle=90}
}
\caption{\label{3c437vlaabc} The HST image of 3C437 using the f785W
filter.  Overlaid are contours of the radio emission at 8.4~GHz from the
A, B and C array VLA observations. Contour levels are $(1,4,8,16,32)
\times 140 \mu$Jy beam$^{-1}$.}
\end{figure}

\begin{figure*}
\centerline{
\psfig{figure=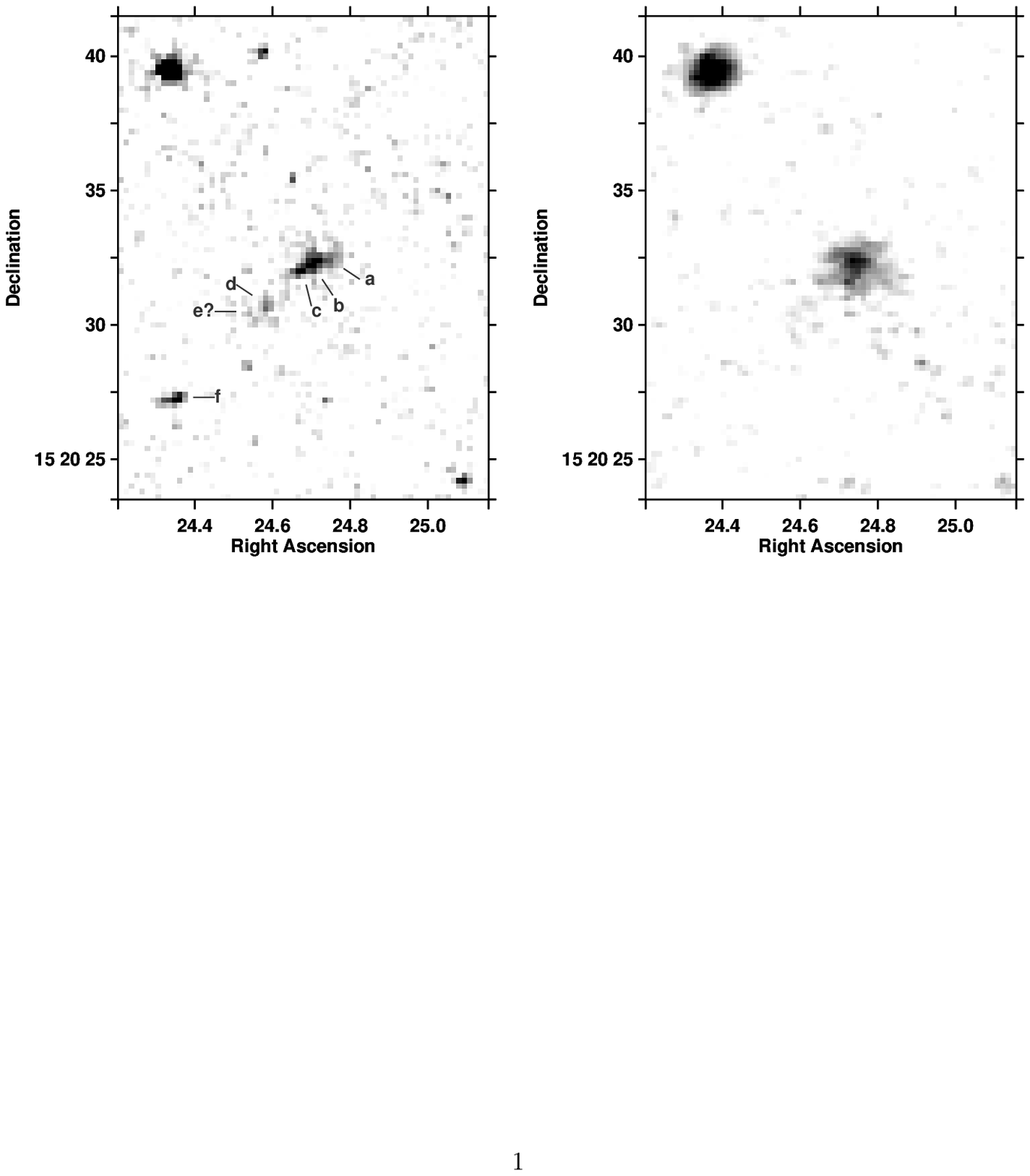,clip=,width=\textwidth}
}
\caption{\label{fig3c437} Enlarged images of the central regions of the
radio galaxy 3C437. (a) The f785LP HST image (b) The UKIRT K--band image. }
\end{figure*}

The radio source 3C437, at redshift $z = 1.480$, is displayed with radio
contours overlaid in Figure~\ref{3c437vlaabc}, and an enlargement of the
central regions of the source is shown in Figure~\ref{fig3c437}. As
with 3C68.2, the HST data have been averaged in blocks on 2 by 2 pixels
and smoothed with a 0.2 arcsecond Gaussian profile to improve the
signal--to--noise ratio. The optical emission from the host galaxy is
highly elongated but misaligned with respect to the axis of the radio
source by about 35 degrees (Figure~\ref{fig3c437}a). Le F\`evre \etal\
\shortcite{fev88b} suggest that the optical emission arises from five
components in two groups. The north--western group of three (`a', `b' and
`c') is detected in the HST image, consisting of two small emission
regions plus diffuse emission to the north--west. Approximately 2 arcsec
to the south--east of these lies a further emission region (`d' and `e?'),
associated with the pair suggested by Le F\`evre \etal , but the low
signal--to--noise ratio makes these difficult to detect.  There is another
emission region (`f'), 7 arcsec away in the same direction; this is
probably too far from the radio axis to be considered part of the radio
source, but may be at the same redshift as the other components.

The J--band image of this galaxy presented by Eisenhardt and Chokshi
\shortcite{eis90} shows a similar morphology to the optical image for the
north--western three emission regions, whilst the south--eastern pair
weren't detected. Our K--band image (Figure~\ref{fig3c437}b) shows
emission from the region of the north--western components on the HST
image; this has a slight east--west extension but does not resolve
individual emission regions. There is a marginal detection of emission
from the south--eastern pair, whilst the possible companion, `f', is not
detected. A narrow--band [OII]~3727 image similarly shows only the
north--western emission region, again with a slight east--west extension
\cite{mcc88}.

\subsection*{3C441}

\begin{figure}
\centerline{
\psfig{figure=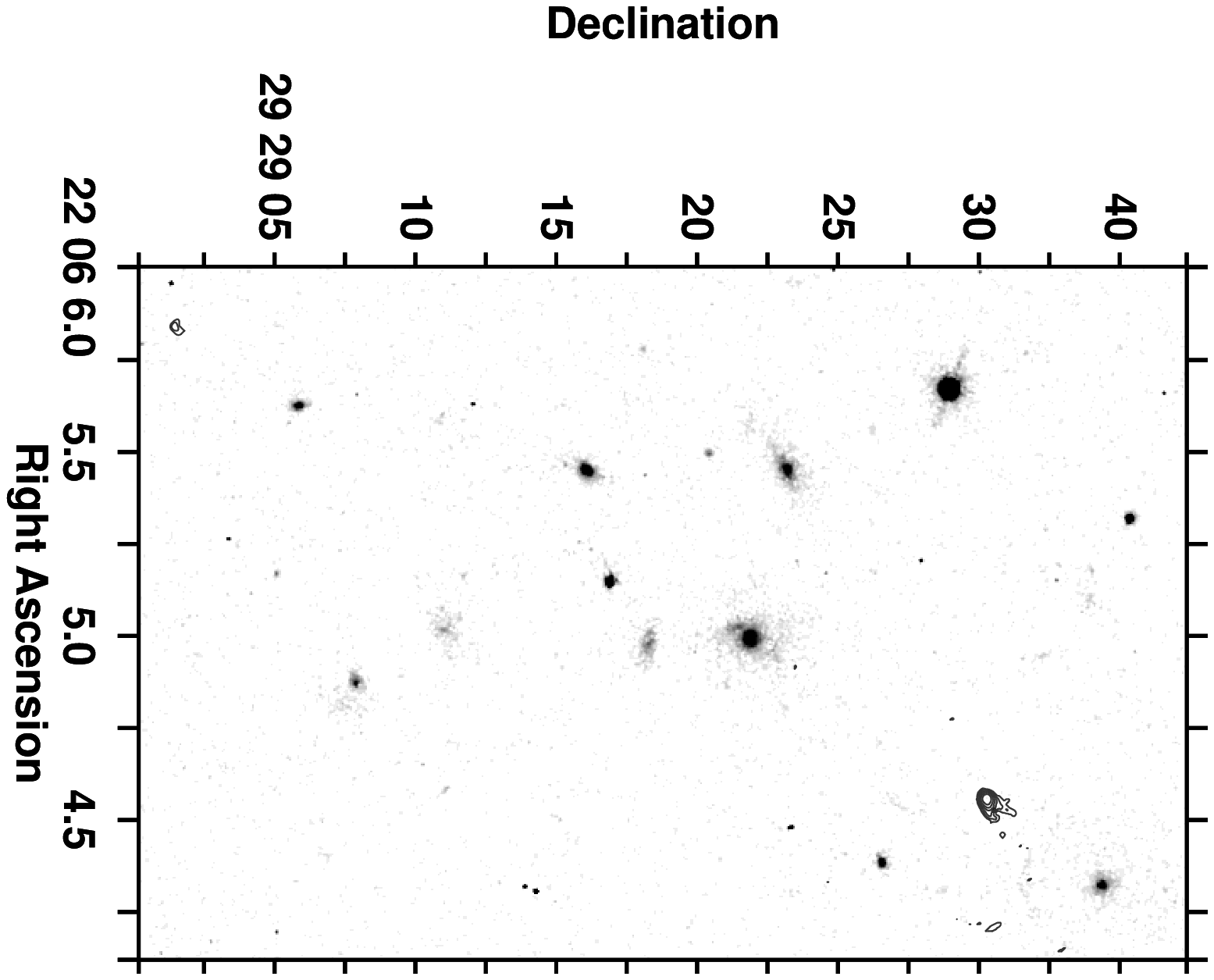,clip=,width=7.8cm,angle=90}
}
\caption{\label{fig3c441rad} The sum of the f555W and f785W HST images of
the radio galaxy 3C441. Overlaid are contours of the radio emission from
the 8.4~GHz A array VLA observation; contour levels are $(1,2,4,8,16,64)
\times 420 \mu$Jy beam$^{-1}$.}
\end{figure}

\begin{figure*}
\centerline{
\psfig{figure=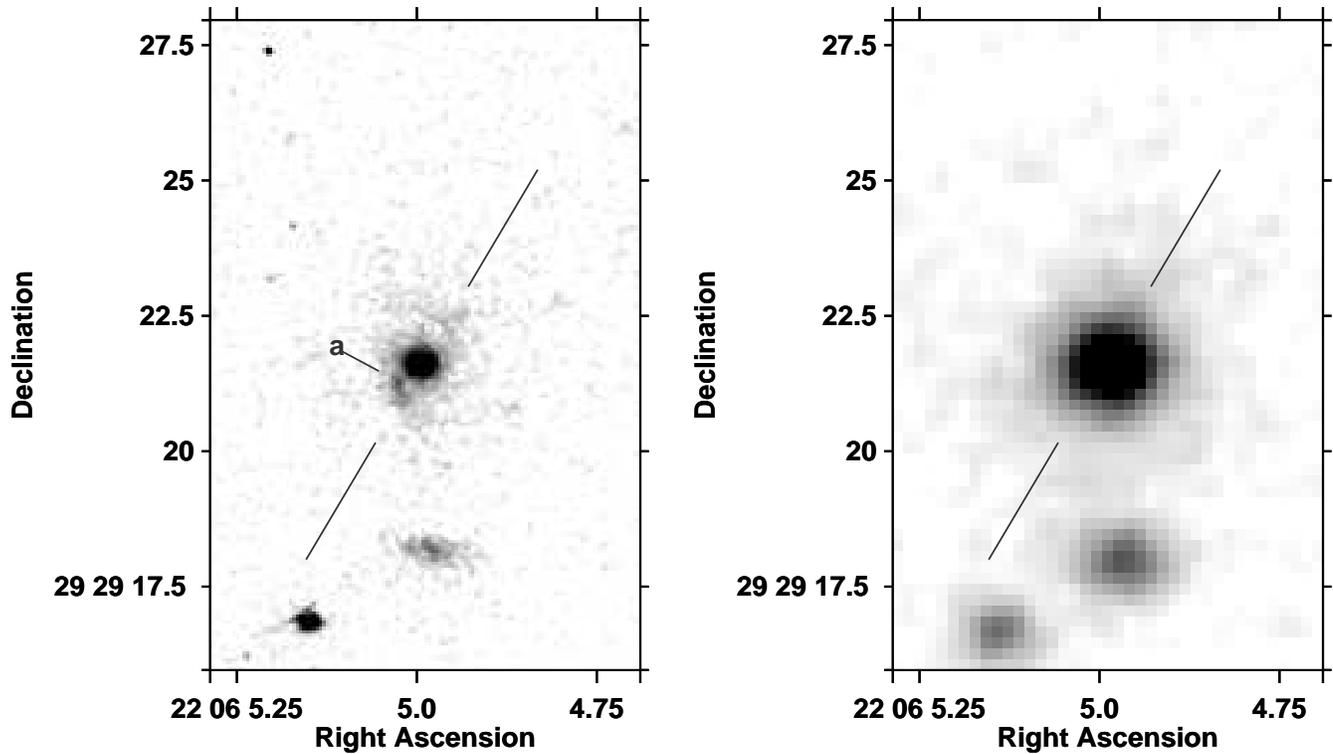,clip=,width=\textwidth}
}
\caption{\label{fig3c441} Enlarged images of the central regions of the
radio galaxy 3C441. (a) The sum of the f555W and f785LP HST images (b) The
UKIRT K--band image to the same scale. }
\end{figure*}

3C441 ($z = 0.707$, RA: 22 06 5.0, Dec: 29 29 22,
Figures~\ref{fig3c441rad},~\ref{fig3c441}) lies in a rather crowded field
and, although a narrow band [OII]~3727 image shows none of these
\cite{mcc88}, the infrared J$-$K colours of all but two of the galaxies
seen in Figure~\ref{fig3c441rad} lie in the range $1.6 \lta$ J$-$K $\lta
1.85$, consistent with them belonging to a cluster at redshift 0.7 (see
Figure~\ref{jkcols}). The only strong [OII] emission other than that of
the radio galaxy arises from a region 12 arcsec north and 8 arcsec west of
the host galaxy, slightly beyond the northern radio lobe. A galaxy (RA: 22
06 04.3, Dec: 29 29 39) can be seen just beyond this point in
Figure~\ref{fig3c441rad}, and has the same infrared colour as the host
radio galaxy. It is plausible that the [OII] emission is produced by an
interaction of the radio jet with this cluster galaxy (deeper radio maps
do show a distortion of the radio structures here) but further
observations would be needed to confirm this.

The HST image of the host galaxy itself (shown with contours of the radio
emission overlaid in Figure~\ref{fig3c441rad}, and shown enlarged in
Figure~\ref{fig3c441}a) reveals a small companion galaxy (`a') just to the
south--east of the central bright emission. This companion is bluer than
the host galaxy, with an f555W$-$f785LP colour of $2.34 \pm 0.22$ as
compared to $2.83 \pm 0.09$. There is also diffuse emission, slightly
elongated along the radio axis. The infrared image
(Figure~\ref{fig3c441}b) is nearly symmetrical, with a slight extension to
the south--west, corresponding to the position of galaxy `a'. The galaxy
has a very low optical polarisation, $P = 1.5 \pm 0.7\%$ measured with no
filter \cite{tad92}.

\subsection*{3C470}

\begin{figure*}
\centerline{
\psfig{figure=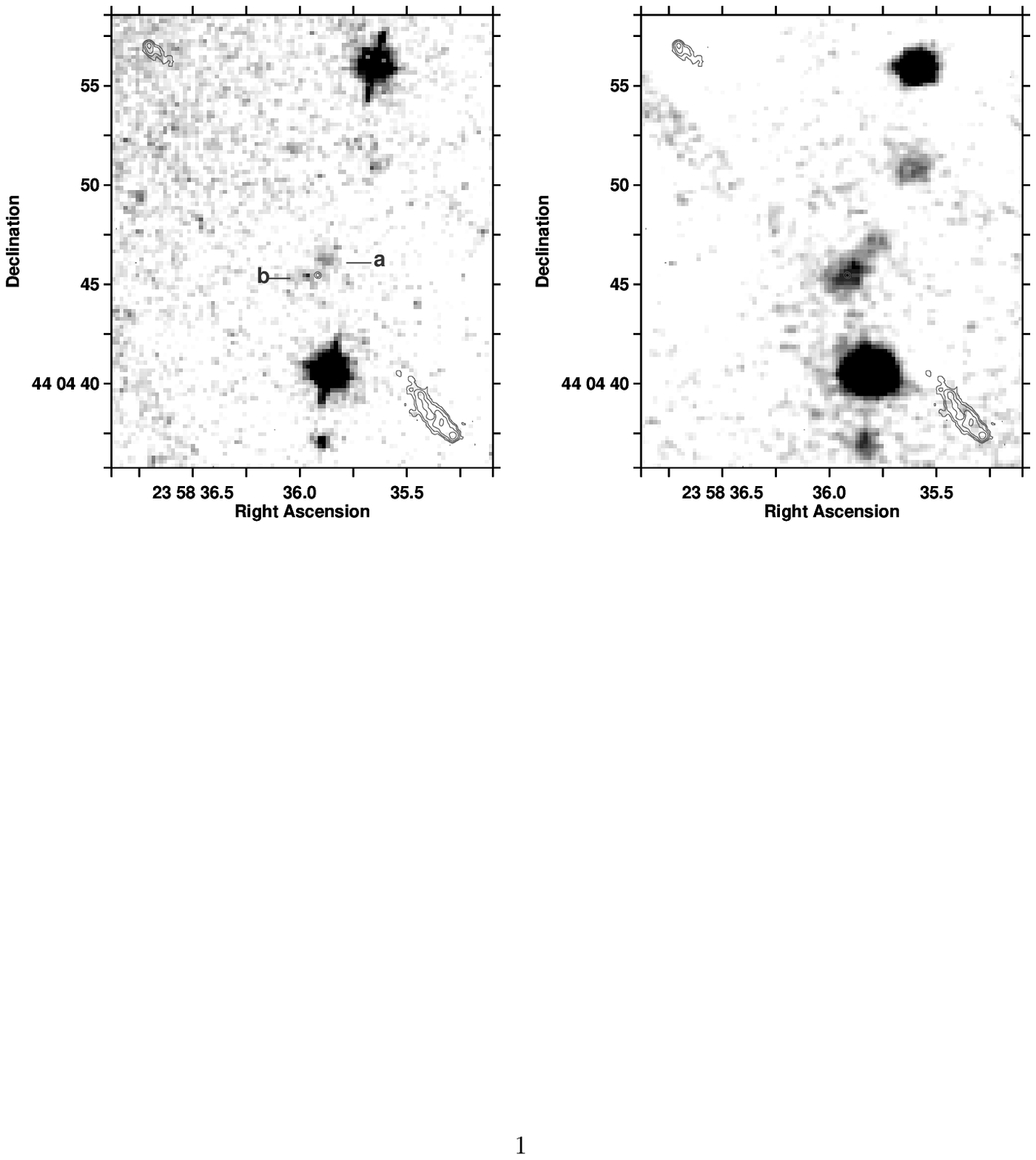,clip=,width=\textwidth}
}
\caption{\label{fig3c470} Images of the radio galaxy 3C470, overlaid with
contours of the radio emission as observed in the A and B array
observations using the VLA at 8.4~GHz. Contour levels are $(1,4,16,64)
\times 160 \mu$Jy beam$^{-1}$. (a) The f785LP HST image. The enhanced
background, particularly to the upper left of the image, is due to
scattered light from a bright star 40 arcsec to the north--east. (b) The
UKIRT K--band image. The `streak' appearing in the top left of the image
is a diffraction spike from the bright star.}
\end{figure*}

3C470 ($z = 1.653$, Figure~\ref{fig3c470}, RA: 23 58 35.9, Dec: 44 04
45.5) is the faintest galaxy in the sample. To make matters worse, a
bright star lies 40 arcsecs north--east of 3C470 which results in an
enhanced background covering the entire upper--left portion of the
image. The HST data have been averaged in blocks of 2 by 2 pixels and
smoothed with a 0.2 arcsec Gaussian profile to improve the
signal--to--noise ratio. The signal--to--noise ratio in the convolved
image is still low, but the galaxy is detected and is seen to consist of
two separate components, `a' and `b', which lie on either side of the
position of the radio core. Previous ground--based imaging \cite{fev88b}
had only shown the galaxy to be slightly extended in this direction. The
K--band observation shows a similar morphology to the HST image, with an
elongation equally misaligned from the radio axis
(Figure~\ref{fig3c470}b).  This elongation is misaligned from the radio
axis by 80 degrees, making this the most misaligned galaxy in the sample.

Being at high redshift, this source is one of the most powerful in the
sample. It possesses a reasonably bright radio core (1.2~mJy). The
south--western radio lobe is long and thin, and elongated along the radio
axis.

\section{Discussion}

We shall defer detailed analysis of the multi--wavelength properties of
the combined sample of sources to later papers in this series. Here, we
make only a few qualitative comments concerning the data presented.

The high resolution of the HST images has provided an unprecedented
opportunity to study the morphologies and colours of the distant 3CR radio
galaxies on sub--galactic scales, and show them to display a wide range of
morphologies. A small number of the galaxies, generally those with
redshifts $z \lta 0.8$, are fairly compact and symmetrical (prime examples
are 3C34 and 3C41), but the majority of the galaxies display some sort of
extended emission or alignment effect. The nature of this alignment is
remarkably varied: in some sources (eg. 3C226 and 3C441) it is associated
with a small companion, close to the host galaxy and positioned along the
radio axis --- in nearly all cases these companions are significantly
bluer than the host galaxy; in other sources (eg. 3C49 and 3C352) there is
evidence for only one central emission region, but this is asymmetrical
and elongated in the direction of the radio emission; finally, a number of
sources show spectacular optical morphologies, with bright strings of
knots tightly aligned along the radio axis (eg. 3C266, 3C324 and 3C368),
or large diffuse regions of emission surrounding the radio galaxy
(eg. 3C239 and 3C265). It is of note, however, that there are a number of
sources in the sample which display extended optical emission, but which
are significantly misaligned from the radio axis (eg. 3C217, 3C267). Any
model to describe the origin of the excess ultraviolet emission must take
account of these misaligned sources as well as those which are tightly
aligned. It is clear that a combination of different alignment mechanisms
will be required to explain all of the properties of these active
galaxies.

The infrared images presented do not show the same complexity as the HST
data, although their much lower angular resolution would wash out some of
the smaller--scale structures. The differences between the optical and
infrared images are not, however, simply a resolution effect. This can be
seen by examining the infrared images for some of the sources with the
most extended optical emission (eg. 3C265, 3C266, 3C324, 3C368): their
optical surface brightnesses remain high well beyond the radii at which
the infrared surface brightnesses are declining.

Whilst a small number of the infrared images do show an alignment effect,
particularly 3C68.2, 3C266 and 3C368, their elongation appears
substantially smaller than at optical wavelengths, even taking account of
the lower resolution of the infrared images. Those galaxies which don't
show a significant ultraviolet excess appear elliptical in the HST images,
consistent with models whereby the spectral energy distribution of these
sources is that of an old giant elliptical galaxy population plus a
variable--strength flat spectrum aligned component (see also Lilly and
Longair 1984, Lilly 1989, Rigler \etal\ 1992, Dunlop and Peacock 1993,
McCarthy 1993 and references
therein).\nocite{lil84a,mcc93,rig92,dun93,lil89} The K--band images are
generally dominated by the old stellar populations of the galaxies. Those
galaxies which do show infrared alignments are amongst the most distant in
the sample and possess large optical alignments: in these cases, the flat
spectrum aligned emission may make a significant contribution at
near--infrared wavelengths, accounting for the observed alignments. We
develop this point further in Paper II, where we investigate the infrared
emission of the galaxies to quantify the proportion that might be
associated with an aligned component, and compare their infrared radial
intensity profiles with de Vaucouleurs' law.

\section*{Acknowledgements}

This work is based on observations made with the NASA/ESA Hubble Space
Telescope, obtained at the Space Telescope Science Institute, which is
operated by AURA Inc., under contract from NASA.  The National Radio
Astronomy Observatory is operated by AURA Inc., under a co-operative
agreement with the National Science Foundation. The United Kingdom
Infrared Telescope is operated by the Royal Observatory Edinburgh. The
authors would like to thank the telescope operators, Thor Wold and Delores
Walther, and the scientific advisors, John Davies and Tom Geballe, for
advice and assistance during our two UKIRT runs. Thanks are also due to
the support scientist team at STScI, in particular Krista Rudloff. PNB
acknowledges support from PPARC. HJAR acknowledges support from an EU
twinning project, a programme subsidy granted by the Netherlands
Organisation for Scientific Research (NWO) and a NATO research grant. This
work was supported in part by the Formation and Evolution of Galaxies
network set up by the European Commission under contract ERB FMRX--
CT96--086 of its TMR programme.

\bibliography{pnb} 
\bibliographystyle{mn} 

\label{lastpage}
\end{document}